\documentclass{cernrep} 
\usepackage{texnames}
\usepackage[T1]{fontenc}
\def\dbar{{/\mkern-13mu{D}}}
\def\ga{\mathrel{\raise.3ex\hbox{$>$\kern-.75em\lower1ex\hbox{$\sim$}}}}
\def\la{\mathrel{\raise.3ex\hbox{$<$\kern-.75em\lower1ex\hbox{$\sim$}}}}

\begin{document}
\title{Higgs Physics}
 
\author{John Ellis}

\institute{Department of  Physics,  King's~College~London, London WC2R 2LS, United Kingdom;\\
Theory Division, CERN, CH-1211 Geneva 23, Switzerland}

\maketitle 

\begin{abstract}
These lectures review the background to Higgs physics, its current status following
the discovery of a/the Higgs boson at the LHC, models of Higgs physics
beyond the Standard Model and prospects for Higgs studies in future runs of the LHC and at
possible future colliders.\\
~\\
KCL-PH-TH/2013-49, LCTS/2013-36, CERN-PH-TH/2013-315

\end{abstract}
 
\section{Motivations and Context}

\subsection{To Higgs or not to Higgs?}

The Standard Model describes all the visible matter in the Universe in terms of a limited
number of fermionic constituents of matter: six quarks, three charged leptons
and three light neutrinos. It also comprises three fundamental gauge interactions between
these constituents, namely the electromagnetic, strong and weak forces, to which should
be added gravitation. The Standard Model is in good agreement with all confirmed
experimental results from particle accelerators. However, there was, until July 4th 2012,
one crucial missing ingredient: the origin of particle masses. Without a mass for the electron
there would be no atoms, as electrons would escape from nuclei at the speed of light, and
the weak interactions would not live up to their name: they would be stronger than
electromagnetism. So discovering the origin of particle masses has been a Big Deal.

In addition to elucidating the origin of particle masses and establishing whether a/the
Higgs boson exists, there are many important open questions beyond the Standard Model.
Why are there so many different types of matter particles? Why do weak interactions mix
them the way they do? How/why do they discriminate between matter and antimatter, and
might this difference explain the dominance of matter over antimatter in the Universe today?
What is the nature of the invisible dark matter that, according to astrophysicists and
cosmologists, dominates over the visible matter described by the Standard Model?
Are the fundamental forces unified, and what is the full quantum theory of gravity?

The key to answering many of these other questions may be provided by finding the
origin of particle masses. For example, decays of the Higgs boson may discriminate
between matter and antimatter, and might lie at the origin of the cosmological
matter-antimatter asymmetry. In many theoretical extensions of the Standard Model,
the Higgs boson is accompanied by other new particles, the lightest of which might
provide the astrophysical dark matter. A possible unified theory of all the elementary
particle interactions might employ a symmetry-breaking mechanism analogous to
that in the Standard Model, and the existence of any light `elementary' scalar boson
would pose a challenge for many quantum theories of gravity.

The equations describing the gauge interactions of the Standard Model do not discriminate 
between matter particles with the same quantum numbers, which differ only in their masses. 
The mass-generation mechanism must discriminate between these otherwise-identical
particles of matter and between the different force-carrying vector bosons: it must break 
the symmetries between them. One way to achieve this might be to break the symmetry
explicitly in the equations, but then the calculability of the theory would be lost. The
alternative is to retain symmetric equations but break the symmetry in their solutions.

The issue then is whether to break the symmetry throughout space, or via boundary
conditions. The latter is not possible in conventional three-dimensional space, since
it has no boundaries. However, it would be possible in theories with additional
dimensions of space: one could postulate different behaviours in the extra dimension(s)
for different particle species. The discovery of a/the Higgs boson at the LHC~\cite{ATLAS,CMS} has somewhat 
deflated interest in extra-dimensional models, unless their spectrum features a low-mass
excitation that resembles closely the Higgs boson of the Standard Model.

This discovery seems to mark the latest success of a long-running theoretical strategy
in particle physics: when in trouble, postulate one of more new particles. A partial list
includes reconciling quantum mechanics and special relativity (antimatter), nuclear
spectra (the neutron), the continuous spectrum in $\beta$ decay (the neutrino),
nucleon-nucleon interactions (the pion), the suppression of $\mu \to e \gamma$
(the second neutrino), flavour SU(3) ($\Omega^-$ and quarks), the suppression of
flavour-changing neutral currents (charm), CP violation (the third generation),
strong dynamics (the gluons), the discovery of the $\tau$ lepton (the $b$ and $t$
quarks), weak interactions (the $W^\pm$ and $Z^0$), and their renormalizability
(a/the Higgs boson)~\footnote{Outstanding examples include dark matter (the axion
or a WIMP?) and the fine-tuning problem (supersymmetry?). Depressed advocates
of supersymmetry should remember that it took 48 years to discover a/the Higgs boson,
whereas at the time of writing four-dimensional supersymmetric gauge theories are
`only' 40 years old.}.

The discovery of a/the Higgs boson marks the completion of the Holy Trinity of
particle types seen in Table~\ref{TabSMPartCont}. It has been known for decades that the only type of field
theory capable of making non-trivial over many magnitudes of energy is a
renormalizable one. Also, it has been a theorem for some 40 years that such a theory
could only contain (i) gauge vector bosons, (ii) spin-1/2 fermions, and (iii) scalar
bosons. Specification of a renormalizable theory is completed by choosing
(a) the gauge group, (b) the fermion representations, and (c) the scalar fields used
to realize the desired symmetry-breaking
pattern. We have long known that the answer (so far) to (a) is SU(3)$\times$SU(2)$\times$U(1)
and that the fermion representations are triplets of SU(3) and singlets and doublets of
SU(2). Finally we have an example of category (iii), a scalar boson, and there is strong
evidence that it is responsible for (c) electroweak symmetry breaking.

The first of these lectures describes the long road towards the discovery of
a/the Higgs boson, the second lecture describes the state of our knowledge
after Run 1 of the LHC, and the third lecture outlines some of the prospects for
future studies, including supersymmetric Higgs bosons and concepts for Higgs
factories.

\subsection{Summary of the Standard Model}

Table~\ref{TabSMPartCont} summarizes the particle content of the Standard Model~\cite{W,Salam,BCE,Boos}.
The electromagnetic and weak interactions are described by an SU(2)$_L\times$U(1)$_Y$ group~\cite{Glashow},
where the subscript $L$ reminds us that the weak SU(2) group 
acts only the left-handed fermions, and $Y$ is the hypercharge. The 
SU(2)$_L\times$U(1)$_Y$ part of the Standard Model Lagrangian may be written as
\begin{eqnarray}\label{EqSMLag}
 \nonumber
 \mathcal{L} &=& -\frac{1}{4} \mathbf{F}_{\mu\nu}^a \mathbf{F}^{a\mu\nu} \\ \nonumber
             &+& i \bar{\psi}  \dbar \psi + h.c. \\ \nonumber
             &+& \psi_i y_{ij} \psi_j \phi + h.c. \\
             &+& \lvert D_\mu \phi \rvert^2 - V\left(\phi\right) ~,
\end{eqnarray}
which is short enough to write on a T-shirt!

\begin{table}
 \label{TabSMPartCont}
 \centering
 \begin{tabular}{|c|c|}
  \hline
  \textbf{Gauge bosons} & \textbf{Scalar bosons} \\ 
  \hline
  $\gamma$, $W^+$, $W^-$, $Z^0$, $g_{1 \ldots 8}$ & $\phi$ (Higgs) \\
  \hline \hline
  \multicolumn{2}{c}{\textbf{Spin-1/2~Fermions}} \\
  \hline
  Quarks (each with 3 colour charges) & Leptons \\
  $\begin{array}{r}
   {\rm Charges}~+2/3: \\
   -1/3:
  \end{array} ~
  \left(\begin{array}{c}
   u \\ d       
  \end{array}\right) ~,
  \left(\begin{array}{c}
   c \\ s       
  \end{array}\right) ~,
  \left(\begin{array}{c}
   t \\ b       
  \end{array}\right)$ 
  &
  $\begin{array}{r}
   {\rm Charges}~~0: \\
   -1:
  \end{array} ~
  \left(\begin{array}{c}
   \nu_e \\ e^-       
  \end{array}\right) ~,
  \left(\begin{array}{c}
   \nu_\mu \\ \mu^-       
  \end{array}\right) ~,
  \left(\begin{array}{c}
   \nu_\tau \\ \tau^-       
  \end{array}\right)$ \\
  \hline
 \end{tabular}
 \caption{\it The particle content of the Standard Model with a minimal Higgs sector.}
\end{table}

The first two lines of (\ref{EqSMLag}) have been confirmed in many different
experiments with a high degree of accuracy. However, consistency with the
precision electroweak measurements made at LEP and other accelerators
agreed with the Standard Model if only if there was a relatively light Higgs boson
weighing $< 180$~GeV or so~\cite{LEPEWWG,Gfitter}. However, until July 2012 there was no direct experimental 
evidence for the last two lines~\cite{ATLAS, CMS}. One of the main objectives of the LHC was to discover 
whether they are right, need modification, or are simply wrong. The most important result of
the first run of the LHC has been to find some evidence that these lines do indeed
contain a grain of truth.

The first line in (\ref{EqSMLag}) is the kinetic term for the gauge sector of the electroweak theory, 
with $a$ running over the total number of gauge fields: 
three associated with SU(2)$_L$, which we call $B_\mu^1$, $B_\mu^2$, $B_\mu^3$, 
and one associated with U(1)$_Y$, which we call $A_\mu$. Their field-strength tensors are
\begin{eqnarray}
 F^a_{\mu\nu} &=& \partial_\nu B_\mu^a - \partial_\mu B_\nu^a + g \varepsilon_{bca} B_\mu^b B_\nu^c  \; \; {\rm for~a = 1,2, 3} ~;
 \label{EqFSTSU2} \\
 f_{\mu\nu} &=& \partial_\nu A_\mu - \partial_\mu A_\nu \label{EqFSTU1} ~.
\end{eqnarray}
In (\ref{EqFSTSU2}), $g$ is the coupling constant of the weak-isospin group SU(2)$_L$, 
and the $\varepsilon_{bca}$ are its structure constants. The last term in this equation stems 
from the non-Abelian nature of SU(2). The gauge fields are massless in the absence of any 
scalar fields, but we will see later how specific linear combinations of the four
electroweak gauge fields can acquire masses by spontaneous symmetry breaking induced
by a scalar field.  

The second line in Eq.~(\ref{EqSMLag}) describes the interactions between the matter fields 
$\psi$, described by Dirac equations, and the gauge fields via covariant derivatives.

The third line is the Yukawa sector and incorporates the interactions between the matter fields 
and the scalar field $\phi$ that is responsible for giving fermions their masses when 
electroweak symmetry breaking occurs. 

The fourth and final line is the engine room of the scalar sector. The first piece is the
kinetic term for $\phi$ with the covariant derivative defined here to be
\begin{equation}
 D_\mu = \partial_\mu + \frac{ig^\prime}{2} A_\mu Y + \frac{ig}{2} 
 \mathbf{\tau} \cdot \mathbf{B}_\mu ~,
\end{equation}
where $g^\prime$ is the U(1) coupling constant, and $Y$ and 
$\mathbf{\tau} \equiv \left(\tau_1, \tau_2, \tau_3\right)$ are the Pauli matrices that generate, respectively, 
U(1) and SU(2). The second piece of the final line of (\ref{EqSMLag}) is
the effective potential $V\left(\phi\right)$ constructed in such a way that its minimization gives rise to
a non-zero v.e.v. for the scalar field, and hence spontaneous electroweak symmetry breaking. 

\subsection{The (NG)AEBHGHKMP Mechanism}

This mechanism is often called the Higgs mechanism, but the history is quite complicated,
with many antecedents to the Higgs papers, with antecedents in condensed-matter physics, 
e.g., in the theories of superfluidity and superconductivity~\cite{SC}~\footnote{The wheel has turned full
circle with their importance for the construction of the LHC!}. Spontaneous global symmetry
breaking was introduced into four-dimensional, relativistic particle physics by Nambu~\cite{NJL},
providing insight into the lightness of the pion and spontaneous chiral symmetry breaking.
Subsequently, a simple
field-theoretical model was formulated by Goldstone~\cite{G}, and the appearance and number
of massless scalar bosons appearing in such theories was specified by a theorem proved in~\cite{GSW}.
In parallel, Nambu~\cite{N} and Anderson~\cite{A} had shown how to interpret superconductivity
in terms of  a spontaneously-broken local U(1) symmetry. Moreover, Anderson had conjectured
that this could occur also in the four-dimensional case, but Gilbert~\cite{Gilbert} argued in early 1964
that this would be impossible in a relativistic theory.

However, later in 1964 several papers successfully introduced spontaneously-broken 
local symmetry into particle physics. The initial paper
by Englert and Brout~\cite{EB} was followed a few weeks later by two papers 
written by Higgs, who did not know about their work at the
time. The first Higgs paper~\cite{H1} pointed out that Gilbert's objection to a four-dimensional extension of Anderson's
approach could be circumvented, and the second proposed a specific four-dimensional model
with a massive scalar boson~\cite{H2}. The subsequent paper by Guralnik, Hagen and Kibble~\cite{GHK} developed the ideas
proposed in these earlier papers. Also of note is a relatively-unknown 1965 paper by Migdal and Polyakov~\cite{MP},
which discusses the partial breaking of a local non-Abelian symmetry, ahead of the influential paper
of Kibble~\cite{K}. 

Of all these authors, Higgs was the only one who mentioned explicitly the existence of a
massive scalar boson (see equation (2b) of his second paper~\cite{H2}), and he went on to write a third
paper in 1966~\cite{H3} that discusses the properties of this `Higgs boson' in surprising detail including, e.g.,
its decays into massive vector bosons.

\subsection{Spontaneous Symmetry-Breaking in a U(1) Model}

As an illustration of spontaneous symmetry breaking~\cite{Quigg},
consider first a single complex scalar field $\phi = \phi_1 + \phi_2$ with the following effective potential:
\begin{equation}
\label{EqHiggsPotU1}
 V\left(\phi^\ast \phi\right) = \mu^2 \left(\phi^\ast \phi\right) + \lambda \left(\phi^\ast \phi\right)^2 ~,
\end{equation}
where $\mu^2$ and $\lambda > 0$ are real constants. 
This Lagrangian is clearly invariant under global U(1) phase transformations
\begin{equation}
 \phi \rightarrow e^{i\alpha} \phi ~,
\end{equation}
where $\alpha$ is a phase (rotation) angle. If the parameter $\mu^2$
in (\ref{EqHiggsPotU1}) is positive, there is a unique vacuum state with $\langle \phi \rangle = 0$. Perturbing 
around this vacuum reveals that, in this case, $\phi_1$ and $\phi_2$, have the same mass. The symmetry 
of the original Lagrangian is explicit in this case.

\begin{figure}[h!]
\centering
\includegraphics[scale=0.5]{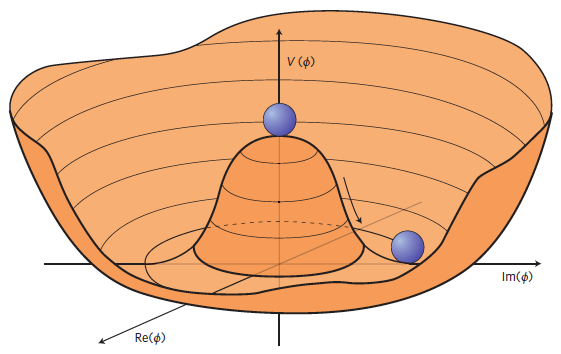}
\caption{\it An illustration of the Higgs potential (\protect\ref{EqHiggsPotU1}) in the case that
$\mu^2 <0$, in which case the minimum is at $|\phi|^2 = - \mu^2/(2 \lambda)$. Choosing any
of the points at the bottom of the potential breaks  spontaneously the rotational U(1) symmetry.}
\label{fig:higgspotential}
\end{figure}

Consider now the case $\mu^2 <0$, corresponding to the `Mexican Hat'
potential illustrated in Fig.\ref{fig:higgspotential}. When
we minimize the potential (\ref{EqHiggsPotU1}) we find a non-zero 
vacuum expectation value, or v.e.v., of the scalar field with:
\begin{equation}
 |\phi|^2 \; = \; \phi_1^2 + \phi_2^2 \; = \; - \frac{\mu^2}{2 \lambda} ~,
\end{equation}
and the phase $\alpha$ undetermined. Thus, when $\mu^2 < 0$ there is a set of equivalent
minima lying around a circle of radius $\sqrt{-\mu^2/\left(2\lambda\right)}$, and choosing one
of them breaks the rotational symmetry spontaneously.

The U(1) symmetry is now implicit, since it relates the different equivalent vacua,
corresponding to the appearance of spontaneous symmetry breaking.
In order to see the particle content, we choose, without loss of generality, a particular ground state
around which to perturb:
\begin{equation}
 \phi_{1,\text{vac}} = \sqrt{-\frac{\mu^2}{2 \lambda}} \equiv \frac{v}{\sqrt{2}} ~~~, ~~~~ \phi_{2,\text{vac}} = 0 ~.
\end{equation}
The perturbations may be parametrized by
\begin{equation}\label{EqPerturb}
\frac{ \eta}{\sqrt{2}} \equiv \phi_1 - \frac{v}{\sqrt{2}} ~~~, ~~~~ \frac{\xi}{\sqrt{2}} \equiv \phi_2 ~,
\end{equation}
so that $\phi = \left(v + \eta + i \xi\right)/\sqrt{2}$,
where $\eta$ and $\xi$ are real fields. In terms of these, the effective potential becomes
\begin{equation}
\label{EqLagGoldstone}
 \nonumber
 \mathcal{L} \; = \;- \frac{\mu^2}{2} \eta^2
                              - \frac{\lambda}{2} \left[ \left(v+\eta\right)^2 + \xi^2 \right]^2 - \mu^2 v \eta - \frac{\mu^2}{2} \xi^2 - \frac{1}{2} \mu^2 v^2 ~.
\end{equation}
The scalar particle corresponding to $\eta$ is massive 
with $m^2_\eta = -\mu^2 > 0$, whereas the scalar particle corresponding to $\xi$ is massless.

This particle is a prototype of a (Nambu-)Goldstone boson. It is massless because there is a
direction in field space, corresponding to changing the phase, in which the potential energy does
not change. Its appearance is a general feature of models with spontaneously-broken global symmetries,
as proven in~\cite{GSW}.
The total number of such massless particles corresponds in general to the number of field directions
in which the potential is flat. Nambu introduced this idea into particle physics in order to
describe the (relatively light) pion of QCD~\cite{NJL}, which he identified as a (pseudo-)Goldstone boson
of chiral symmetry that would have no mass if the up and down quarks were exactly massless.
The simple field-theoretical model is due to Goldstone~\cite{G}.

We now discuss how this spontaneous symmetry breaking of symmetry manifests
itself in the presence of a U(1) gauge field~\cite{EB,H2,GHK}. In order to construct a theory
that is invariant under local U(1) phase transformations, i.e.
\begin{equation}
 \phi \rightarrow e^{i\alpha\left(x\right)} \phi ~,
\end{equation}
we introduce a gauge field $\mathcal{A}_\mu$
that transforms under U(1)  as follows:
\begin{equation}
\mathcal{A}_\mu^\prime \rightarrow \mathcal{A}_\mu + \frac{1}{q} \partial_\mu \alpha\left(x\right) ~.
\label{gaugeU1}
\end{equation}
The space-time derivatives appearing in the kinetic term for the scalar field $\phi$
are replaced by covariant derivatives
\begin{equation}
 D_\mu = \partial_\mu + i q \mathcal{A}_\mu ~,
\end{equation}
where $q$ is the conserved charge. Including kinetic terms for both
the scalar field and the $\mathcal{A}_\mu$ field: $\left(1/4\right) F^{\mu\nu}F_{\mu\nu}$ where 
$F_{\mu\nu} \equiv \partial_\nu \mathcal{A}_\mu - \partial_\mu \mathcal{A}_\nu$, 
which is invariant under the U(1) gauge transformation (\ref{gaugeU1}), we have the Lagrangian
\begin{equation}\label{EqU1Lag2}
 \mathcal{L} = \left[\left(\partial_\mu - iq\mathcal{A}_\mu\right)\phi^\ast\right] \left[\left(\partial^\mu + iq\mathcal{A}^\mu\right)\phi\right]
             - V\left(\phi^\ast \phi\right) - \frac{1}{4} F^{\mu\nu}F_{\mu\nu} ~,
\end{equation}
which we now analyze.

We minimize the potential $V\left(\phi\right)$ as before,
and write the Lagrangian in terms of the perturbations around the ground state (\ref{EqPerturb}):
\begin{eqnarray}
 \label{L}
 \mathcal{L} 
  &=& \left\{
      \frac{1}{2} \left[ \left(\partial^\mu \eta\right) \left(\partial_\mu \eta\right) - \mu^2 \eta^2 \right]
      + \frac{1}{2} \left(\partial^\mu \xi\right) \left(\partial_\mu \xi\right) 
      -\frac{1}{4}F^{\mu\nu}F_{\mu\nu} + \frac{1}{2} q^2 v^2 \mathcal{A}^\mu \mathcal{A}_\mu \right\} \nonumber \\ \nonumber
  &+& v q^2 {A}^\mu \mathcal{A}_\mu \eta  + \frac{q^2}{2} \mathcal{A}^\mu \mathcal{A}_\mu \eta^2 + q \left(\partial^\mu \xi\right) \mathcal{A}_\mu \left(v+\eta\right) 
      - q \left(\partial^\mu \eta\right) \mathcal{A}_\mu \xi \\ 
  &-& \mu^2 v \eta - \frac{\mu^2}{2} \xi^2 - \frac{\lambda}{2} \left[ \left(v+\eta\right) + \xi^2 \right]^2 - \frac{\mu^2 v}{2} ~.
\end{eqnarray}
As before, the first three terms describe a (real) scalar particle, $\eta$, with
mass $\sqrt{-\mu^2}$ and a massless Goldstone boson, $\xi$. The fourth term describes the free U(1)
gauge field. However, whereas previously the Lagrangian (\ref{EqU1Lag2})
apparently described a massless gauge boson field, we now see in the spontaneously-broken
phase (\ref{L}) a term proportional to $\mathcal{A}_\mu \mathcal{A}^\mu$, corresponding to a mass for the gauge field:
\begin{equation}
 m_\mathcal{A} = q v ~,
\end{equation}
that is proportional to the vacuum expectation value of the Higgs field.

The other terms in (\ref{L}) 
describe couplings between the fields $A^\mu, \eta$ and $\xi$,
including a bilinear interaction coupling $\propto A^\mu \partial_\mu \xi$.
The correct particle interpretation of
(\ref{L}) is obtained by diagonalizing the bilinear terms, which 
is easily done by using the gauge freedom of $\mathcal{A}_\mu$ to replace
\begin{equation}
 \mathcal{A}_\mu \rightarrow \mathcal{A}_\mu^\prime = \mathcal{A}_\mu + \frac{1}{qv} \partial_\mu \xi ~,
\end{equation}
and making the local phase transformation
\begin{equation}
 \phi \rightarrow \phi^\prime = e^{-i\xi\left(x\right)/v} \phi = \frac{v + \eta}{\sqrt{2}} ~.
\end{equation}
Following this transformation, the field $\xi$ disappears, and
(\ref{L}) takes the simple form
\begin{equation}
\label{eaten}
 \mathcal{L} = \frac{1}{2} \left[ \left(\partial^\mu \eta \right) \left(\partial_\mu \eta \right) - \mu^2 \eta^2 \right] 
               - \frac{1}{4} F^{\mu\nu} F_{\mu\nu} + \frac{q^2 v^2}{2} \mathcal{A}^{\mu ~ \prime} \mathcal{A}_\mu^\prime + \ldots ~.
\end{equation}
where the $\ldots$ represent trilinear and quadrilinear interactions.

The Goldstone boson $\xi$
that appeared when the global U(1) symmetry was broken spontaneously by the choice of
ground state when $\mu^2 < 0$ has been absorbed (or `eaten') by the gauge field 
$\mathcal{A}_\mu$, which thereby acquired a mass. Remember
that, whereas a massless gauge boson has only two degrees of freedom (transverse polarization
states), a massive gauge boson has a third (longitudinal) polarization
state that is supplied by the Goldstone boson of the spontaneously-broken
U(1) global symmetry. This is the
Englert-Brout-Higgs mechanism.

In order for this mechanism to work, the magnitude of the v.e.v. of the scalar field must be
fixed dynamically, which occurs in this model because the potential varies non-trivially
in the radial ($|\phi|)$ direction. The mass term for the $\eta$ field in (\ref{eaten}) is a reflection of this
variation in the potential. The appearance of such a massive scalar boson is an unavoidable
signature of spontaneous symmetry breaking.

\subsection{Spontaneous Symmetry Breaking in the Standard Model}

As already mentioned, the gauge group of the Standard Model is SU(2)$_L\times$U(1)$_Y$~\cite{Glashow,W, Salam},
and the Lagrangian can be written in the form
\begin{eqnarray}
 \mathcal{L} &=& \mathcal{L}_{\text{gauge}} + \mathcal{L}_{\text{leptons}}  \nonumber \\
 \mathcal{L}_{\text{gauge}} &=& -\frac{1}{4} F_{\mu\nu}^a F^{a\mu\nu} - \frac{1}{4} f_{\mu\nu} f^{\mu\nu}  \nonumber \\
 \mathcal{L}_{\text{leptons}} &=& \overline{\text{R}} \left( \partial_\mu + i\frac{g^\prime}{2} \mathcal{A}_\mu Y \right) \text{R}
                                  + \overline{\text{L}} i\gamma^\mu \left( \partial_\mu + i \frac{g^\prime}{2} \mathcal{A}_\mu Y 
                                    + i \frac{g}{2} \mathbf{\tau} \cdot \mathbf{B_\mu} \right) \text{L} ~,
\label{LSM}
\end{eqnarray}
where the field-strength tensors, $F_{\mu\nu}$ and $f_{\mu\nu}$, were defined in (\ref{EqFSTSU2}) and (\ref{EqFSTU1}), 
respectively, $g$ is the SU(2) coupling and $g^\prime$ is the U(1) hypercharge coupling. The symbol $\text{L}$
represents doublets of left-handed fermions and $\text{R}$ represents right-handed fermions.
As written in (\ref{LSM}), the theory contains four massless bosons 
($\mathcal{A}_\mu$, $B_\mu^1$, $B_\mu^2$, $B_\mu^3$). 

We now introduce a scalar field that is a complex doublet of SU(2)~\cite{W,Salam}:
\begin{equation}
 \phi = 
 \left(\begin{array}{c}
  \phi^+ \\
  \phi^0 
 \end{array}\right) ~,
 \label{half}
\end{equation}
and add to the Lagrangian
\begin{equation}\label{EqLagHiggs}
 \mathcal{L}_{\text{Higgs}} = \left(D_\mu \phi\right)^\dagger \left(D^\mu \phi\right) - V\left(\phi^\dagger \phi\right) ~,
\end{equation}
with an effective potential of similar form to (\ref{EqHiggsPotU1}):
\begin{equation}\label{EqScalarPot}
 V\left(\phi^\dagger \phi\right) = \mu^2 \left( \phi^\dagger \phi \right) + \lambda \left(\phi^\dagger \phi\right)^2 ~,
\end{equation}
with $\mu^2 < 0$ and $\lambda > 0$. We also include Yukawa interactions between this scalar field and the matter fermions:
\begin{equation}\label{EqYukawa}
 \mathcal{L}_{\text{Yukawa}} = 
 - G_e \left[ \overline{\text{R}} \phi^\dagger \text{L} + \overline{\text{L}} \phi \text{R} \right] ~,
\end{equation}
which yield masses for the matter fermions, as we see later.

As in the U(1) case when $\mu^2 < 0$, $\langle \phi \rangle = 0$ is an unstable local maximum of the effective potential.
The minimum has $\langle \phi \rangle \ne 0$ with an arbitrary SU(2)$\times$U(1) orientation, 
leading to spontaneous symmetry breaking. 
Minimizing the effective potential as in the U(1) case, we obtain
\begin{equation}
 \frac{\partial}{\partial\left(\phi^\dagger \phi\right)} V\left(\phi^\dagger \phi\right)
 = \mu^2 + 2 \lambda \langle\phi\rangle_0 = \mu^2 + 2 \lambda \left[ \left(\phi_{\text{vac}}^+\right)^2 + \left(\phi_{\text{vac}}^0\right)^2 \right] = 0 ~.
\end{equation}
Without loss of generality, we may set $\phi_{\text{vac}}^+ = 0$ and take
$\phi_{\text{vac}}^0 = \sqrt{-\mu^2/\left(2\lambda\right)}$. This choice breaks
both the SU(2) and U(1) symmetries, but preserves
invariance under a residual U(1) gauge symmetry that we may identify with electromagnetism.
Since three of the four generators are broken spontaneously by the v.e.v., the spectrum of the global
theory would contain three massless (Nambu-)Goldstone bosons.

To see how these are `eaten' by three of the gauge bosons,
we consider perturbations around the chosen vacuum, representing the
scalar field as
\begin{equation}
 \phi = \exp\left(\frac{i \mathbf{\xi} \cdot \mathbf{\tau}}{2v}\right) 
        \left(\begin{array}{c}
         0 \\
         \left( v + \eta \right) / \sqrt{2}
        \end{array}\right) ~.
\end{equation}
Just as in the U(1) case discussed in the previous section where we rotated away the Goldstone boson $\xi$, 
we are able in this case to make the following gauge transformation on the scalar $\phi$ and the gauge and matter fields:
\begin{eqnarray}
 \phi &\rightarrow& \phi^\prime 
 = \exp\left(\frac{-i \mathbf{\xi} \cdot \mathbf{\tau}}{2v}\right) \phi 
 = \left(\begin{array}{c}
    0 \\
    \left( v + \eta \right) / \sqrt{2}
   \end{array}\right) ~. \\
 \mathbf{\tau} \cdot \mathbf{B}_\mu &\rightarrow& \mathbf{\tau} \cdot \mathbf{B}_\mu^\prime \\
 \text{L} &\rightarrow& \text{L}^\prime = \exp\left(\frac{-i \mathbf{\xi} \cdot \mathbf{\tau}}{2v}\right) \text{L} ~,
\end{eqnarray} 
where $\tau$ is an SU(2) matrix, which leaves $\mathcal{A}_\mu$ and R invariant.

In this unitary gauge, and henceforward simplifying the notation: $\phi^\prime \to \phi$, etc.,
so that
\begin{equation}
\label{defeta}
\phi = 
 \left(\begin{array}{c}
  0 \\
  \left(v + \eta\right)/\sqrt{2}
 \end{array}\right) ~,
\end{equation}
and
\begin{equation}
 \phi^\dagger \phi = \left( \frac{v+\eta}{\sqrt{2}} \right)^2 ~,
\end{equation}
we see that
\begin{equation}
 V\left( \phi^\dagger \phi \right)
 = \mu^2 \left( \frac{v+\eta}{\sqrt{2}} \right)^2 + \lambda \left( \frac{v+\eta}{\sqrt{2}} \right)^4 ~.
\end{equation}
We also have
\begin{equation}
 D_\mu \phi = \partial_\mu \phi + \frac{ig^\prime}{2} \mathcal{A}_\mu Y \phi + \frac{ig}{2} \mathbf{\tau} \cdot \mathbf{B}_\mu \phi ~,
\end{equation}
which may be written in the form
\begin{equation}
 D_\mu \phi =
 \left(\begin{array}{c}
  \frac{ig}{2} \left(\frac{v+\eta}{\sqrt{2}}\right) \left(B_\mu^1 - iB_\mu^2\right) \\
  \frac{1}{\sqrt{2}} \partial_\mu \eta + \left(\frac{v+\eta}{\sqrt{2}}\right) \frac{i}{2}\left( ig^\prime \mathcal{A}_\mu - ig B_\mu^3 \right)
 \end{array}\right) 
\end{equation}
and hence
\begin{equation}
 \left(D^\mu\phi\right)^\dagger \left(D_\mu \phi\right) =
 \frac{g^2}{8} \left(v+\eta\right)^2 \lvert B_\mu^1 - i B_\mu^2 \rvert ^2 
 + \frac{1}{2} \left(\partial_\mu \eta\right) \left(\partial^\mu \eta\right)
 + \frac{1}{8} \left(v+\eta\right)^2 \left( g^\prime \mathcal{A}_\mu - g B_\mu^3 \right)^2 ~.
\end{equation}
The final form of the scalar Lagrangian is therefore
\begin{eqnarray}\label{EqLagHiggsFinal}
 \mathcal{L}_{\text{Higgs}} &=&
 \left\{
 \frac{1}{2} \left(\partial_\mu \eta\right) \left(\partial^\mu \eta\right) - \frac{\mu^2}{2} \eta^2
 + \frac{v^2}{8} \left[ g^2 \lvert B_\mu^1 - i B_\mu^2 \rvert^2 + \left( g^\prime \mathcal{A}_\mu - g B_\mu^3 \right)^2 \right]
 \right\} \nonumber \\
 &+&
 \left\{
 \frac{1}{8} \left(\eta^2+2v\eta\right) \left[ g^2 \lvert B_\mu^1 - i B_\mu^2 \rvert^2  
                                               + \left( g^\prime \mathcal{A}_\mu - g B_\mu^3 \right)^2 \right] \right\} \nonumber \\
 &-& \left\{ \frac{1}{4} \eta^4 - \lambda v \eta^3 - \frac{\mu^2}{2} \eta^2 - \left( \lambda v^3 + \mu^2 v \right) \eta
 - \left( \frac{\lambda v^4}{4} + \frac{\mu^2 v^2}{2} \right)
 \right\} ~,
\end{eqnarray}
whose interpretation we now discuss.

The second term on the first line of (\ref{EqLagHiggsFinal}) is a mass term for the $\eta$ field: this is the Higgs boson,
which appears in the same way as in the previous U(1) case.
{\it A priori}, there is no theoretical prediction within the Standard Model for the Higgs mass
\begin{equation}
 m_H = - 2 \mu^2 ~,
\end{equation}
since $\mu$ is not determined by any of the known parameters of the Standard Model.
The following terms on the first line of (\ref{EqLagHiggsFinal}) are mass terms
for the massive vector bosons, to which we return later. The second line includes
interactions of the Higgs boson with these massive gauge bosons, and the last line describes 
self-interactions of the Higgs boson.

We define the charged gauge fields $W_\mu^{\pm}$ as the combinations
\begin{equation}
 W_\mu^{\pm} \equiv \frac{B_\mu^1 \mp i B_\mu^2}{\sqrt{2}} ~,
\end{equation}
and identify the following neutral gauge boson mass eigenstates:
\begin{eqnarray} 
 Z_\mu &\equiv& \frac{-g^\prime \mathcal{A}_\mu + g B_\mu^3}{\sqrt{g^2+g^{\prime~2}}} \label{EqDefA} ~, \\ 
 A_\mu &\equiv& \frac{g \mathcal{A}_\mu + g^\prime B_\mu^3}{\sqrt{g^2+g^{\prime~2}}} \label{EqDefZ} ~.
\end{eqnarray}
Substituting these expressions in the Lagrangian (\ref{EqLagHiggsFinal}), we find
\begin{eqnarray}
 \mathcal{L}_{\text{Higgs}}
 = \left[\frac{1}{2} \left( \partial^\mu \eta \right) \left( \partial_\mu \eta \right) - \frac{\mu^2}{2} \eta^2\right] 
   &+& \frac{v^2 g^2}{8} W^{+ ~ \mu}W^+_\mu + \frac{v^2 g^2}{8} W^{- ~ \mu}W^-_\mu 
   + \frac{\left( g^2 + g^{\prime ~ 2} \right) v^2}{8} Z^\mu Z_\mu \nonumber \\
   &+& ... ~,
\end{eqnarray}
and it is evident that the field $\mathcal{A}_\mu$ is massless. This is due to the unbroken 
U(1) symmetry (i.e., the symmetry under $e^{iQ\alpha\left(x\right)}$ rotations) that we identify
with electromagnetism. On the other hand, the charged
vector bosons $W^\pm$ and the neutral vector boson $Z^0$ have masses 
\begin{equation}
 m_W = \frac{gv}{2} ~~~, ~~~~ m_Z = \frac{v}{2} \sqrt{g^2 + g^{\prime ~ 2}} ~.
\end{equation}
We see that the couplings of the Higgs boson to the
$W^\pm$ and $Z^0$ in (\ref{EqLagHiggsFinal}) are related to these masses.
They are related through
\begin{equation}
 m_Z = m_W \sqrt{1 + g^{\prime ~ 2} / g^2} ~.
 \label{MWMZ}
\end{equation}
It is convenient to introduce the angle $\theta_W$ to parametrize the mixing of the neutral gauge bosons, defined by
\begin{equation}
 \tan \theta_W \; = \; \frac{g^\prime}{g} ~,
\end{equation}
so that
\begin{equation}
 \cos\theta_W = \frac{g}{\sqrt{g^2+g^{\prime ~ 2}}} ~~~, ~~~~
 \sin \theta_W = \frac{g^\prime}{\sqrt{g^2+g^{\prime ~ 2}}} ~.
\end{equation}
Eqs.~(\ref{EqDefA}) and (\ref{EqDefZ}) can then be written as
\begin{eqnarray}
 Z_\mu &=& -\sin \theta_W \mathcal{A}_\mu + \cos \theta_W B_\mu^3 ~,\\
 A_\mu &=& \cos \theta_W \mathcal{A}_\mu + \sin \theta_W B_\mu^3 ~,
\end{eqnarray}
and the relation (\ref{MWMZ}) between the masses of $W^\pm$ and $Z^0$ becomes
\begin{equation}
 m_W = m_Z \cos \theta_W ~.
\end{equation}
The ratio
\begin{equation}\label{EqRhoDef}
 \rho \; \equiv \; \frac{m_W^2}{m_Z^2 \cos^2 \theta_W}
\end{equation}
is equal to unity at the tree level in the Standard Model. This is a direct consequence of
the choice of isospin 1/2 for the Higgs field (\ref{half}).

This choice also enables the Higgs field to give masses to the Standard Model fermions~\cite{W},
as we now discuss.
Looking at the fermion Lagrangian (\ref{EqYukawa}) in the unitary gauge, it becomes
\begin{equation}\label{EqLagYukawa}
 \mathcal{L}_{\text{Yukawa}}
 = - G_e \left[ \overline{e}_R \phi^\dagger 
   \left(\begin{array}{c}
    \nu_L \\
    e_L
   \end{array}\right)
   +
   \left( \overline{\nu}_L ~ \overline{e}_L \right) \phi e_R \right]
 = - G_e \frac{v+\eta}{\sqrt{2}} \left( \overline{e}_R e_L + \overline{e}_L e_R \right) ~.
\end{equation}
In terms of $\overline{e} \equiv \left( \overline{e}_R,  \overline{e}_L \right)$ 
and $e \equiv \left( e_L,  e_R \right)^T$, (\ref{EqLagYukawa}) becomes
\begin{equation}
 \mathcal{L}_{\text{Yukawa}} = -\frac{G_e v}{\sqrt{2}} \overline{e}e - \frac{G_e \eta}{\sqrt{2}} \overline{e}e ~,
\end{equation}
and we see that the electron has a mass
\begin{equation}
 m_e = G_e \frac{v}{\sqrt{2}} ~.
\end{equation}
The same holds for all the Standard Model fermions, and their couplings to the Higgs
boson are proportional to their masses. 
Thus, the Higgs boson prefers to decay into the heaviest
fermions $f$ that are kinematically accessible, i.e., have $m_f < m_H/2$.

At first sight, the Lagrangian (\ref{EqLagHiggsFinal}) may look rather artificial.
However, as described above, the spontaneous symmetry breaking mechanism
for giving masses to vector bosons and fermions is in fact very generic.
It suffices to have a scalar field $\phi$ and choose the coefficient $\mu^2$
of the quadratic term in its effective potential to be negative, and the
mass-generation mechanism follows automatically. The original Lagrangian
(\ref{EqLagHiggs}) is in fact very symmetric, and this symmetry is still present,
though hidden, in (\ref{EqLagHiggsFinal}). 

\begin{figure}[h!]
\centering
\includegraphics[scale=0.5]{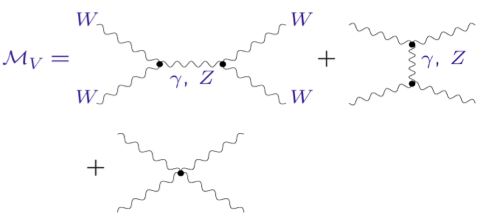}
\includegraphics[scale=0.5]{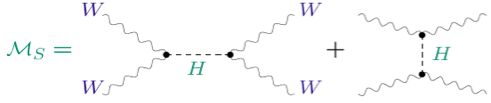}
\caption{\it Contributions to $WW$ scattering from diagrams due to vector boson exchange and the four-point gauge
interaction in the absence of a Higgs boson (upper two rows), and
diagrams due to Higgs boson exchange (lowest row).}
\label{fig:WWscattering}
\end{figure}

Since the theory (\ref{EqLagHiggsFinal}) still possesses symmetry, it is a
renormalizable theory at the quantum level~\cite{tHV}, which enables many detailed
calculations to be compared with precise experimental measurements. In
fact, not only is it {\it a} renormalizable theory, it is {\it the only} way to
construct a renormalizable theory of interacting massive vector bosons~\cite{CLT,B,CLS}.
In order to get some flavour why this is the case, consider $W^+ W^- \to W^+ W^-$
scattering. At the tree level, the combination of $\gamma$ and $Z^0$ exchanges
in the direct and crossed channels with the point-like quartic coupling, shown in
the two upper rows of Fig.~\ref{fig:WWscattering}, yields
a scattering amplitude that grows quadratically with energy:
\begin{equation}
{\cal M}_V \; = - g^2 \frac{E^2}{m_W^2} + {\cal O}(E^0) \, .
\label{E2}
\end{equation}
This is a problem when one calculates loop diagrams, since the integral
over the loop momenta is uncontrollably divergent. If one now includes
a scalar with a coupling $g_{HWW}$ to the vector bosons, the direct and
crossed-channel scalar exchanges, shown in
the bottom row of Fig.~\ref{fig:WWscattering}, yield an amplitude
\begin{equation}
{\cal M}_S \; = + g^2_{HWW} \frac{E^2}{m_W^2} + {\cal O}(E^0) \, .
\label{SE2}
\end{equation}
Adding these contributions, we find
\begin{equation}
{\cal M} \; = \; {\cal M}_V + {\cal M}_S \; = \; \frac{m_H^2}{v^2} \left( 2 + \frac{m_H^2}{s - m^2_H} + \frac{m_H^2}{t - m^2_H} \right) + \dots \, ,
\label{ME2}
\end {equation}
where the $\dots$ represent terms that are subdominant at high energies, {\it iff}
the $HWW$ coupling coincides with the prediction from spontaneous symmetry
breaking. The fact that the resultant amplitude ${\cal M}$ is asymptotically constant
for this particular choice of $g_{HWW}$ ensures that the integration over the loop momentum
is controllable and permits the theory to be renormalizable~\footnote{A similar
argument applies to fermion-$W$ scattering and the scalar-fermion coupling, which
must also coincides with the prediction from spontaneous symmetry
breaking.} - which it is if the other
particle couplings also coincide with the spontaneously-broken gauge theory.

\subsection{A Phenomenological Profile of the Higgs Boson}

\subsubsection{Before the LHC}

For a decade after the original papers after spontaneously-broken gauge
theories were formulated, and even for several years after they were proven
to be renormalizable, rather few people took seriously Higgs' prediction of the boson
that bears his name. Indeed, only a handful of papers discussed its possible
experimental signatures before the paper that Mary Gaillard, Dimitri Nanopoulos
and I wrote in 1975~\cite{EGN} with the same title as this Section heading. 
In the last sentence of our paper we wrote that ``{\it we do not want to encourage big 
experimental searches for the Higgs boson}". Fortunately, the experimental community did not pay attention to this caveat,
and the ATLAS and CMS experiments announced the discovery of a candidate for a (the?) Higgs boson
on July 4th 2012~\cite{ATLAS,CMS}.

\begin{figure}[h!]
\centering
\includegraphics[scale=0.6]{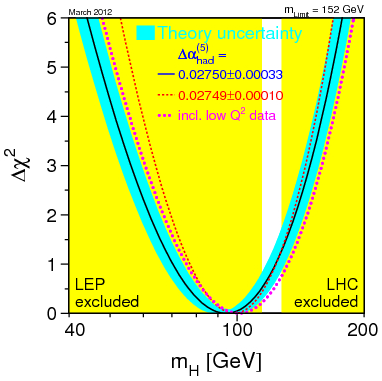}
\caption{\it The status of the Higgs search in March 2012~\protect\cite{LEPEWWG}. The left-hand yellow-shaded region
is the LEP exclusion, and the right-hand yellow-shaded region is the Tevatron exclusion at that time~\protect\cite{TevH}.}
\label{fig:BlueBand}
\end{figure}

The search for the Higgs boson at LEP was advertized in the first survey of
LEP physics made in 1976~\cite{EG} and featured strongly in the subsequent LEP
experimental programme~\cite{LEPH}. These direct searches for the Higgs boson
resulted in the lower limit~\cite{LEPlimit}
\begin{equation}
m_H \; > \; 114.4~{\rm GeV} \, ,
\label{LEP}
\end{equation}
shown as the left-hand of the two yellow excluded regions in Fig.~\ref{fig:BlueBand}.
The search for the Higgs boson was also advertized at the first LHC physics workshop
in 1984 and grew subsequently to become one of the major objectives of the
LHC experimental programme.

Although LEP found no direct evidence for the Higgs boson, the precision of
electroweak measurements at LEP and elsewhere provided indirect indications
on the Higgs mass through the sensitivity of electroweak observables
to quantum loop corrections. For example, there is a one-loop correction
$\Delta r$ to the $W^\pm$ and $Z^0$ masses:
\begin{equation}
m_W^2 \sin^2 \theta_W \; = \; m_Z^2 \cos^2 \theta_W \sin^2 \theta_W \; = \; \frac{\pi \alpha}{\sqrt{2} G_F} \left( 1 + \Delta r \right) \, .
\label{Deltar}
\end{equation}
The one-loop correction $\Delta r$ and other electroweak 
radiative corrections are in turn sensitive to the masses of heavy virtual
particles, in particular $m_t$ and $m_H$:
\begin{equation}
\propto \; \frac{3 G_F}{8 \pi^2 \sqrt{2}} m_t^2 \; , \; \frac{\sqrt{2} G_F}{16 \pi^2} \left( \frac{11}{3} \ln \frac{m_H^2}{m_Z^2} + \dots \right)
\label{mtmH}
\end{equation}
in the limits of large $m_t$ and $m_H \gg m_{W,Z}$. Note that the sensitivity to $m_t$ is quadratic~\cite{quadV},
whereas that to $m_H$ is only logarithmic~\cite{logV}. The large-mass divergences
in (\ref{mtmH}) reflect the fact that the electroweak theory would become
renormalizable if these particles were absent. In the case of the top quark, its
absence would leave us with an incomplete fermion doublet, and the problems
arising in the absence of the Higgs boson were discussed at the end of
the previous Section.

First attempts to use precision measurements to constrain $m_H$ were made
before the discovery of the top quark~\cite{EF}, and already indicated that $m_H = {\cal O}(m_W)$.
The discovery of the top quark with a mass consistent with predictions based on
electroweak data and (\ref{Deltar}) enabled the prediction of $m_H$ to be
sharpened, as has the inclusion of QCD and higher-order electroweak effects,
as illustrated by the blue band in Fig.~\ref{fig:BlueBand}. 

A conservative estimate of the current estimate of $m_H$ on the
basis of precision electroweak data alone is~\cite{LEPEWWG,Gfitter}
\begin{equation}
m_H \; = \; 100 \pm 30~{\rm GeV} \, ,
\label{estmH}
\end{equation}
which is quite compatible with the direct lower limit (\ref{LEP}) and the exclusion by
the Tevatron of $m_H \sim 160$ to 170~GeV. Combining the LEP and Tevatron exclusions
with the precision electroweak data led in mid-2011 to the prediction~\cite{Gfitter2011}:
\begin{equation}
m_H \; = \; 125 \pm 10~{\rm GeV} \, .
\label{Gfitter}
\end{equation}
This has been an impressively successful prediction!

\subsubsection{Higgs Production at the LHC}

\begin{figure}[h!]
\centering
\includegraphics[scale=0.5]{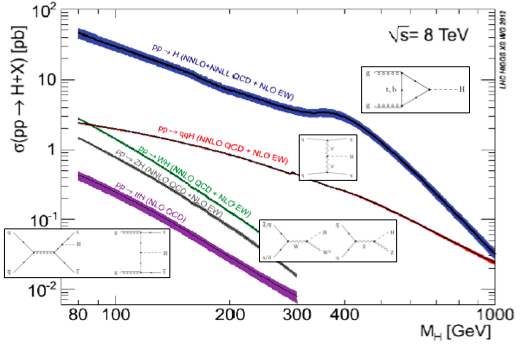}
\caption{\it The principal Higgs production cross sections at the LHC at 8~TeV. The insets depict
the corresponding fundamental production subprocesses.}
\label{fig:sigmaH8TeV}
\end{figure}

\begin{figure}[h!]
\centering
\includegraphics[scale=0.45]{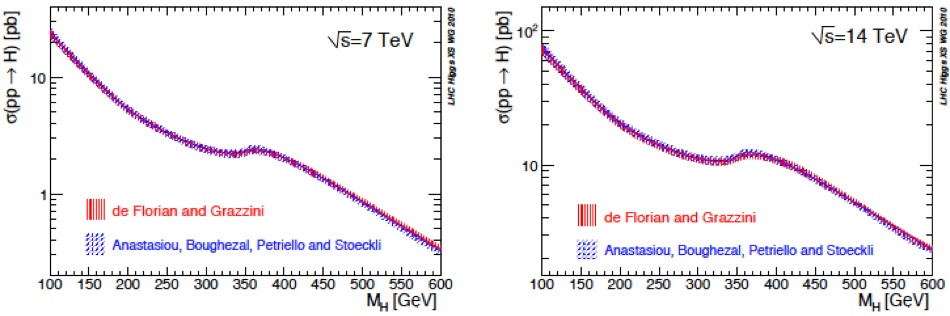}
\caption{\it The cross sections for  $gg \to H$ production at the LHC at 7 and 14~TeV, comparing two
next-to-next-leading order (NNLO) calculations that also include leading higher-order logs (NNLL)~\protect\cite{LHXSWG}.}
\label{fig:ggtoH}
\end{figure}

Several production modes are measurable at the LHC for a Higgs boson weighing $\sim 125$~GeV,
as displayed in Fig.~\ref{fig:sigmaH8TeV} for Higgs production at the LHC at 8~TeV.
The dominant mechanism is calculated to be gluon-gluon fusion: $gg \to H$ via loops of heavy coloured
particles~\cite{GGMN}, of which the most important in the Standard Model is the top quark. Since the leading-order
contribution to the production amplitude is ${\cal O}(\alpha_s)$, making an accurate calculation is
particularly challenging. However, the cross section has been calculated by different theoretical
groups at next-to-next-to-leading order (NNLO), including also the most important logarithms at
higher orders (NNLL), with good agreement between the calculations as seen in Fig.~\ref{fig:ggtoH}~\cite{LHXSWG}. The main remaining
uncertainties in the calculation are associated with the choice of renormalization scale in the
calculation, and in the gluon parton distribution function within the proton. The overall theoretical uncertainty
is currently estimated to be about 10\%, and a significant rise in the cross section is expected when the LHC
reaches 13/14~TeV in 2015.

The second-largest contribution to Higgs production at the LHC is due to vector-boson fusion (VBF)~\cite{Dawson}.
This has now been calculated at NNLO in $\alpha_s$ and including electroweak corrections at NLO,
and the perturbation expansion is converging well. In this case, the uncertainties in quark parton distribution
functions are relatively small, so this cross section is known more accurately, and it grows with energy more
rapidly than gluon-gluon fusion~\cite{LHXSWG}.

\begin{figure}[h!]
\centering
\includegraphics[scale=0.45]{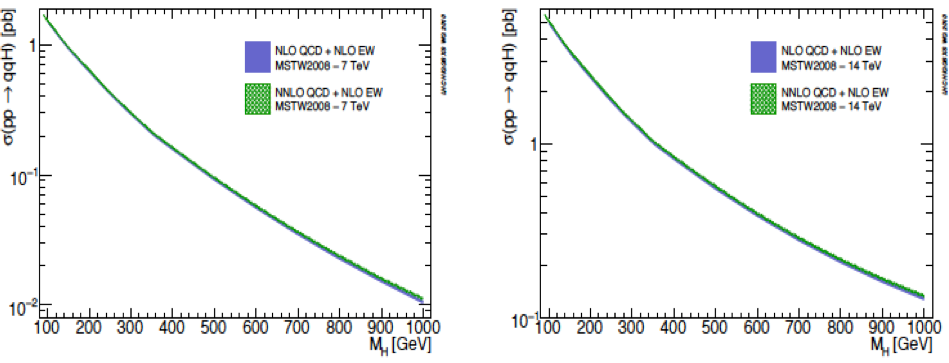}
\caption{\it The cross sections for VBF production of the $H$ at the LHC at 7 and 14~TeV, comparing
calculations at NLO in the electroweak interactions and at NLO and NNLO in QCD~\protect\cite{LHXSWG}.}
\label{fig:VBFtoH}
\end{figure}

\begin{figure}[h!]
\centering
\includegraphics[scale=0.45]{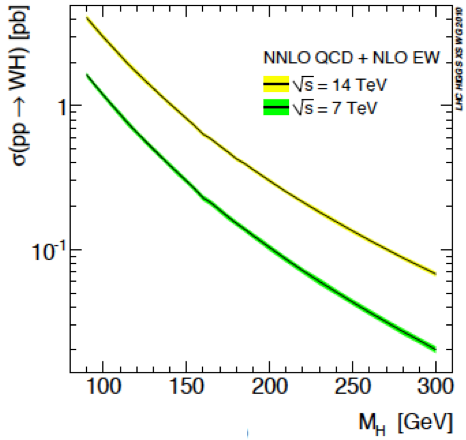}
\includegraphics[scale=0.47]{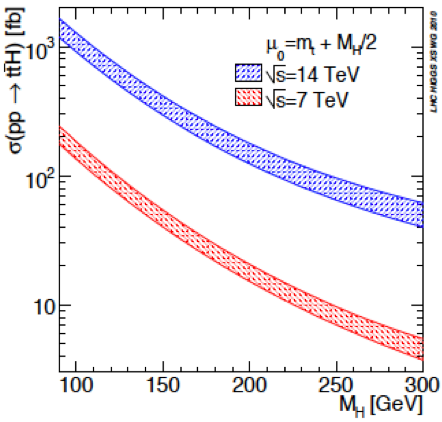}
\caption{\it The cross sections for production of the $H$ in association with a massive
vector boson $V$ (left panel) and in association with ${\bar t} t$ (right panel) at the LHC at 7 and 14~TeV~\protect\cite{LHXSWG}.}
\label{fig:assocH}
\end{figure}

The third-largest contribution is associated production of the Higgs with a massive vector boson $V = W^\pm, Z^0$~\cite{GNY}.
This has also been calculated at NNLO in $\alpha_s$ and including electroweak corrections at NLO,
as seen in the left panel of Fig.~\ref{fig:assocH}.
and the perturbation expansion again converges well. Here the rate of growth of the cross section is less
rapid.

The next contribution is from associated ${\bar t} t + H$ production, which is currently known less accurately:
it has been calculated at NLO in $\alpha_s$, so there are larger uncertainties in the perturbation expansion,
and the choice of parton distributions is also an important uncertainty~\cite{LHXSWG}. This process has the most rapid
cross section increase with energy, as seen in the right panel of Fig.~\ref{fig:assocH},
offering interesting prospects for measurement at LHC 13/14.

Finally, interest has recently been attracted by $H$ production in association with a single $t$
or ${\bar t}$~\cite{Maltoni}. This has a relatively small cross section in the Standard Model, but it may be
enhanced or suppressed significantly in models where the $H {\bar t}t$ coupling differs from its Standard Model
value~\cite{Biswas,EHST}, as we discuss later.

LHC experimentalists are indeed fortunate that all of these mechanisms are potentially measurable
at the LHC for $m_H \sim 125$~GeV! This would not have been the case if the Higgs mass had been 400~GeV, say,
in which case only $gg \to H$ and VBF could have been measurable.

\subsubsection{Higgs Decays~\cite{LHXSWG3}}

\begin{figure}[h!]
\centering
\includegraphics[scale=0.45]{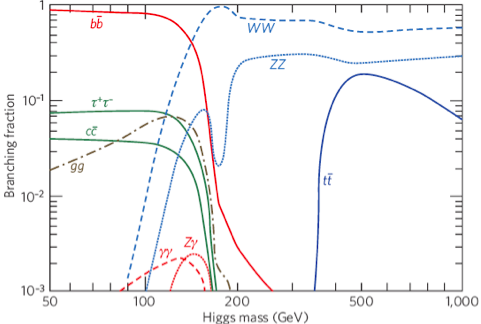}
\includegraphics[scale=0.63]{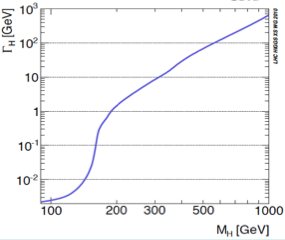}
\caption{\it The most important decay branching fractions for the decays of a Standard Model Higgs boson (left panel),
and the total decay rate $\Gamma_H$ (right panel)~\protect\cite{LHXSWG3}.}
\label{fig:Hdecays}
\end{figure}

The Higgs decay rate into a pair of fermions is given at the tree level by
\begin{equation}
\Gamma (H \to {\bar f}f) \; = \; N_c \frac{G_F m_H}{4 \pi \sqrt{2}} m_f^2 \, ,
\label{fdecay}
\end{equation}
where $N_c = 3 (1)$ for decays into quarks (leptons).
Since the tree-level Higgs couplings to other particles are proportional to their masses (squared
in the cases of massive vector bosons), the dominant Higgs decays are into the heaviest particles
that are kinematically accessible. This implies that the dominant two-body decays of the Higgs weighing
$\sim 125$~GeV are expected to be into ${\bar b}b, {\bar c}c$ and $\tau^+ \tau^-$, as seen in
the left panel of~Fig.~\ref{fig:Hdecays}. However, decays  into ${\bar b}b$ have
yet to confirmed, there is no evidence for ${\bar c}c$, and $H \to \tau^+ \tau^-$ decay has only recently
has the been observed unambiguously~\cite{tautau}.

The $H$ decay rate into a pair of $W^\pm$ bosons, one on- and one off-shell, is given by
\begin{equation}
\Gamma (H \to W W^*) \; = \; \frac{G_F m_H^3}{8 \pi \sqrt{2}} F(r) \, ,
\label{bdecay}
\end{equation}
where $F(r \equiv m_W/m_H)$ is a kinematic factor, and the corresponding decay into a pair
of $Z$ bosons is given by a similar formula with $m_W \to m_Z$ and a symmetry factor of 1/2 ~\cite{LHXSWG3}.
The decays $H \to$ virtual $W^+ W^-$ and $Z^0 Z^0$ are important for $m_H \sim 125$~GeV,
as seen in the left panel of~Fig.~\ref{fig:Hdecays}, despite the fact that $m_H < 2 m_W$ and $2 m_Z$. 

Moreover,
although the decays $H \to gg$ and $\gamma \gamma$ are absent at the tree level, they are
generated by quantum loops, as discussed above in connection with $gg \to H$ production.
The dominant contributions to the $H \to \gamma \gamma$ decay amplitude are due to
massive charged particles~\cite{EGN}, the most important in the Standard Model being the $t$ quark and
the $W^\pm$ boson, whose contributions interfere destructively. At the one-loop level
\begin{equation}
\Gamma (H \to \gamma \gamma) \; = \; \frac{G_F \alpha^2 m_H^3}{128 \pi^3 \sqrt{2}}
\left| \Sigma_f N_c Q_f^2 A_{1/2} (r_f) + A_1 (r_W) \right|^2 \, ,
\label{bgammadecay}
\end{equation}
where $A_{1/2}$ and $A_1$ are known functions of $r_f \equiv m_f / m_H$ and 
$r_f \equiv m_W / m_H$ that have opposite signs~\cite{LHXSWG3}.

Decays into strongly-interacting final states have been evaluated at NNNLO in $\alpha_s$,
while electroweak decays have been evaluated at NLO. The total Higgs decay rate in the
Standard Model is expected to be $\sim 4.2$~MeV for $m_H \sim 125$~GeV, as seen in
the right panel of Fig.~\ref{fig:Hdecays}~\cite{LHXSWG3}.

Once again, Nature has been kind in her choice of the Higgs mass, with half-a-dozen Higgs
decays being observable at the LHC for $m_H \sim 125$~GeV. If the Higgs mass had been
300~GeV, say, only the decays $H \to W^+ W^-$ and $Z^0 Z^0$ would have
been measurable.

\subsubsection{From Discovery to Measurement}

Following the dramatic announcements of the discoveries of the Higgs boson by
the ATLAS and CMS Collaborations on July 4th, 2012~\cite{ATLAS,CMS}, the emphasis is now on
measurements and the information they provide about physics within and beyond
the Standard Model.

The strengths of the signals observed in many channels are compatible with the
Standard Model predictions~\cite{Pages}. However, the bulk of the evidence concerns production
by gluon-gluon fusion. There have been several observations of VBF channels at
the 2-$\sigma$ level, and the overall significance of the evidence for VBF is some
3 $\sigma$. So far the best evidence for production in association with massive
vector bosons comes from the CDF and D0 experiments at the Fermilab Tevatron,
and there are only upper limits on production in association with ${\bar t} t$.

The primary evidence for $H$ decays is in final states involving vector bosons:
$Z Z^*, \gamma \gamma$ and $W W^*$, whereas the direct evidence for decays
into fermions is much weaker. Concerning leptons, although evidence for $H \to \tau^+ \tau^-$
decay is emerging~\cite{tautau}, but there are only upper limits on the decay into $\mu^+ \mu^-$~\cite{Pages}
(which may be the only second-generation final state accessible at the LHC)~\footnote{It
would also be interesting to search for the flavour-changing decays $H \to \tau \mu$ and $\tau e$.
These are highly suppressed in the Standatd Model, but model-independent upper bounds
from low-energy flavour-changing processes allow these decays to occur at rates similar
to $H \to \tau^+ \tau^-$~\cite{BEI}.}, and the prospects for measuring the $H e^+ e^-$ coupling look very dim.
Concerning quarks, evidence for $H \to {\bar b}b$ decay is also emerging, but
there is only indirect evidence for an $H {\bar t}t$ coupling via measurements of
$gg \to H$ production and $H \to \gamma \gamma$ decay. In the future, more
information could be provided by $H \to Z^0 \gamma$ decay and $H$ production
in association with a single $t$ or ${\bar t}$~\cite{Maltoni,Biswas,EHST}, as well as $H {\bar t}t$ production.

\subsubsection{The Higgs Mass - Evidence for Physics beyond the Standard Model?}

\begin{figure}[h!]
\centering
\includegraphics[scale=0.5]{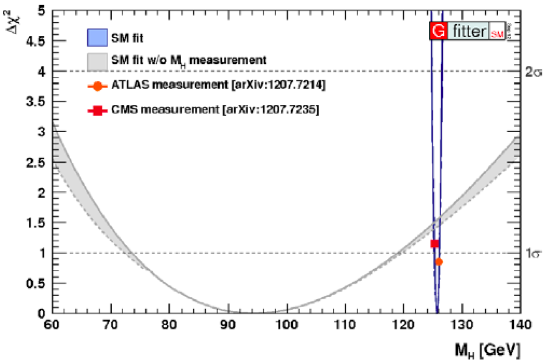}
\caption{\it Comparison of the indirect estimate of the Higgs mass based on precision
electroweak data with the direct measurement by ATLAS and CMS~\protect\cite{Gfitter}.}
\label{fig:Gfitter}
\end{figure}

There are two ways to measure the Higgs mass accurately with the present data: using $H \to Z Z^* \to 4 \ell^\pm$
and $\gamma \gamma$ decays. In the case of CMS, these two final states yield very similar
masses, with $m_{\gamma \gamma}$ slightly lower. In the case of ATLAS, there is some tension
between the measurements in the two channels, with $m_{\gamma \gamma}$ higher by $\sim 2$~GeV,
corresponding to a $\sim 2$-$\sigma$ discrepancy. However, the CMS and ATLAS measurements~\cite{Pages} are
quite consistent, and a naive global average is
\begin{equation}
m_H \; = \; 125.6 \pm 0.4 \; {\rm GeV} \, .
\label{averagemH}
\end{equation}
As seen in Fig.~\ref{fig:Gfitter}, this is consistent at the $\Delta \chi^2 \sim 1.5$ level with the estimate of $m_H$
provided by precision electroweak data~\cite{Gfitter}. A victory for the Standard Model at the
quantum (loop) level! 

However, issues arise when we consider the effective Higgs potential. There are two
important sources of renormalization of the quartic Higgs self-coupling $\lambda$:
that due to the Higgs self-coupling itself:
\begin{equation}
\lambda (Q) \; = \; \frac{\lambda(v)}{1 - \frac{3}{4 \pi^2}\lambda(v) \ln \frac{Q^2}{v^2}} + \dots \, ,
\label{selfrenn}
\end{equation}
where $Q$ is some renormalization scale above the electroweak scale $v$, and
that due to the $H {\bar t}t$ coupling:
\begin{equation}
\lambda (Q) \; = \; \lambda(v) - \frac{3 m_t^4}{4 \pi^2 v^4} \ln \frac{Q^2}{v^2} + \dots \, ,
\label{trenn}
\end{equation}
where in each case the $\dots$ represent subleading terms in the solution of the renormalization-group equation.
We see in (\ref{selfrenn}) that the self-renormalization tends to increase $\lambda$ as $Q$ increases, leading
to an apparent singularity (a so-called Landau pole). On the other hand, the renormalization by the $t$ quark tends to
reduce $\lambda$ as $Q$ increases, potentially driving it negative at some scale above $v$.

\begin{figure}[h!]
\centering
\includegraphics[scale=0.33]{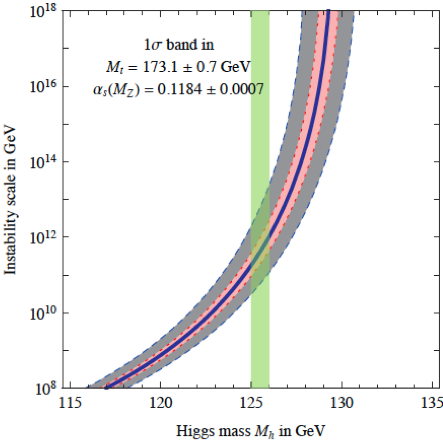}
\includegraphics[scale=0.4]{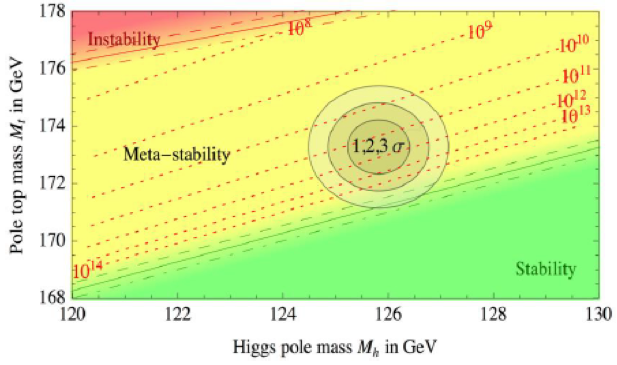}
\caption{\it Negative renormalization of the Higgs self-coupling by the top quark is calculated within the Standard Model
to lead to an instability in the effective Higgs potential for field values $\sim 10^{13}$ to $10^{14}$~GeV
(left panel). The current estimates of $m_t$ and $m_H$ suggest that the current electroweak vacuum is in fact metastable
(right panel), though a definite conclusion must wait a more accurate measurement of $m_t$, in particular.
Figures taken from~\protect\cite{DDEEGIS}.}
\label{fig:unstable}
\end{figure}

This would imply an instability in the electroweak vacuum if~\cite{DDEEGIS}
\begin{equation}
m_H \; < \; \left[129.4 + 1.4 \left(\frac{m_t - 173.1~{\rm GeV}}{0.7} \right) - 0.5 \left(\frac{\alpha_s (m_Z) - 0.1184}{0.0007} \right) \pm 1.0_{\rm TH} \right]~{\rm GeV} \, .
\label{unstableH}
\end{equation}
The measured values of $m_t \sim 173$~GeV and $m_H$ (\ref{averagemH})
would drive the quartic Higgs self-coupling negative at some scale
$\sim 10^{10}$ to $10^{14}$~GeV, as seen in the left panel of Fig.~\ref{fig:unstable},
if no physics beyond the Standard Model intervenes at some lower energy scale. (One example of
possible new physics is supersymmetry, to which we return later.) However, the lifetime of the
vacuum is estimated to be probably much longer than the age of the Universe, as seen in the right panel of Fig.~\ref{fig:unstable},
so it is not an immediate
issue for the future of humanity, leading some people to suggest that this instability is not a problem.
My own point of view is that such an instability would make it much more difficult to understand why the
current vacuum energy (cosmological constant) is so close to zero in natural units. Why should our
present vacuum energy be small if we are in a temporary state on the way to a state with vacuum
energy much larger in magnitude than now (and negative)?

It should be emphasized, however, that the conclusion that the electroweak vacuum is unstable
is not definite, even within the Standard Model. The stability or otherwise of the electroweak vacuum
depends sensitively on $m_t$ as well as $m_H$ (and. to a lesser extent, $\alpha_s$). In addition to the
quoted experimental error in $m_t$, there is also a theoretical uncertainty associated with the way
$m_t$ is defined and introduced into experimental Monte Carlo programmes~\cite{Skands}, which warrants more study.

\subsubsection{The Higgs Discovery is a Big Deal}

As already mentioned,
without the Higgs boson (or something to replace it), there would be no atoms, because massless
lectrons would escape from nuclei at the speed of light without forming atoms, and
the weak interactions would not be weak, everything would be radioactive, and life would be impossible.
The discovery of a/the Higgs boson tells us how gauge symmetry is broken and whether there is such a
thing as an elementary scalar field. it is likely to be the portal to new physics such as dark matter.
The switch-on of the Higgs v.e.v. would have caused a
phase transition in the Universe when it was about $10^{-12}$ seconds old,
and may have played a role then in generating the matter in the Universe via electroweak baryogenesis.
A related inflaton might have made the Universe expand exponentially when it was about $10^{-35}$ seconds old,
and might contribute $10^{60}$ too much to today's dark energy!

\section{What we know now}

\subsection{The Particle Jigsaw Puzzle}

\begin{figure}[h!]
\centering
\includegraphics[scale=0.5]{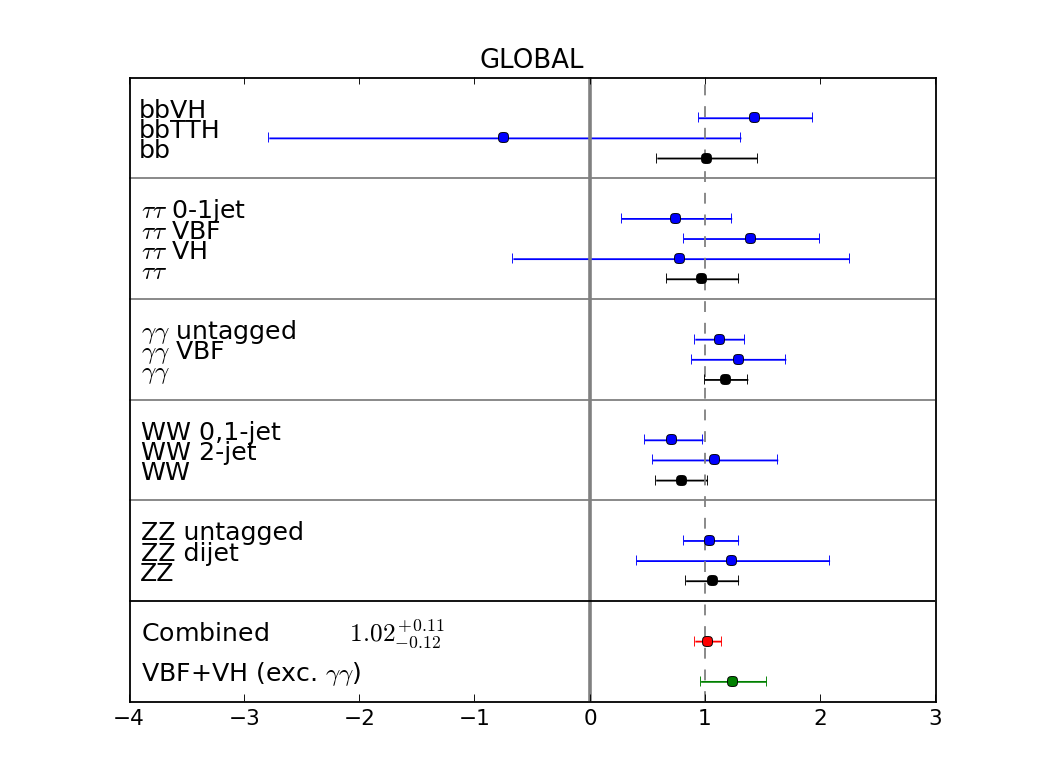}
\caption{\it A compilation of the Higgs signal strengths measured by the ATLAS, CDF, D0 and CMS
Collaborations in the ${\bar b} b$, $\tau^+ \tau^-$, $\gamma \gamma$, $WW^*$ and $ZZ^*$ final states.
We display the combinations of the different channels for each final state, and also the 
combination of all these measurements, with the result for the VBF and VH channels (excluding the $\gamma\gamma$ final state) shown separately in the bottom line. Figure taken from~\protect\cite{EY3}.}
\label{fig:EY1}
\end{figure}

Fig.~\ref{fig:EY1} summarizes our state of knowledge concerning the $H$
signal strengths in the various different final states: ${\bar b} b, \tau^+ \tau^-,
\gamma \gamma, W W^*$ and $ZZ^*$, averaged over the results of ATLAS,
CMS and the Tevatron experiments. The signal strengths in different
production channels are presented (in blue), followed by the global
combinations in each final state (in black). The combined mean signal
strength in all channels (in red) is
\begin{equation}
\mu \; = \; 1.02^{+0.11}_{-0.12} \, ,
\label{mu}
\end{equation}
and finally the combination of the signal strengths in the VBF and
associated $V + H$ channels is shown (in green). We see that there
is no indication of any significant deviation from the Standard Model predictions:
Peter Higgs should be smiling!

The current situation of particle physicists resembles that of someone who
has spent more than 100 years putting together a jigsaw puzzle, and has
finally (after 48 years) discovered what may be the last missing piece,
hidden away in the back of the sofa and with the picture rubbed off.
Have the LHC experiments really discovered the missing piece, or is it
an impostor? Does it have the right shape to fit into the empty space in the puzzle,
and does it have the right size?

The rest of this lecture is devoted to answering these questions as best we can
on the basis of the present data.

\subsection{Is it the Missing Piece?}

\subsubsection{Does it have Spin 0 or 2?}

The question `does the newly-discovered $H$ particle have the right shape
to be the missing piece of the puzzle?' can be parsed as the question `what is
its spin?' Since the $H$ particle decays into pairs of photons - identical spin-1 bosons - it
must have some spin $\ne 1$. The simplest possibilities are spin 0 and 2, the
Standard Model being an example of the former case, a Kaluza-Klein graviton
being an example of the second case. A higher spin cannot be excluded {\it a priori},
but I am unaware of any model with spin $> 2$.

Several ways to diagnose the $H$ spin have been proposed, including the
characteristics of production in association with $W^\pm$ or $Z^0$~\cite{EHSY,ESY2}, the
angular distribution of $\gamma \gamma$ decays~\cite{EH,EFHSY}, and the kinematic correlations
of leptons in $W W^*$ and $Z Z^*$ decays~\cite{Bol}. In general, a massive spin-2
particle has many possible couplings to Standard Model particles, and
distinguishing the spin-0 and general spin-2 hypotheses is difficult. Here
we consider the simplest spin-2 case (see~\cite{ESY1} and references therein), in which it has minimal graviton-like
couplings, as in simple models with extra dimensions:
\begin{equation}
{\cal L}_{int} \; = \; \Sigma_i \frac{c_i}{M_{eff}} G^{\mu \nu} T^i_{\mu \nu} \, ,
\label{KKG}
\end{equation}
where the sum is over Standard Model particle types $i$ and the overall
mass scale and the individual coefficients $c_i$ are model-dependent.

\begin{figure}[h!]
\centering
\includegraphics[scale=0.5]{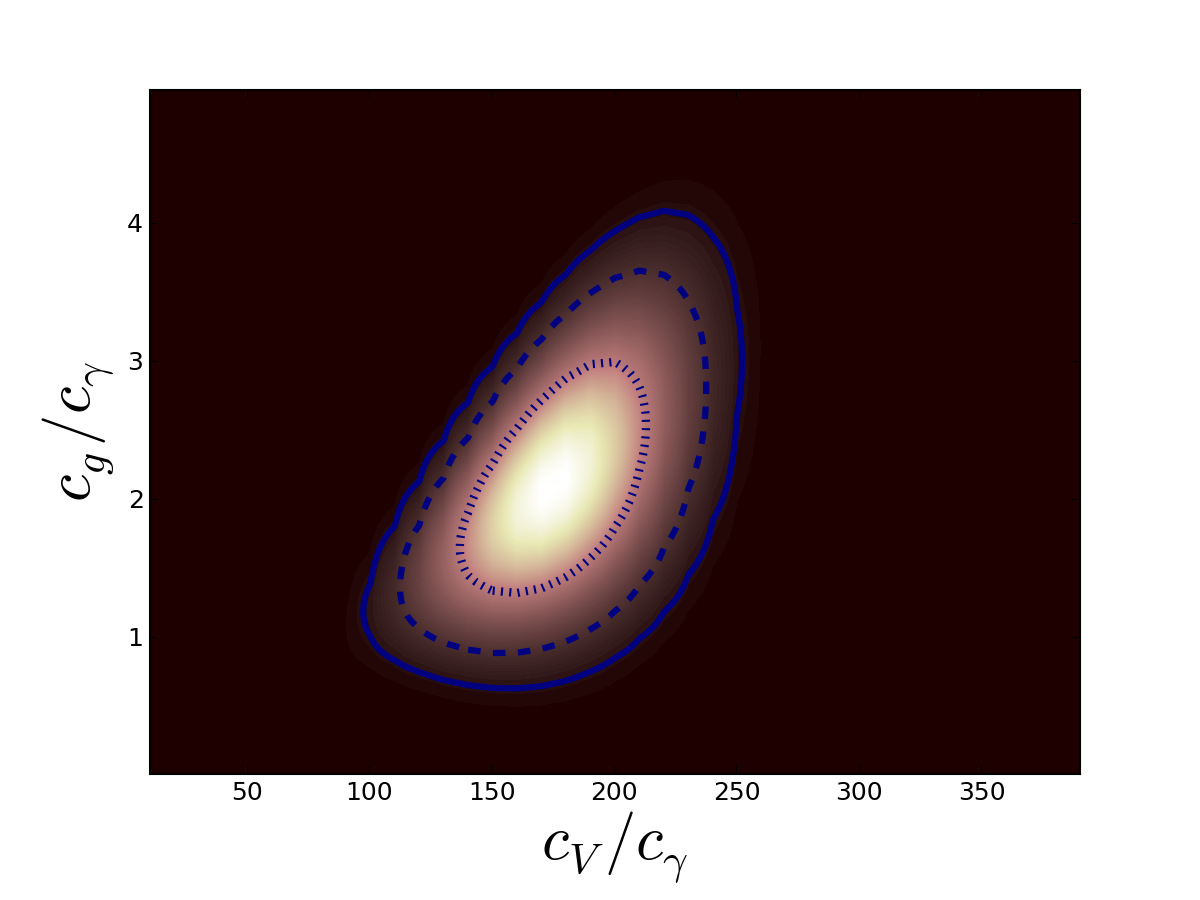}
\caption{\it The correlation between the values of $c_W/c_\gamma$ (horizontal axis) and
$c_g/c_\gamma$ (vertical axis) found in a global fit to the current experimental data under the spin-
two hypothesis~\protect\cite{ESY1}.}
\label{fig:ESY}
\end{figure}

For definiteness, we can consider warped compactifications of 5-dimensional
theories~\cite{ESY1}, in which the metric takes the form:
\begin{equation}
ds^2 \; = \; w(z)^2 \left( \eta_{\mu \nu} dx^\mu dx^\nu - dz^2 \right) \, .
\label{warp}
\end{equation}
In such a scenario we expect identical coefficients for the couplings 
of a spin-2 particle $X$ to the massless vector bosons $g$ and $\gamma$:
\begin{equation}
c_g \; = c_\gamma \; = \; 1/ \int_{z_{UV}}^{z_{IR}} w(z) dz \, .
\label{cgcgamma}
\end{equation}
since their wave functions are uniform in the extra dimension.
This implies the following simple relation between the decay rates of a spin-2 particle $X$ 
into photons and gluons:
\begin{equation}
\Gamma (X \to gg ) \; = \; 8 \Gamma (X \to \gamma \gamma) \, ,
\label{factor8}
\end{equation}
which is disfavoured by the data on the $H(126)$ decay branching ratios,
as seen in Fig.~\ref{fig:ESY}. The couplings of the other
Standard Model particles are non-universal, reflecting their different
wave functions in the extra dimension. In simple warped
compactifications one expects
\begin{equation}
c_b \; \simeq \; c_t \ga \; c_W \simeq \; c_Z \; = {\cal O}(35) \times \left(c_g = c_\gamma > c_{u, d} \right) \; .
\label{allthecs}
\end{equation}
The experimental data on $H(126)$ decays also disfavour the expected hierarchy
(\ref{allthecs}) between $c_W \simeq c_Z$ and $c_g = c_\gamma$, as also seen in Fig.~\ref{fig:ESY}~\footnote{Another
problem for this scenario is that one expects, as in QCD, that the tensor boson should have a higher mass than
the lightest Kaluza-Klein vector boson, which has not been seen.}.

\begin{figure}[htbp!]
\centering
\includegraphics[scale=0.38]{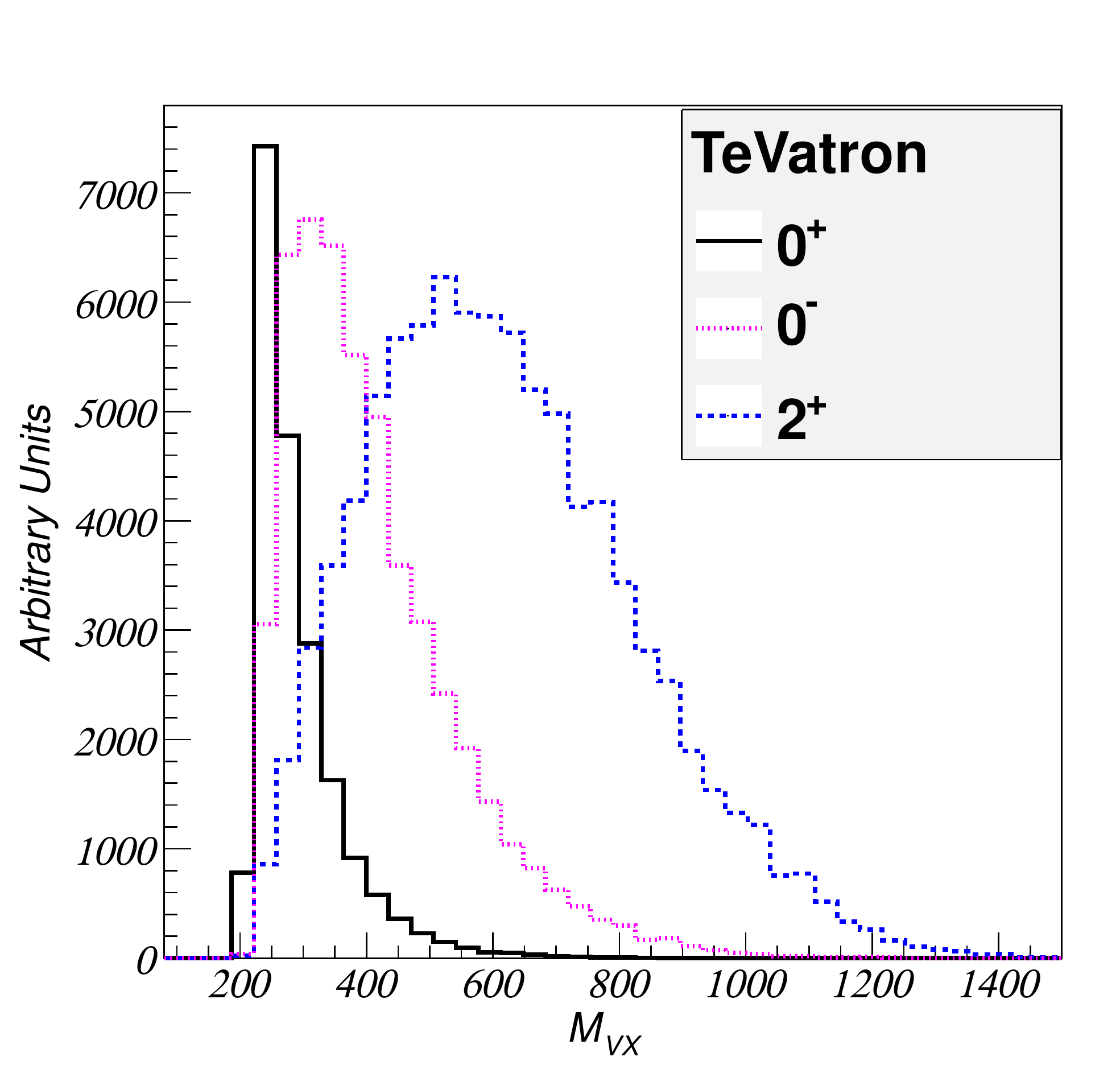}
\includegraphics[scale=0.38]{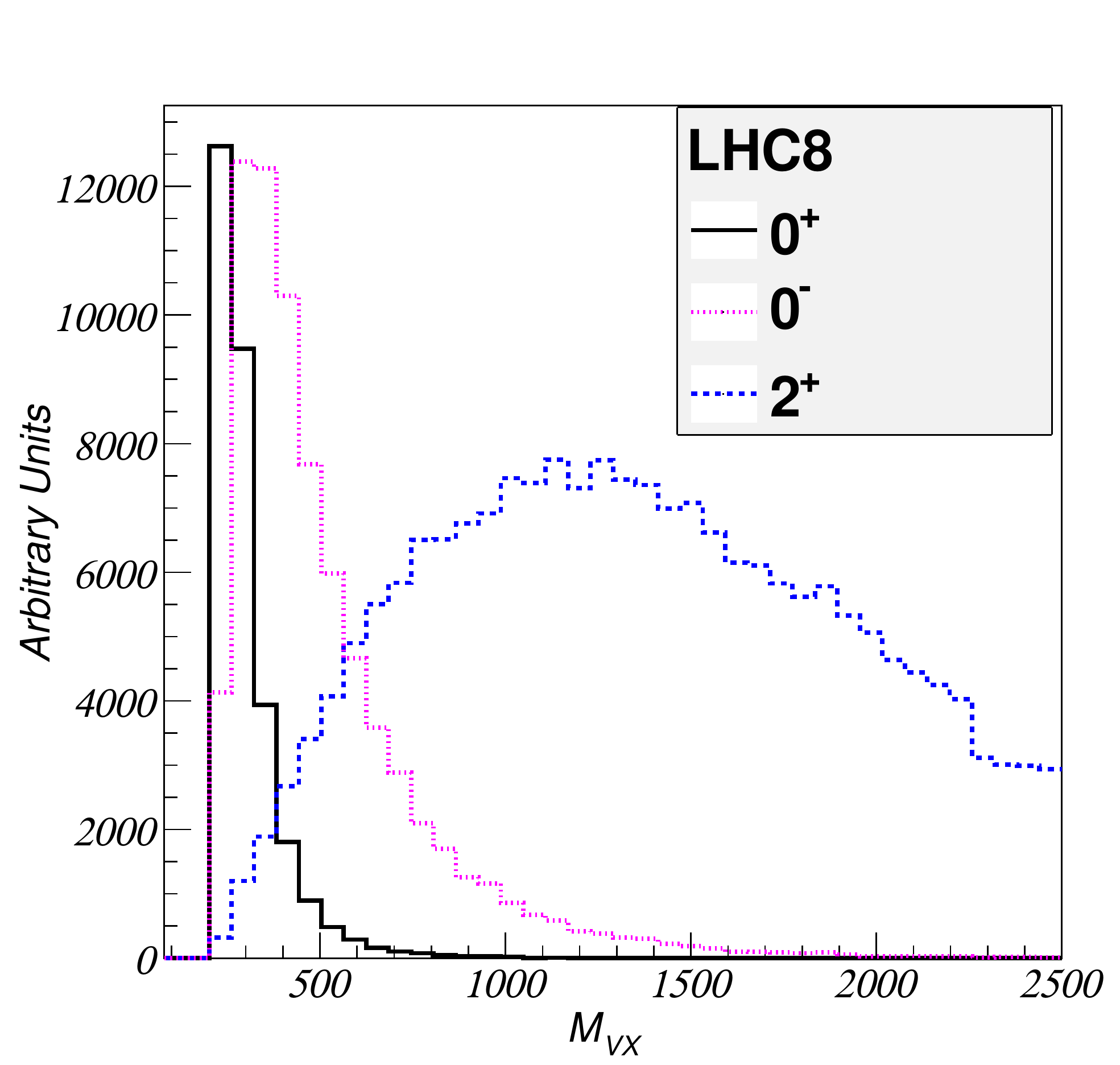}
\caption{\it The distributions in the $Z + H$ invariant mass $M_{ZH}$ for the $0^{+}$ (solid black), 
$0^{-}$ (pink dotted) and $2^+$ (blue dashed) assignments for the $H$ particle 
discovered by ATLAS~\cite{ATLAS} and CMS~\cite{CMS}, calculated for the reaction ${\bar p} p \to Z + H$ at the TeVatron
(left) and for the reaction $p p \to Z + H$ at the LHC at 8 TeV (right)~\protect\cite{EHSY}.}
\label{fig:EHSY}
\end{figure}

\begin{figure}[htbp!]
\centering
\includegraphics[scale=0.4]{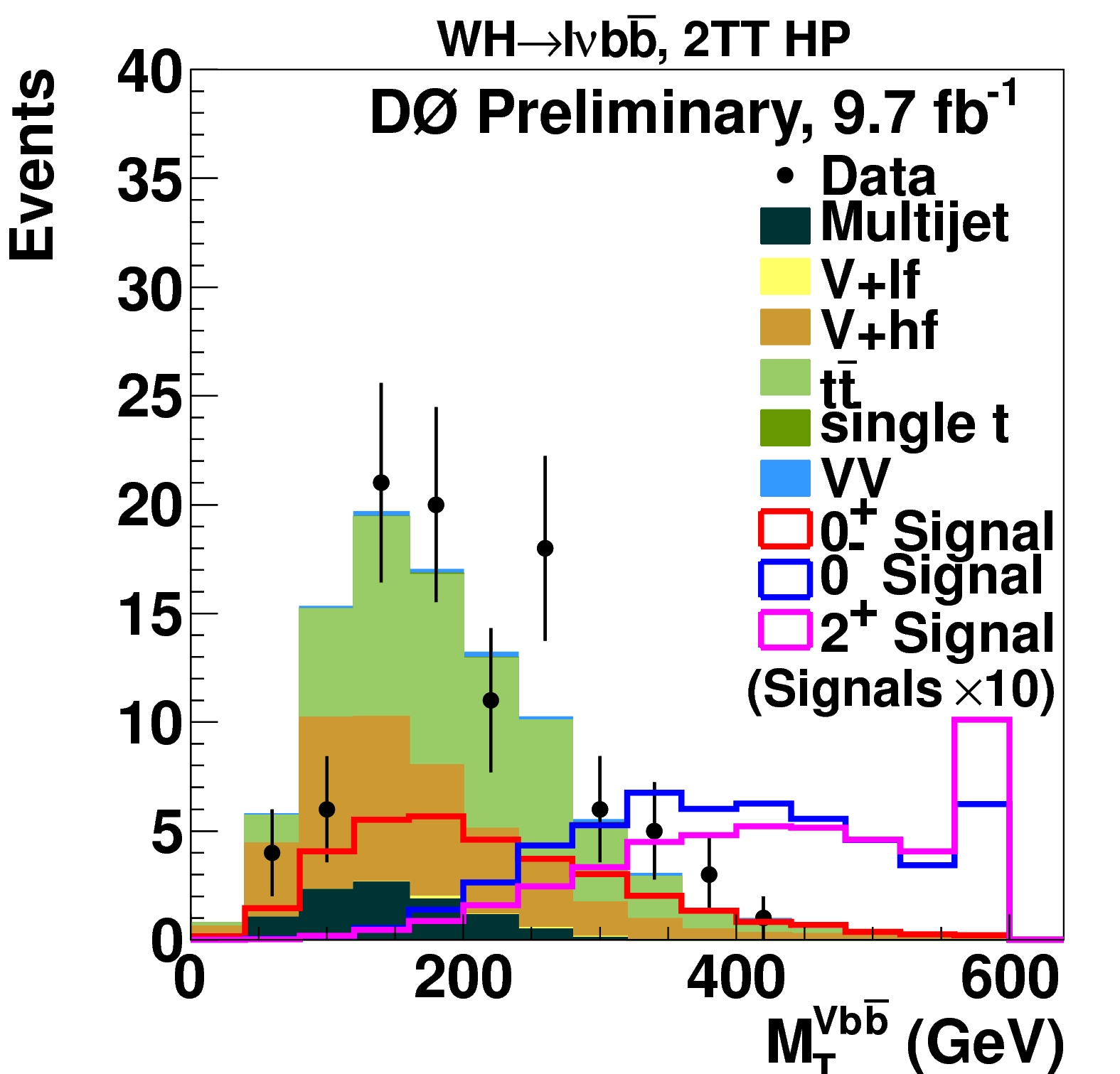}
\caption{\it The distribution in the $Z + H$ transverse mass $M_{ZH}$ measured by the D0 Collaboration
compared with simulations for the $0^{+}$ (red), 
$0^{-}$ (blue) and $2^+$ (mauve) hypotheses for the $H$ particle~\protect\cite{D0mass}.}
\label{fig:D0}
\end{figure}

The difference between the spin-2 couplings (\ref{warp}) and spin-0 couplings
also affect the kinematics of production in association with massive vector bosons
$W^\pm$ and $Z^0$. For example, the $VX$ invariant mass distributions for
spin 2 and a scalar $0^+$ particle are very different, as seen in Fig.~\ref{fig:EHSY}~\cite{EHSY},
as also are the invariant mass distributions for a pseudoscalar $0^-$ particle.
The Tevatron experiments have studied the related transverse-mass distribution,
as seen in Fig.~\ref{fig:D0}, and have found that the spin-2 and $0^-$ hypotheses
are disfavoured at the 99\% CL~\cite{D0mass} - assuming that these experiments have indeed observed the
same particle as discovered by ATLAS and CMS.

\begin{figure}[htbp!]
\vskip 0.5in
\begin{minipage}{8in}
\hspace*{-0.7in}
\centerline{{\includegraphics[height=6.5cm]{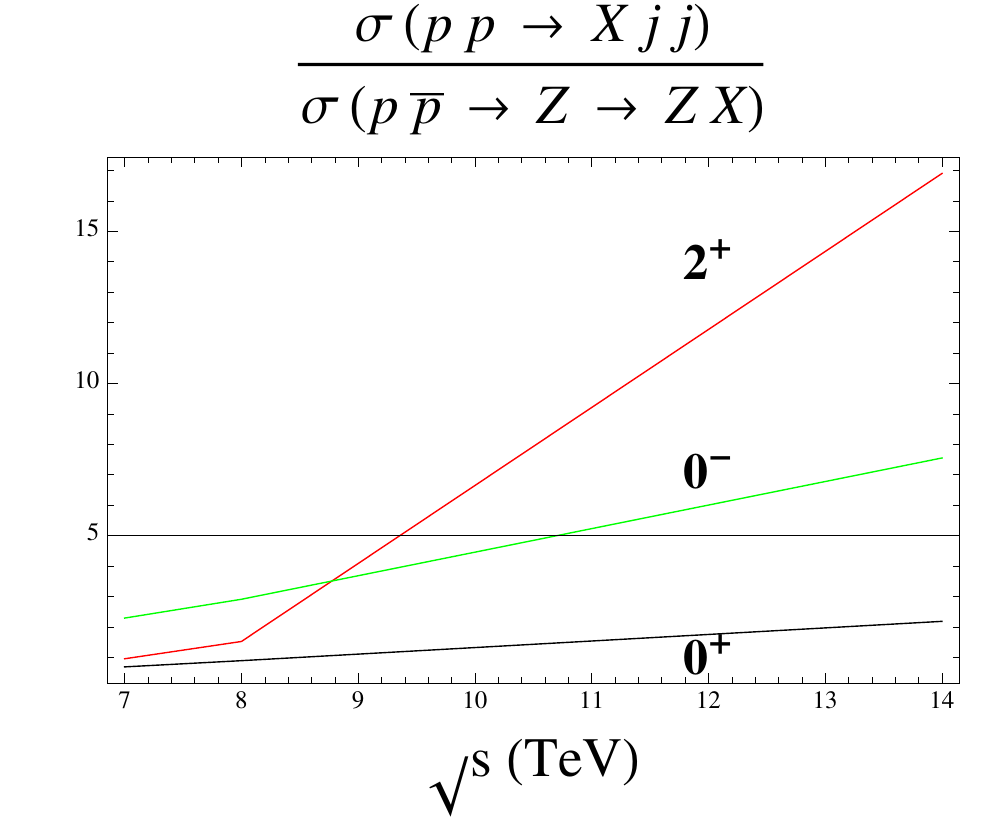}}
{\includegraphics[height=6cm]{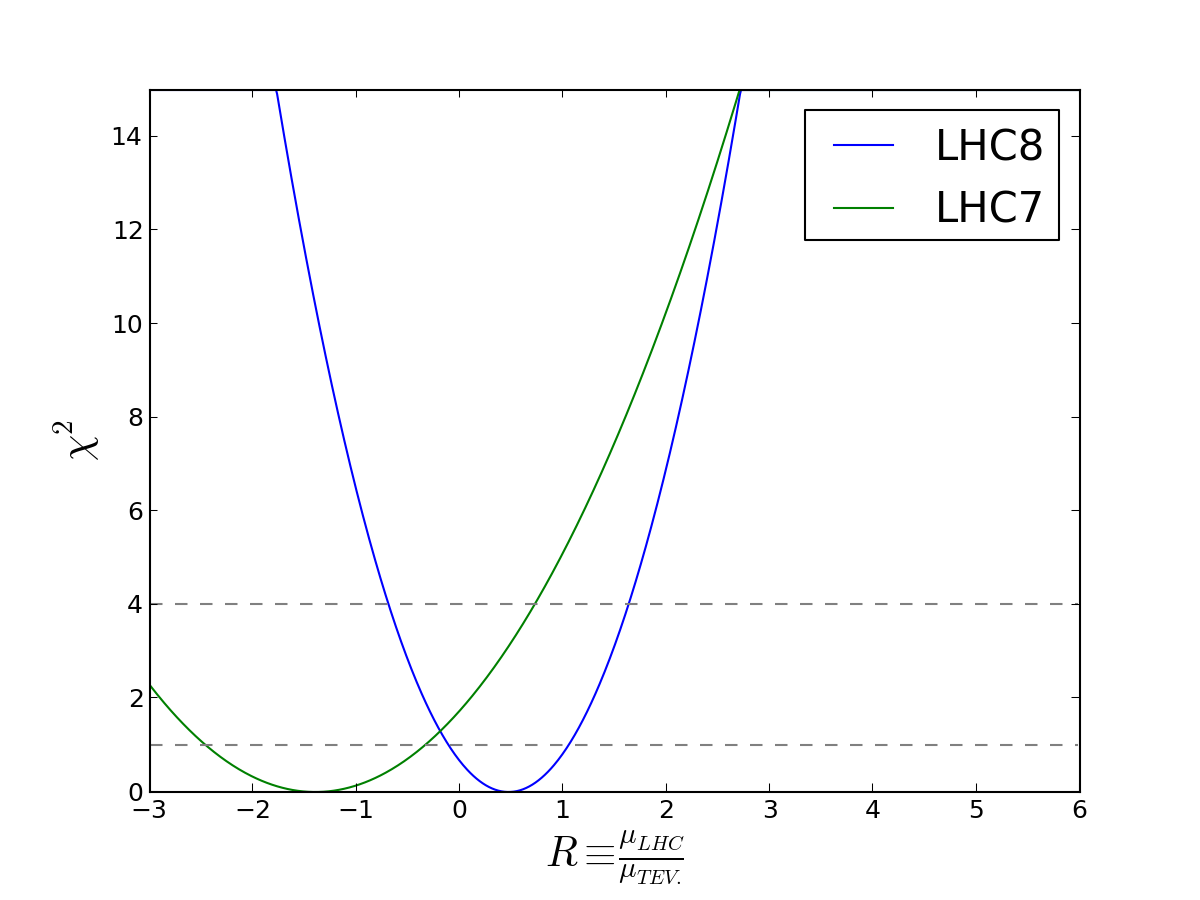}}}
\hfill
\end{minipage}
\caption{
{\it
Left - The energy dependence of the cross section for production of $h$ in association with a $Z$ boson
under different hypotheses for the $J^P$ of $H$: $0^+$ (black), $2^+$ (red) and $0^-$ (green).
Right - The likelihood for the ratio $R_\text{data} = \mu_\text{LHC~8}/\mu_\text{TeVatron}$ extracted from the experimental data
at 8~TeV (blue) and 7~TeV (green). 
The spin-two expectations $R_\text{Spin~2} = 5.4$ and $6.7$ for 7 and 8 TeV, respectively, are excluded,
and the $0^-$ expectations $R_{0^-} = 3.1$ and $2.7$ for 7 and 8 TeV, respectively, are highly disfavoured,
whereas the $0^+$ expectation $R = 1$ is quite consistent with the data~\protect\cite{ESY2}.
} 
}
\label{fig:Edep}
\end{figure}

The differences in the couplings also lead to different energy dependences
in the spin-2, $0^+$ and $0^-$ cases, as seen in the left panel of Fig.~\ref{fig:Edep}.
If one accepts the Tevatron evidence for $H(126)$ production, the ratio of the cross
section to that at the LHC is also strong evidence against the spin-2 and $0^-$ hypotheses,
as seen in the right panel of Fig.~\ref{fig:Edep}.

The polar angle distributions in the $H(126)$ centre-of-mass frame under the spin-2
and -0 hypotheses are also expected to be  easily distinguishable. In the case of
$gg \to H \to \gamma \gamma$ production, one expects the initial state to be an
incoherent superposition of parallel and antiparallel gluon spins along the
proton-proton collision axis. This knowledge of the initial state enables the final-state
$\gamma \gamma$ polar-angle distribution to be calculated: it is expected to be
non-uniform and peaked in the forward and backward directions, whereas the
angular distribution would be isotropic in the spin-0 case. The ATLAS Collaboration 
has found that the spin-2 case is disfavoured at more than the 99\% CL~\cite{ATLASspin}, and has
extended the analysis to include an arbitrary admixture of ${\bar q}q$ initial states
(which would be suppressed in the warped compactification scenario discussed above).

\begin{figure}[htbp!]
\vskip 0.5in
\begin{minipage}{8in}
\hspace*{-0.7in}
\centerline{{\includegraphics[height=6cm]{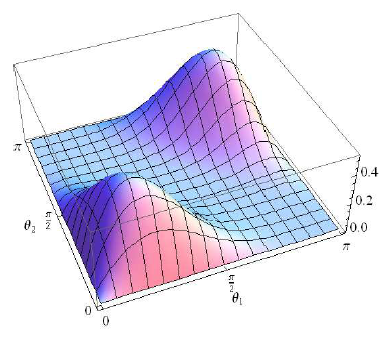}}
{\includegraphics[height=6cm]{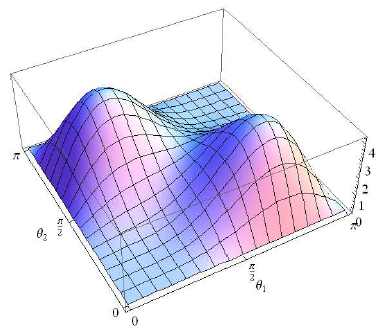}}}
\hfill
\end{minipage}
\caption{
{\it
Correlated distributions for the lepton polar angles in $H \to W^\pm W^{\mp*} \to \ell^+ \ell^- + \dots$ decays
for the $0^+$ assignment (left panel) and for a graviton-like $2^+$ particle (right panel)~\protect\cite{EH}.
} 
}
\label{fig:EH}
\end{figure}

The azimuthal and polar angle distributions of the charged leptons in 
$H \to W^\pm W^{\mp*} \to \ell^+ \ell^- + \dots$ decays also provide significant
power to distinguish between the spin-2 and -0 hypotheses~\cite{EH,Bol}, as seen in Fig.~\ref{fig:EH}, and also disfavour
the interpretation of $H(126)$ as a spin-2 particle with graviton-like couplings.
Finally, the multiple kinematical observables in $H \to Z^0 Z^{0*} \to 2 \ell^+ 2 \ell^-$
also provide many powerful ways to distinguish between different spin-parity
assignments for the $H(126)$ particle.

\begin{figure}[h!]
\centering
\includegraphics[scale=0.6]{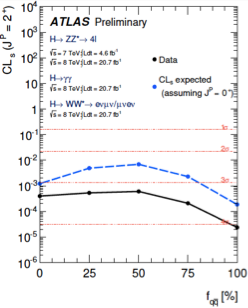}
\caption{\it Combining measurements in the $H \to \gamma \gamma, Z Z^* \to 4 \ell^\pm$
and $WW^* \to 2 \ell^\pm \nu {\bar \nu}$ final states, the ATLAS Collaboration excludes the
spin-2 hypothesis for $H$ at more than the 99.9\% CL for any combination of $H$ production
via $gg$ and ${\bar q}q$ collisions~\protect\cite{ATLASspin}.}
\label{fig:ATLASspin}
\end{figure}

It is on the basis of a combination of $\gamma \gamma, W^\pm W^{\mp*}$ and $ Z^0 Z^{0*}$
measurements, the ATLAS Collaboration has been able to exclude the graviton-like
spin-2 hypothesis at more than the 99.9\% CL, as seen in 
Fig.~\ref{fig:ATLASspin}~\cite{ATLASspin}. Peter Higgs can continue smiling!

\subsubsection{Is it Scalar or Pseudoscalar?}

As has been discussed above, many of the analyses that discriminate between the
spin-2 and $0^+$ hypotheses can also be used to discriminate between $0^-$ and
$0^+$. For example, the $V+H$ invariant mass distributions in associated production
are different, and already disfavour $0^-$ quite strongly. The angular distribution of
the $\gamma \gamma$ final state does not distinguish, but the angular and
kinematic distributions in $H \to W^\pm W^{\mp*} \to \ell^+ \ell^- + \dots$ and
$H \to Z^0 Z^{0*} \to 2 \ell^+ 2 \ell^-$ do offer discrimination between $0^+$ and $0^-$~\cite{Bol}.
Thus the possibility of a pure $0^-$ spin-parity assignment can also be excluded
beyond the 99\% CL~\cite{ATLASspin}. 

On the other hand, in the presence of CP violation the $H$
particle could decay as a mixture of scalar and pseudoscalar, and the fractions could
be different in different final states. At the moment, the admixture of a substantial fraction of
pseudoscalar final states in $H \to W^\pm W^{\mp*}$ and $H \to Z^0 Z^{0*}$ decays cannot 
be excluded. It is important to extend probes of a possible $0^-$ admixture to final states
involving fermions, and measurements in $\tau^+ \tau^-$ final states,
${\bar t} t + H$ and single $t + H$ production (see later) have been proposed~\cite{EHST}.
But, for the time being, Peter Higgs can continue smiling!

\subsubsection{Is it Elementary or Composite?}

This question may be addressed by constructing a phenomenological
Lagrangian ${\cal L}$ with free parameters to describe the interactions of the `Higgs' boson, and
constraining the parameters using data on $H$ production and decay.
Motivated by the success of the Standard Model relation $\rho \equiv m_W/m_Z \cos \theta_W = 1$,
it is usually assumed that this phenomenological Lagrangian possesses a custodial
symmetry: SU(2)$\times$SU(2) $\to$ SU(2). In this case, one may parameterize the
leading-order terms in ${\cal L}$ as follows~\cite{Some}:
\begin{eqnarray}
{\cal L} & = & \frac{v^2}{4} {\rm Tr} D_\mu \Sigma^\dagger D^\mu \Sigma \left( 1 + 2 a \frac{H}{v} + b \frac{H^2}{v^2} + \dots \right) \nonumber \\
& - & {\bar \psi}^i_L \Sigma \left(1 + c \frac{H}{v} + \dots \right) \nonumber \\
& + & \frac{1}{2} \left(\partial_\mu H \right)^2 + \frac{1}{2} m_H^2 H^2 + d_3 \frac{1}{6} \left(\frac{3 m_H^2}{v} \right) H^3 + d_4 \frac{1}{24} \left(\frac{3 m_H^2}{v} \right) H^4 + \dots \, ,
\label{calL}
\end{eqnarray}
where
\begin{equation}
\Sigma \; \equiv \; {\rm exp} \left( i \frac{\sigma^a \pi^a}{v} \right) \, ,
\label{Sigma}
\end{equation}
and the effective interaction with massless gauge bosons is written as
\begin{equation}
{\cal L}_\Delta \; = \; - \left( \frac{\alpha_s}{8 \pi} c_g G_{a \mu \nu} G^{\mu \nu}_a + \frac{\alpha_{em}}{8 \pi} c_\gamma F_{\mu \nu} F^{\mu \nu} \right) \left(\frac{H}{v} \right)^2 \, .
\label{anomL}
\end{equation}
The free coefficients $a, b, c, d_3, d_4, c_g$ and $c_\gamma$ are all normalized such that they are unity
in the Standard Model: composite models may give observable deviations from these values.
For example, in the composite model known as MCHM4 one has~\cite{EY3}
\begin{equation}
a \; = c \; = \; \sqrt{1 - \xi} \; : \; \xi \; \equiv \; \left( \frac{v}{f} \right)^2 \, ,
\label{MCHM4}
\end{equation}
where $f$ is an analogue of the pion decay constant in QCD. On the other hand, in the MCHM5 composite
model, one has 
\begin{equation}
a \; = \; \sqrt{1 - \xi} \; , \; c \; = \; \frac{1 - 2 \xi}{\sqrt{1 - \xi}} \, ,
\label{MCHM5}
\end{equation}
and in a pseudo-dilaton model one has
\begin{equation}
a \; = c \; = \frac{v}{V} \, ,
\label{pseudoD}
\end{equation}
where $V$ is the dilaton v.e.v. that breaks scale invariance. On may also consider an `anti-dilaton'
scenario in which $a = - c$.

The signal strengths $R$ in various channels relative to the Standard Model values
are related in an obvious way to the parameters in (\ref{calL}). For example,
for vector boson fusion and production in association with $V = W, Z$ one has
\begin{equation}
R_{VBF} \; = \; R_{VH} \; = \; a^2 \, ,
\label{RVBF}
\end{equation}
and for production in association with ${\bar t} t$ and the rates for decays into ${\bar b} b$ and $\tau^+ \tau^-$ one has
\begin{equation}
R_{{\bar f}f} \; = \; c^2 \, .
\label{Rffbar}
\end{equation}
The corresponding ratio for the $ggH$ coupling strength depends on the ${\bar t} t$ coupling:
\begin{equation}
R_{gg} \; = \; c_g^2 \; = \; c^2 + \dots \, ,
\label{Rgg}
\end{equation}
where the $\dots$ represent possible contributions from particles beyond the Standard Model,
and the ratio for the $H \gamma \gamma$ coupling depends on both $a$ and $c$
as well as possible non-Standard Model contributions:
\begin{equation}
R_{\gamma \gamma} \; = \; c_\gamma^2 \; = \; \frac{\left( - \frac{8}{3} c F_t + a F_W \right)^2}{\left( - \frac{8}{3} F_t + F_W \right)^2} + \dots \, ,
\label{Rgammagamma}
\end{equation}
where $F_{t, W}$ are form factors that depend on the ratios $m_H/m_t$ and $m_H/m_W$, respectively.
It is apparent from these expressions that only $R_{\gamma \gamma}$ is sensitive to the relative sign of
the $H {\bar f}f$ and $HVV$ couplings. The principal dependences of the signal strengths in various channels on
the $a$ and $c$ parameters in (\ref{calL}) are summarized in Table~\ref{table:channeleffparamssensitivity}.

\begin{table}[h!]
	\center
	\begin{tabular}{ | c | c | c | c | c |}
		\hline
		 & \multicolumn{2}{|c|}{Production sensitive to} & \multicolumn{2}{|c|}{Decay sensitive to} \\ 
		channel & \quad $a$ \quad & \quad $c$ \quad & \quad $a$ \quad & \quad $c$ \quad  \\ \hline
		$\gamma\gamma$ & $\checkmark$ & $\checkmark$ & $\checkmark$ & $\checkmark$ \\ \hline
		$\gamma\gamma$ VBF & $\checkmark$ & $\times$ & $\checkmark$ & $\checkmark$ \\ \hline
		$WW$ & $\checkmark$ & $\checkmark$ & $\checkmark$ & $\times$ \\ \hline 
		$WW$ + 2 jets & $\checkmark$ & $\times$ & $\checkmark$ & $\times$ \\ \hline
		$WW$ + 0,1 jet & $\times$ & $\checkmark$ & $\checkmark$ & $\times$ \\ \hline
		$b\bar{b}$ & $\checkmark$ & $\times$ & $\times$ & $\checkmark$  \\ \hline
		$ZZ$ & $\checkmark$ & $\checkmark$ & $\checkmark$ & $\times$ \\ \hline
		$\tau\tau$ & $\checkmark$ & $\checkmark$ & $\times$ & $\checkmark$ \\ \hline 
		$\tau\tau$ VBF, VH & $\checkmark$ & $\times$ & $\times$ & $\checkmark$ \\ \hline 
	\end{tabular}
	\caption{\it The dominant dependences on the model parameters $(a, c)$ (\protect\ref{calL}) of the $H$
	signal strengths in various channels, from~\protect\cite{EY2}.}
	\label{table:channeleffparamssensitivity}
\end{table}

\begin{figure}[h!]
\centering
\includegraphics[scale=0.35]{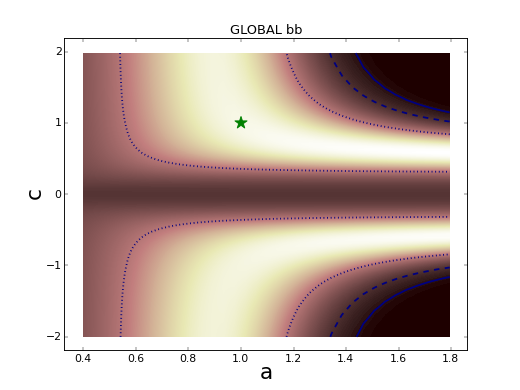}
\includegraphics[scale=0.35]{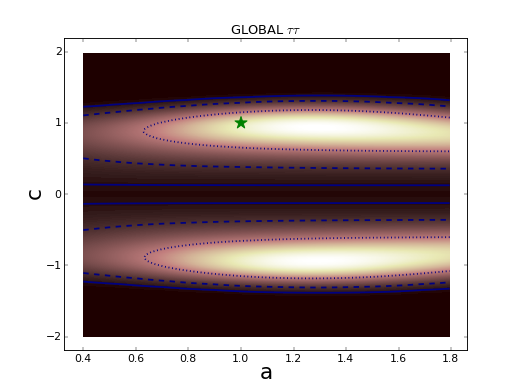}\\
\includegraphics[scale=0.35]{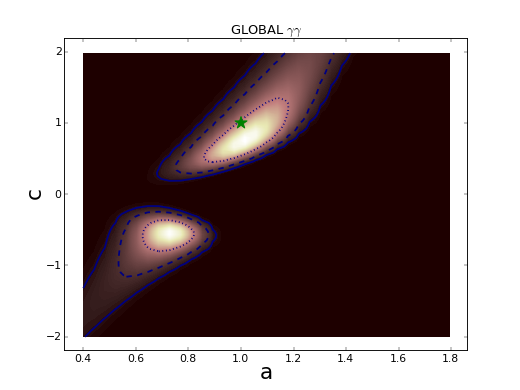}
\includegraphics[scale=0.35]{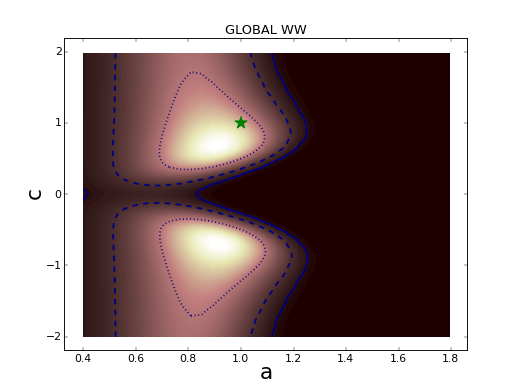}\\
\includegraphics[scale=0.35]{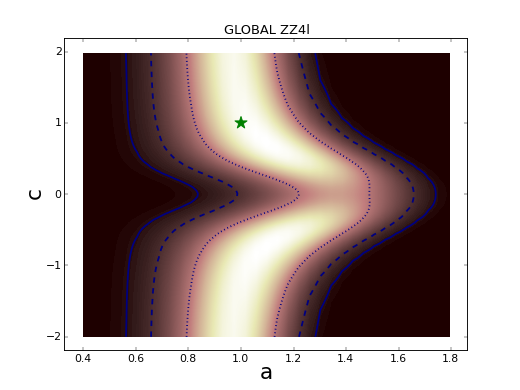}
\includegraphics[scale=0.35]{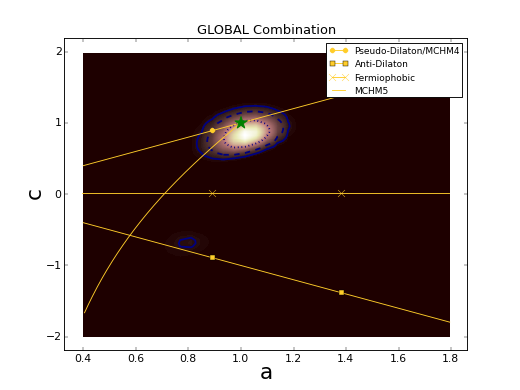}
\caption{\it The constraints in the $(a, c)$ plane imposed by the measurements in Fig.~\ref{fig:EY1}
 of the ${\bar b} b$ final state (top left), of the $\tau^+ \tau^-$ final state (top right),
 of the $\gamma \gamma$ final state (middle left), of the $WW$ coupling (middle right)
 and in the $ZZ$ coupling (bottom left). The combination of all these constraints is
 shown in the bottom right panel~\protect\cite{EY3}.}
\label{fig:ac}
\end{figure}

Fig.~\ref{fig:ac} shows how measurements of these various channels at the
Tevatron collider and the LHC combine to constrain the parameters
$(a, c)$~\cite{EY3}. We see in the top left panel that the data on ${\bar b}b$ final states already disfavour
leptophobic models in which the $H$ particle has no couplings to fermions - here the
Tevatron experiments play an important r\^ole. In the top right panel we see that
data on the $\tau^+ \tau^-$ final state also disfavour leptophobic models. However, as expected
on the basis of (\ref{Rffbar}), these measurements by themselves offer no information about the
sign of the fermion coupling coefficient $c$. The middle left panel shows the constraint imposed
by the data on the $\gamma \gamma$ final state. As seen in (\ref{Rgammagamma}), this final
state gives a constraint that is not symmetric between the signs of $c$, since there is
interference between the virtual ${\bar t}t$ and $W^+ W^-$ intermediate states that may be
either constructive of destructive, depending on the sign of $c$. The middle right panel
of Fig.~\ref{fig:ac} shows the constraint imposed by measurements of $WW^*$ final states
as well as $H$ production via $W^+W^-$ VBF and production in association with $W^\pm$,
and the bottom left panel shows the corresponding constraint on the $H ZZ$ coupling.
These measurements are highly consistent with custodial symmetry: SU(2)$\times$SU(2) $\to$ SU(2),
as assumed in writing (\ref{calL}).

\begin{figure}[htbp!]
\centering
\includegraphics[scale=0.35]{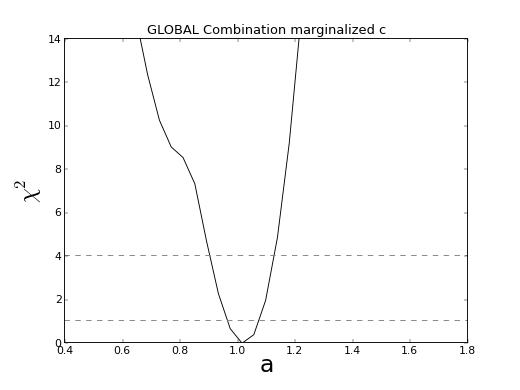}
\includegraphics[scale=0.35]{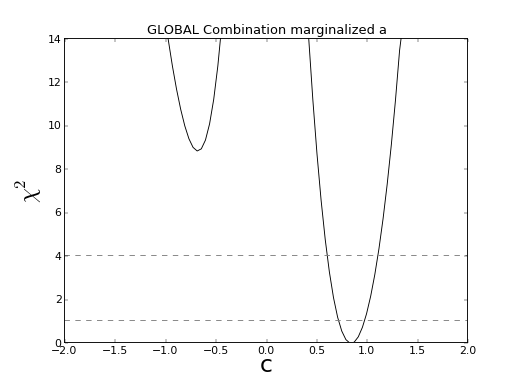}
\caption{\it The one-dimensional likelihood functions for the boson coupling parameter $a$ (left panel) and 
the fermion coupling parameter $c$ (right panel),
as obtained by marginalizing over the other parameter in the bottom right panel of Fig.~\ref{fig:ac}~\protect\cite{EY3}.}
\label{fig:sign}
\end{figure}

Finally, the bottom right panel of Fig.~\ref{fig:ac} displays the constraints in
the $(a, c)$ plane obtained in a global combination of these measurements~\cite{EY3}.
We see that the positive sign of $c$, as expected in the Standard Model, is
strongly favoured. This point is made explicitly in Fig.~\ref{fig:sign}, where we
see that the data favour $a \sim 1$ and disfavour $c < 0$ by $\Delta \chi^2 \sim 9$,
i.e., 3 standard deviations. The continuous yellow lines the bottom right panel of Fig.~\ref{fig:ac} show the predictions of
various composite alternatives to the Standard Model. As already mentioned,
leptophilic models (represented by the horizontal line) are strongly
disfavoured, as are `anti-dilaton' models with $a = - c$ (downwards-sloping line). The global analysis
is compatible with the MCHM4 and dilaton models {\it iff} they are tuned to
resemble the Standard Model, with $\xi \sim 0, f \sim v$ in the MCHM4 (\ref{MCHM4}) or
$V \sim v$ in the pseudo-dilaton (\ref{pseudoD}) model (upwards-sloping line). Likewise,
the MCHM5 model is compatible with the data only if $\xi \sim 0$ in (\ref{MCHM5}).
Clearly, there is no evidence for any significant deviation from the Standard Model,
and Peter Higgs may continue to smile!

\begin{figure}[htbp!]
\centering
\includegraphics[scale=0.6]{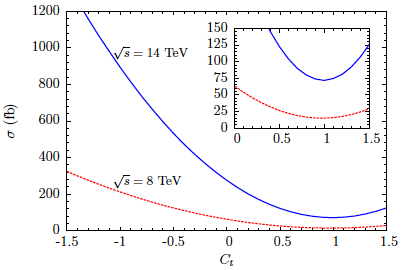}
\includegraphics[scale=0.35]{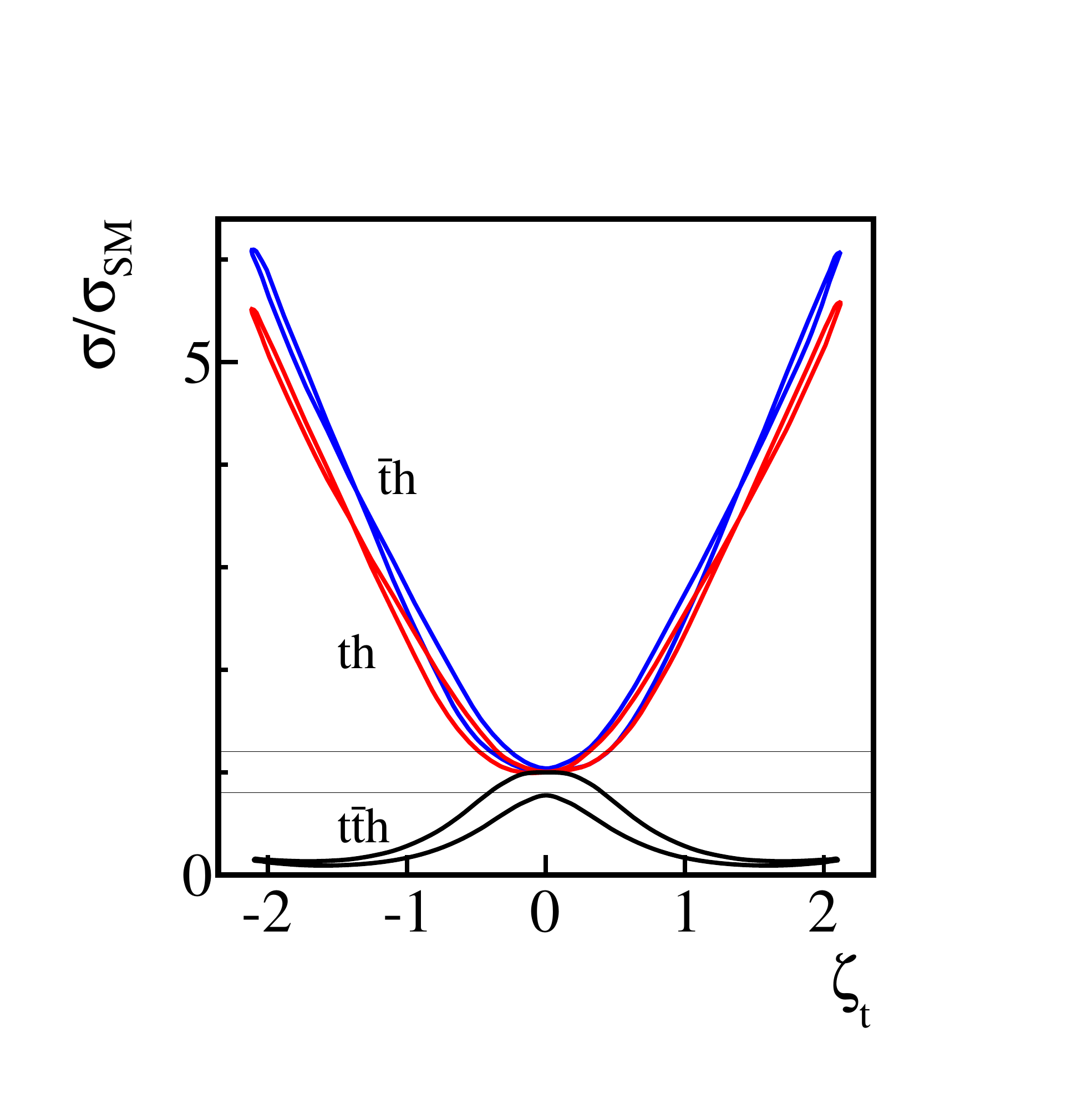}
\caption{\it Left panel: The cross section for single $t + H$ production as a function of the scalar top-$H$
coupling $c_t$ normalized to its Standard Model value~\protect\cite{Biswas}. Right panel: The cross sections
for ${\bar t}t + H$, single $t$ and ${\bar t} +H$ production relative to their Standard Model values for
ranges of the scalar and pseudoscalar couplings $(c_t, {\tilde c_t})$ compatible with current data on
$gg \to H$ production and $H \to \gamma \gamma$ decay~\protect\cite{EY3}, 
as functions of $\zeta_t \equiv {\rm arc~tan} (c_t/{\tilde c_t)}$~\protect\cite{EHST}.}
\label{fig:signt}
\end{figure}

Before concluding this Section, it is interesting to discuss in more detail a Higgs production channel that
could give direct information on the sign and magnitude of $c$, namely single $t$
(or ${\bar t}$) production in association with $H$~\cite{Maltoni,Biswas}. The two dominant amplitudes are
due to Higgsstrahlung from an exchanged $W$ boson and the final-state $t$ quark.
In the Standard Model with $c > 0$, these diagrams interfere destructively, as a
precursor of the good high-energy behaviour expected in a spontaneously-broken
gauge theory, whereas if $c < 0$ the production cross section may be much larger.
Even establishing an upper limit on single $t$ (or ${\bar t}$) production in association with $H$
may be sufficient to determine the sign of $c$, independently of the $\gamma \gamma$ measurement~\cite{Biswas}.
One may also consider the possibility of a CP-violating ${\bar t} t H$ vertex ${\tilde c_t}$ in addition to a 
conventional scalar vertex with coefficient $c_t$ relative to the Standard Model value. The right panel of
Fig.~\ref{fig:signt} shows the dependences of the ${\bar t}t H$, $t H$ and ${\bar t}H$ cross sections on
$\zeta_t \equiv {\rm arc tan} ({\tilde c_t}/c_t)$~\cite{EHST} for choices of the $c_t$ and ${\tilde c_t}$ that are compatible with the
constraints on the $H gg$ and $H \gamma \gamma$ couplings shown in Fig.~\ref{fig:cgammacg}~\cite{EY3}.
We see that measurements of the ${\bar t}t H$, $t H$ and ${\bar t}H$ cross sections could provide interesting
information on the top-$H$ couplings.

\subsubsection{Are its Couplings Proportional to Particle Masses?}

\begin{figure}[h!]
\centering
\includegraphics[scale=0.35]{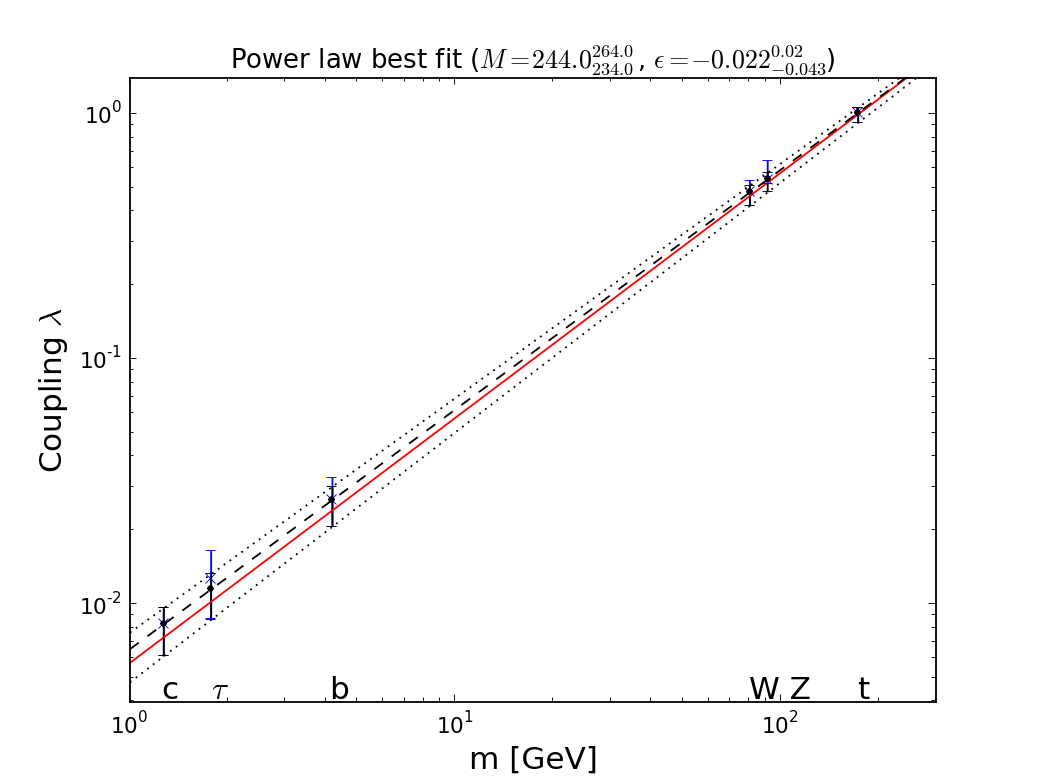}
\includegraphics[scale=0.35]{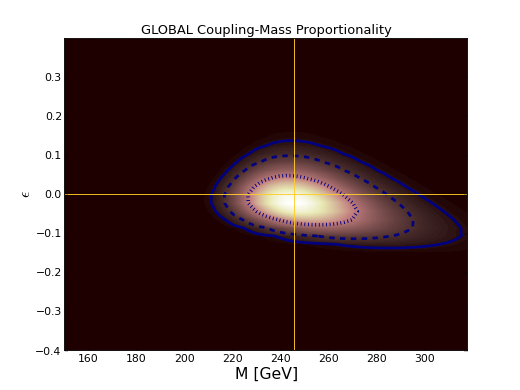}
\caption{\it The constraints on $M$ and $\epsilon$ (\ref{Mepsilon}) imposed by the measurements in Fig.~\ref{fig:EY1}.
The left panel shows the strengths of the couplings to different fermion flavours and massive bosons predicted
by this two-parameter $(M, \epsilon)$ fit. The red line is the Standard Model prediction, the
black dashed line is the best fit, and the dotted lines are the 68\% CL ranges. For each particle species,
the black error bar shows the range predicted by the global fit, and the blue error bar shows the range predicted for
that coupling if its measurement is omitted from the global fit. The right panel displays the fit constraint in the
$(M, \epsilon)$ plane~\protect\cite{EY3}.}
\label{fig:Mepsilon}
\end{figure}

\begin{figure}[h!]
\centering
\includegraphics[scale=0.35]{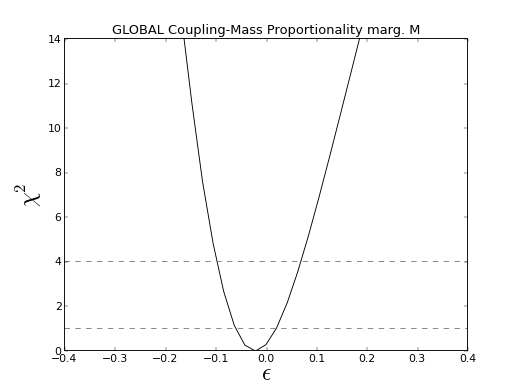}
\includegraphics[scale=0.35]{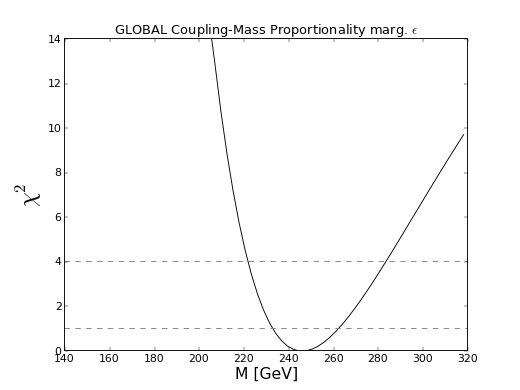}
\caption{\it The one-dimensional $\chi^2$ functions for $\epsilon$ (left panel) and $M$ (right panel),
as obtained by marginalizing over the other fit parameter~\protect\cite{EY3}.}
\label{fig:epsilonandM}
\end{figure}

It is a key property of the Higgs boson of the Standard Model that its couplings to other
particles should be proportional to their masses, and this is verified implicitly by the
type of analysis reviewed in the previous Section. In order to verify it more explicitly,
we may consider a parametrization of the $H$ couplings to fermions $\lambda_f$
and massive bosons $g_V$ of the form~\cite{EY2,EY3}
 \begin{equation}
\lambda_f \; = \; \sqrt{2} \left(\frac{m_f}{M}\right)^{1 + \epsilon}, \; g_V \; = \; 2 \left(\frac{m_V^{2(1 + \epsilon)}}{M^{1 + 2\epsilon}}\right) \, .
\label{Mepsilon}
\end{equation}
In the Standard Model, one would expect the power $\epsilon = 0$ and the scaling
coefficient $M = v = 246$~GeV. The results of a fit in terms of the two parameters $(M, \epsilon)$
is shown in Fig.~\ref{fig:Mepsilon}.  It is represented in the left panel by the dashed line,
with the one-$\sigma$ excursions shown as dotted lines. The solid red line is the
prediction of the Standard Model, and the points with error bars are the predictions of
the two-parameter fit. We see that these are completely compatible with the Standard
Model predictions.
In the right panel of Fig.~\ref{fig:Mepsilon} we see the 68 and 95\% CL regions given
by the fit in the $(M, \epsilon)$ plane. Here the solid horizontal and vertical lines
represent the Standard Model predictions $\epsilon = 0$ and $M = 246$~GeV.
The data are quite close to the bull's eye!
We display in the left panel of Fig.~\ref{fig:epsilonandM} the one-dimensional $\chi^2$ function 
for $\epsilon$, marginalized over $M$, and in
the right panel the one-dimensional $\chi^2$ function for $M$, marginalized over $\epsilon$.
The central values and the 68\% CL ranges of $M$ and $\epsilon$ are:
\begin{equation}
M \; = \; 244^{+ 20}_{- 10}~{\rm GeV} \; , \; \epsilon \; = \; -0.022^{+ 0.042}_{- 0.021} \, .
\label{Mepsilonvalues}
\end{equation}
As we wrote in~\cite{EY2}: {\it "It walks and quacks like a Higgs boson."}

\subsubsection{Are there Extra Contributions to its Loop Couplings?}

\begin{figure}[h!]
\centering
\includegraphics[scale=0.5]{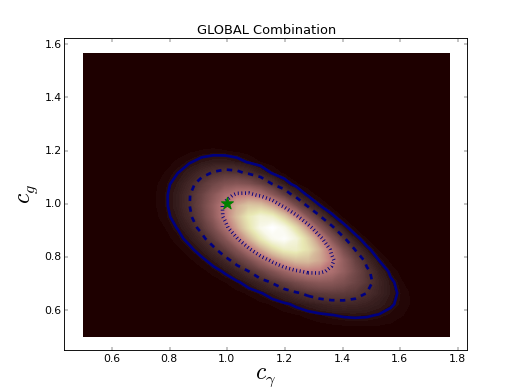}
\caption{\it The constraints in the $(c_\gamma, c_g)$ plane imposed by the measurements in Fig.~\ref{fig:EY1},
assuming that $a = c = 1$, i.e., the Standard Model values for the tree-level couplings to massive 
bosons and fermions~\protect\cite{EY3}.}
\label{fig:cgammacg}
\end{figure}

\begin{figure}[h!]
\centering
\includegraphics[scale=0.35]{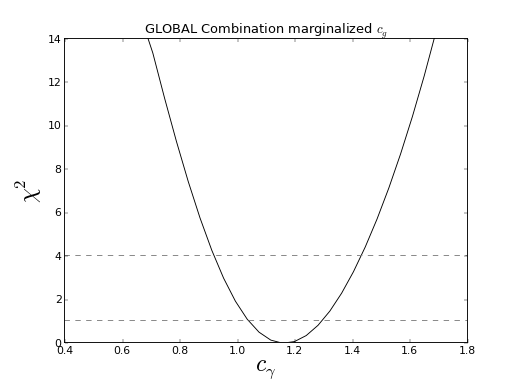}
\includegraphics[scale=0.35]{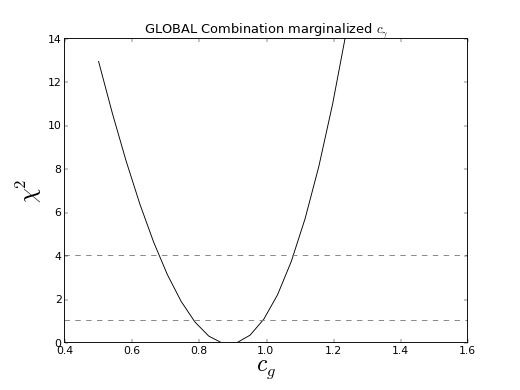}
\caption{\it The one-dimensional $\chi^2$ functions for $c_\gamma$ (left panel) and $c_g$ (right panel),
assuming that $a = c = 1$, so that the tree-level couplings to massive bosons and fermions
have the Standard Model values~\protect\cite{EY3}.}
\label{fig:cgammaandcg}
\end{figure}

The previous two Sections show that the tree-level $H$ couplings are similar to
those of a Standard Model Higgs boson. What can one say about its loop couplings
to $gg$ and $\gamma \gamma$? Here we assume that its couplings to fermions and
massive vector bosons are indeed Standard Model-like, so that $a = c = 1$, and
investigate whether there is any evidence for other coloured (in the case of the $gg$
coupling) or charged (in the case of the $\gamma \gamma$ coupling) particles
contributing via triangular loop diagrams, so that $c_g$ and/or $c_\gamma \ne 1$ in (\ref{anomL}).
We see in Fig.~\ref{fig:cgammacg} that the central value of $c_\gamma > 1$ and the central value of
$c_g < 1$~\cite{EY3}:
\begin{equation}
c_\gamma \; = \; 1.18 \pm 0.12 \; , \; c_g \; = \; 0.88 \pm 0.11 \, .
\label{cgammacgvalues}
\end{equation}
However, the data are compatible with the Standard Model at the 68\% CL, as seen by the
location of the green star in Fig.~\ref{fig:cgammacg}. Thus, there is no good evidence for
new particles circulating in loop diagrams. Fig.~\ref{fig:cgammaandcg} displays
the one-dimensional $\chi^2$ functions for $c_\gamma$ (left panel) and $c_g$ (right panel),
assuming, as above that $a = c = 1$, so that the the tree-level couplings to massive bosons and fermions
have the Standard Model values.

\subsubsection{What is its Total Decay Rate?}

\begin{figure}[h!]
\centering
\includegraphics[scale=0.35]{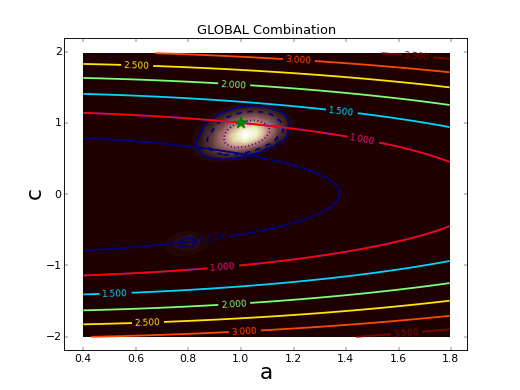}
\includegraphics[scale=0.35]{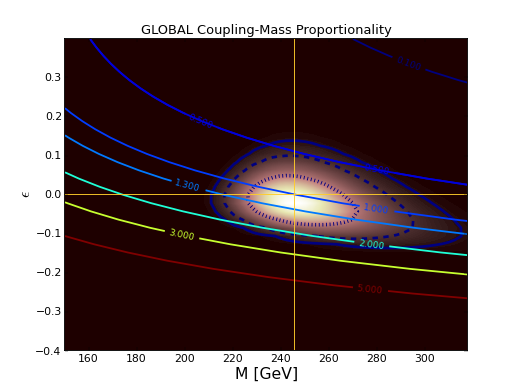}
\caption{\it Contours of the total Higgs decay rate relative to the Standard Model prediction
in the $(a, c)$ plane shown in the bottom right panel of Fig~\ref{fig:ac} (left) and the $(M, \epsilon)$ plane 
shown in the right panel of Fig.~\ref{fig:Mepsilon} (right)~\protect\cite{EY3}.}
\label{fig:Rtotal}
\end{figure}

\begin{figure}[h!]
\centering
\includegraphics[scale=0.5]{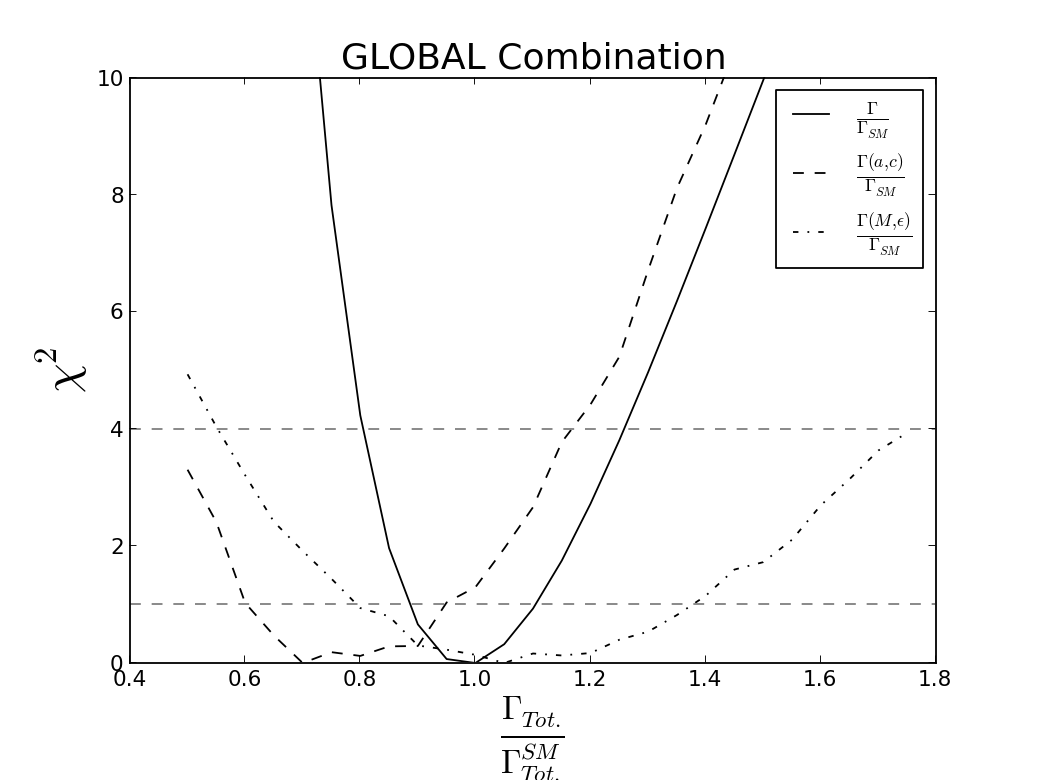}
\caption{\it The one-dimensional $\chi^2$ function for the
total Higgs decay width relative to its value in the Standard Model, $R \equiv
\Gamma/\Gamma_{SM}$, assuming decays into Standard Model particles alone and
assuming $a = c$ or equivalently $\epsilon = 0$ (solid line), marginalizing over $(a, c)$
(dashed line) and marginalizing over $(M, \epsilon)$ (dot-dashed line)~\protect\cite{EY3}.}
\label{fig:Rtotalmarginalized}
\end{figure}

We now assume that the Higgs has no other decays beyond those in the Standard Model,
and discuss the total Higgs decay rate in the two global fits discussed above,
in terms of the parameters $(a, c)$ and $(M, \epsilon)$ and assuming no 
contributions from non-Standard-Model particles.
The left panel of Fig.~\ref{fig:Rtotal} displays contours of the Higgs decay width relative to the
Standard Model prediction in the $(a, c)$ plane shown in the bottom right panel of Fig.~\ref{fig:ac},
and the right panel of Fig.~\ref{fig:Rtotal} displays analogous contours in the
$(M, \epsilon)$ plane. We see that in each case the best fit has a total decay rate
close to the Standard Model value. Fig.~\ref{fig:Rtotalmarginalized} displays
the one-dimensional $\chi^2$ function for the total Higgs decay width
relative to its Standard Model value. The solid line is obtained assuming that 
$a = c$ (or, equivalently, that $\epsilon = 0$
but $M$ is free), the dashed line is obtained by marginalizing over $(a, c)$, and the dot-dashed line
is obtained by marginalizing over $(M, \epsilon)$. In all cases, we see that the total
$H$ decay width is compatible with the Standard Model prediction~\cite{EY3}.

In the absence of an assumption about $H$ decays into non-Standard Model particles,
it is difficult to obtain an accurate measurement of the total Higgs decay rate $\Gamma_H$. The CMS
Collaboration has given a model-independent upper limit of 3~GeV, based on the width
of the $H \to \gamma \gamma$ signal peak that they observe, which is dominated by
the experimental resolution. It has also been suggested~\cite{CM}
that one could establish an upper limit on $\Gamma_H$ using measurements of
$ZZ$ final states mediated by off-shell $H$ bosons. Using CMS data for $m_{ZZ} \in (100, 800)$~GeV,
it was estimated in~\cite{CM} that $\Gamma_H < 163$~MeV at the 95\% CL, it was suggested
that restricting to $M_{ZZ} > 300$~GeV this bound could be improved to $\Gamma_H < 88$~MeV, 
and it was suggested that the ultimate LHC sensitivity would be to $\Gamma_H \sim 40$~MeV (see also~\cite{RKE1}).
Similar sensitivity may be obtained from an analysis of off-resonance $W^+ W^-$ final states~\cite{RKE2}.

\begin{figure}[h!]
\centering
\includegraphics[scale=0.5]{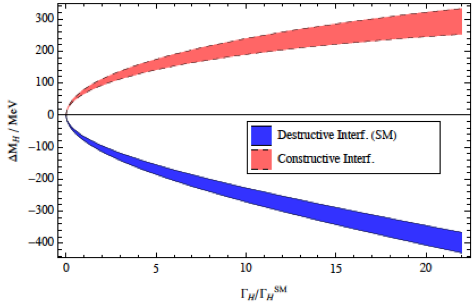}
\caption{\it The shift $\Delta M_H$ between measurements of the $H$ mass in the $\gamma \gamma$
and $4 \ell^\pm$ final states due to interference with QCD processes yielding $\gamma \gamma$ final states,
calculated in NLO QCD as a function of $\Gamma_H$ relative to its value in the Standard Model~\protect\cite{DL}.}
\label{fig:DL}
\end{figure}

Another way to constrain or measure $\Gamma_H$ may be via interference effects between
the QCD and $H$ contributions to the $\gamma \gamma$  final state, which could shift the $\gamma \gamma$
peak relative to its position in the $Z Z^* \to 4 \ell^\pm$ final state. (These are the only two 
observed $H$ states where the invariant mass can be measured accurately.) This mass shift is
sensitive to the sign and magnitude of the $H \gamma \gamma$ coupling, by an amount that
depends on the production kinematics. For $\Gamma_H$ similar to the Standard Model value, the
mass shift $\sim 70$~MeV, as seen in Fig.~\ref{fig:DL}~\cite{DL}, so this is not a measurement for the faint-hearted! The published
mass measurements have the problems that ATLAS and CMS find opposite signs for the
$\gamma \gamma$ and $4 \ell^\pm$ final states, though they are compatible within the
experimental uncertainties, and their sensitivity is not yet very interesting.

\subsubsection{The Story so Far}

The discovery of the $H$ particle has opened a new chapter in particle physics.
Alternatives to the expectation that it is a scalar boson have been excluded with
a high degree of confidence, and its couplings are consistent with those of a
Standard Model Higgs boson. In particular, they exhibit the expected
correlation with the masses of other particles. This is why we wrote in~\cite{EY3}: {\it ``Beyond any
reasonable doubt, it is a Higgs boson."}~\footnote{This phrase was quoted by the
Royal Swedish Academy of Sciences in the Advanced Information it released about
the award of the 2013 Nobel Physics Prize~\cite{Nobel}. Ironically, this phrase had been removed 
from the published version of~\cite{EY3} at the request of the referee, who found the
phrase {\it ``unscientific"}.}.

Experiments are placing severe constraint on composite models,
pushing upwards the possible scale of compositeness. On the other
hand, an elementary scalar is a challenge for theorists, in particular
because of the issue of quadratic divergences, that are symptomatic
of extreme sensitivity to details of the ultraviolet completion.
However, an elementary Higgs boson fits naturally within supersymmetry
and, as we shall see in the next Section, simple supersymmetric models
predict a Higgs mass below $\sim 130$~GeV, as observed, and also predict
that its couplings should be very similar to those of a Standard Model Higgs boson.
As yet, there are no signs of supersymmetric particles, and we wait with
interest to see what the LHC will reveal at 13/14~TeV, and in its high-luminosity
incarnation.

In 1982 I was introduced to Mrs. Thatcher
when she visited CERN, and she asked me what I did. I replied that {\it ``My job as a theorist
is to think of thing for the experiments to look for, but then we hope they find
something different."} Mrs. Thatcher liked things to be the way she wanted them to be, so
she asked {\it ``Wouldn't it be better if they found what you predicted?".} I responded along the lines
that {\it ``If they just find exactly what we predict, we would have no clues how to progress."}
In this spirt, let us all hope that the Higgs boson is not exactly that of the
Standard Model, and that higher-energy LHC running will reveal other new
physics beyond the Standard Model. In the next lecture I will discuss some of the
prospects for these hopes.

\section{What may the Future hold?}

\subsection{Theoretical Confusion}

So far, though experiments at the LHC have discovered the Higgs boson,
as yet they have found no direct hint of any new physics beyond the Standard Model
such as supersymmetry or compositeness.
The combination of these facts has caused a high mortality rate among theories,
though not among theorists!
As discussed previously, however, the fact that the measured values of
$m_H$ and $m_t$ lie in a region where the electroweak vacuum would be
unstable (OK, metastable) has led to suggestions that there should be new physics
below $10^{10}$~GeV to stabilize the vacuum. There have also been suggestions that the
proximity of $(m_H, m_t)$ to the stability boundary may be an indirect hint for some new
physical principle~\footnote{This is also sometimes linked to the fact that the cosmological
constant (dark energy) lies close to the upper bound proposed by Weinberg~\cite{WeinbergCC}.}.

Supersymmetrists are among the most confused theorists. Motivated by the apparent
fine-tuning of the electroweak scale and several phenomenological considerations
such as dark matter, many of them had expected supersymmetry to appear during
the first LHC run. In my view, this should be understood as the first run of the LHC
at or close to its design energy of 14~TeV, so we should wait a while before
jumping to conclusions. However, voices have been heard favouring
very high-scale supersymmetry or split supersymmetry. The faint-hearted are asking
whether we should modify or abandon the principle of naturalness. It is a
reasonable question whether Nature needs to care about the naturalness of the electroweak scale,
as long as she can find one set of parameters that includes the Standard Model.
This is one possibility opened up by the string landscape, which comprises an
exceedingly large number of possible vacua that all seem consistent with our
current understanding. My own point of view is that supersymmetry
anywhere would be better than nowhere, in terms of reducing the required
amount of fine-tuning. In any case, supersymmetry alone could not explain the hierarchy
between the electroweak and gravitational scales: another mechanism would be
needed to establish the hierarchy. New ideas are clearly needed!

In the absence of signatures of physics beyond the Standard Model at the LHC,
there has been a tendency among some physicists to wonder whether the
Standard Model is all there is, despite the persistence of a few loose ends such
as the hierarchy, dark matter, the origin of matter, quantum gravity, etc.
History gives many examples where such pessimism has turned out to
be unwarranted: consider the examples of Albert Michelson (1894) {\it ``The 
more important fundamental laws and facts of physical science have all been discovered"}
or Lord Kelvin (1900) {\it ``There is nothing new to be discovered in physics now. 
All that remains is more and more precise measurement"}, not to mention a
Spanish Royal Commission, rejecting the proposal of Christopher Columbus to sail west
(before 1492) {\it ``So many centuries after the Creation, it is unlikely that anyone could find 
hitherto unknown lands of any value"}.

Perhaps we should rather follow the approach of Sherlock Holmes in the ``Silver Blaze" story who,
when asked by a policeman {\it ``Is there any other point to which you would wish to draw my attention?"},
responded {\it ``To the curious incident of the dog in the night-time."} The policeman then remarked that
{\it ``The dog did nothing in the night-time"}, to which Holmes replied {\it ``That was the curious incident."}
In our case, the {\it "curious incident"} is that no beyond the Standard Model dog has yet barked. Nevertheless,
experiments have already provided theorists with many other clues: perhaps we need next to examine them
more carefully, as well as planning ambitious future experiments. These are the themes
of this Lecture.

\subsection{Additional Topics in Higgs Studies}

\subsubsection{Higher-Dimensional Operators}

\begin{figure}[t]
\centering
 \includegraphics[width=.46\columnwidth]{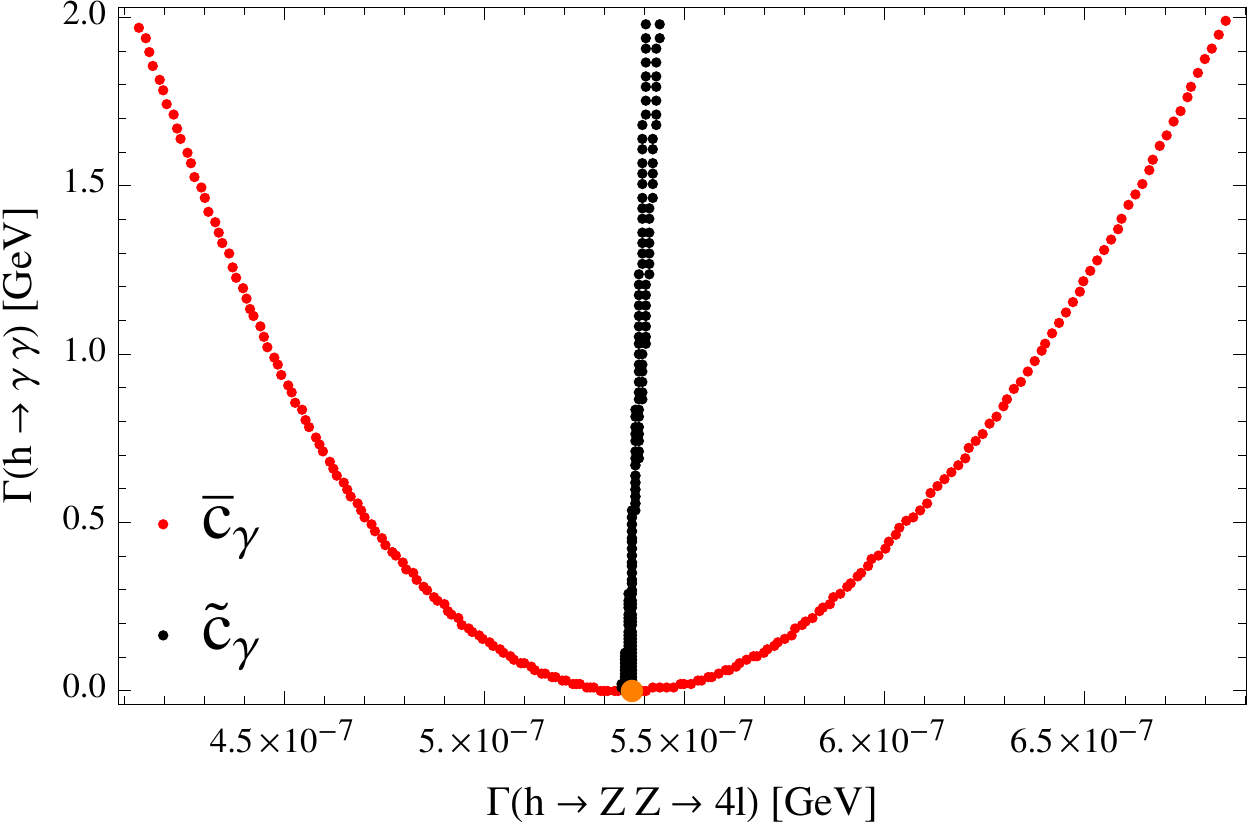}
 \includegraphics[width=.52\columnwidth]{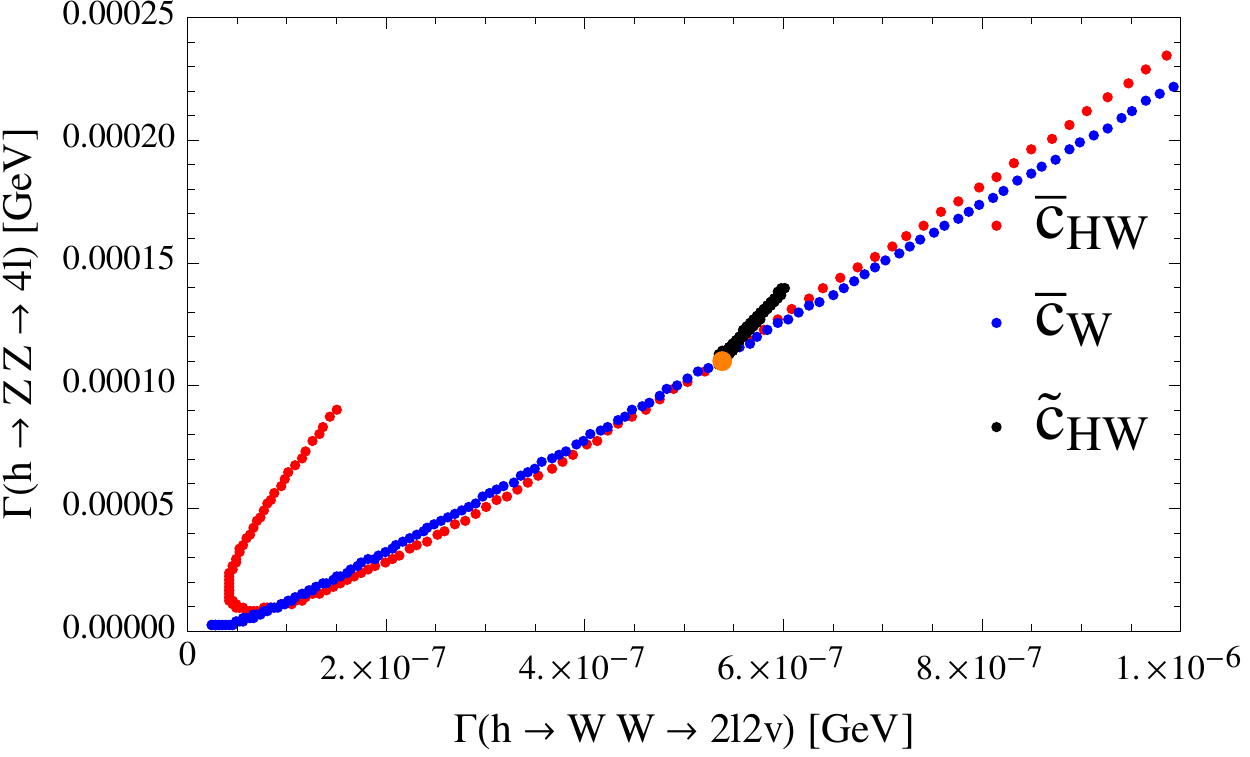}
\caption{\it Effects of the dimension-6 operators ${\bar c_{\gamma}}, { \tilde c_{\gamma}}, {\bar c_{HW}}, {\bar c_{W}}$ and ${ \tilde c_{HW}}$
from (\ref{SILH}, \protect\ref{SILHCPodd}) on the 
$H \to Z Z^*$ and $\gamma \gamma$ partial widths (left panel) and the
$H \to W W^*$ and $H \to Z Z^*$ partial widths (right panel). In each case, the Standard Model prediction
is indicated by an orange dot~\protect\cite{AFS}.}
\label{fig:AFS}
\end{figure}

A powerful way to probe indirectly possible physics beyond the Standard Model is to consider
additional higher-dimensional operators that might be generated by new physics such as
the exchanges of heavy particles, use data to constrain their coefficients, and thereby
constrain the possibilities for physics beyond the Standard Model. In principle, this
offers a way to constrain new physics in a coherent and effective way using a formalism that is consistent
with all the established gauge and other symmetries. As an example, the CP-conserving
dimension-6 operators in an effective Lagrangian involving just boson fields may be written in the form~\cite{AFS}:
\begin{eqnarray}
{\cal L}_{CP+6} & = &
    \frac{\bar c_{H}}{2 v^2} \partial^\mu\big[\Phi^\dag \Phi\big] \partial_\mu \big[ \Phi^\dagger \Phi \big]
  + \frac{\bar c_{T}}{2 v^2} \big[ \Phi^\dag {\overleftrightarrow{D}}^\mu \Phi \big] \big[ \Phi^\dag {\overleftrightarrow{D}}_\mu \Phi \big]
  - \frac{\bar c_{6} \lambda}{v^2} \big[H^\dag H \big]^3 \\
    & + & \frac{i g\ \bar c_{W}}{m_W^2} \big[ \Phi^\dag T_{2k} \overleftrightarrow{D}^\mu \Phi \big]  D^\nu  W_{\mu \nu}^k
  + \frac{i g'\ \bar c_{B}}{2 m_W^2} \big[\Phi^\dag \overleftrightarrow{D}^\mu \Phi \big] \partial^\nu  B_{\mu \nu} \\
  & + & \frac{2 i g\ \bar c_{HW}}{m_W^2} \big[D^\mu \Phi^\dag T_{2k} D^\nu \Phi\big] W_{\mu \nu}^k
  + \frac{i g'\ \bar c_{HB}}{m_W^2}  \big[D^\mu \Phi^\dag D^\nu \Phi\big] B_{\mu \nu} \\
  & + & \frac{g'^2\ \bar c_{\gamma}}{m_W^2} \Phi^\dag \Phi B_{\mu\nu} B^{\mu\nu}
   +\frac{g_s^2\ \bar  c_{g}}{m_W^2} \Phi^\dag \Phi G_{\mu\nu}^a G_a^{\mu\nu}\ .
   \label{SILH}
   \end{eqnarray}
The coefficients ${\bar c_{H}}, {\bar c_{T}}, {\bar c_{W}}, {\bar c_{B}}, {\bar c_{HW}}, {\bar c_{HB}}, {\bar c_{\gamma}}$
and ${\bar  c_{g}}$ may then be constrained using precision electroweak data, measurements of Higgs production
and decays, triple-gauge-boson couplings, etc. These constraints may then be compared with calculations in
specific extensions of the Standard Model such as supersymmetry or composite models. One can also consider a
CP-violating set of dimension-6 bosonic operators:
\begin{eqnarray}
\label{SILHCPodd}
  {\cal L}_{CP-6} & = &
    \frac{i g\ \tilde c_{HW}}{m_W^2}  D^\mu \Phi^\dag T_{2k} D^\nu \Phi {\widetilde W}_{\mu \nu}^k
  + \frac{i g'\ \tilde c_{HB}}{m_W^2} D^\mu \Phi^\dag D^\nu \Phi {\widetilde B}_{\mu \nu}
  + \frac{g'^2\  \tilde c_{\gamma}}{m_W^2} \Phi^\dag \Phi B_{\mu\nu} {\widetilde B}^{\mu\nu}\\
 & + &  \frac{g_s^2\ \tilde c_{g}}{m_W^2}      \Phi^\dag \Phi G_{\mu\nu}^a {\widetilde G}^{\mu\nu}_a
  \!+\!  \frac{g^3\ \tilde c_{3W}}{m_W^2} \epsilon_{ijk} W_{\mu\nu}^i W^\nu{}^j_\rho {\widetilde W}^{\rho\mu k}
  \!+\!  \frac{g_s^3\ \tilde c_{3G}}{m_W^2} f_{abc} G_{\mu\nu}^a G^\nu{}^b_\rho {\widetilde G}^{\rho\mu c} \ ,
\end{eqnarray}
where the dual field strength tensors are defined by
\begin{equation}
  \widetilde B_{\mu\nu} = \frac12 \epsilon_{\mu\nu\rho\sigma} B^{\rho\sigma} \ , \quad
  \widetilde W_{\mu\nu}^k = \frac12 \epsilon_{\mu\nu\rho\sigma} W^{\rho\sigma k} \ , \quad
  \widetilde G_{\mu\nu}^a = \frac12 \epsilon_{\mu\nu\rho\sigma} G^{\rho\sigma a} \ . 
\end{equation}
Two specific examples of possible effects on $H$ decays due to higher-dimensional
operators are shown in Fig.~\ref{fig:AFS}. In the left panel, we see the effects of the terms 
$\propto {\bar c_{\gamma}}$ and ${ \tilde c_{\gamma}}$ on $H \to ZZ^*$ and $\gamma \gamma$ decays,
and in the right panel we the effects of the $\propto {\bar c_{HW}}, {\bar c_{W}}$ and ${ \tilde c_{HW}}$
on $H \to ZZ^*$ and $W W^*$ decays,.

\begin{figure}[hbt]
  \centering
  \includegraphics[width=.55\columnwidth]{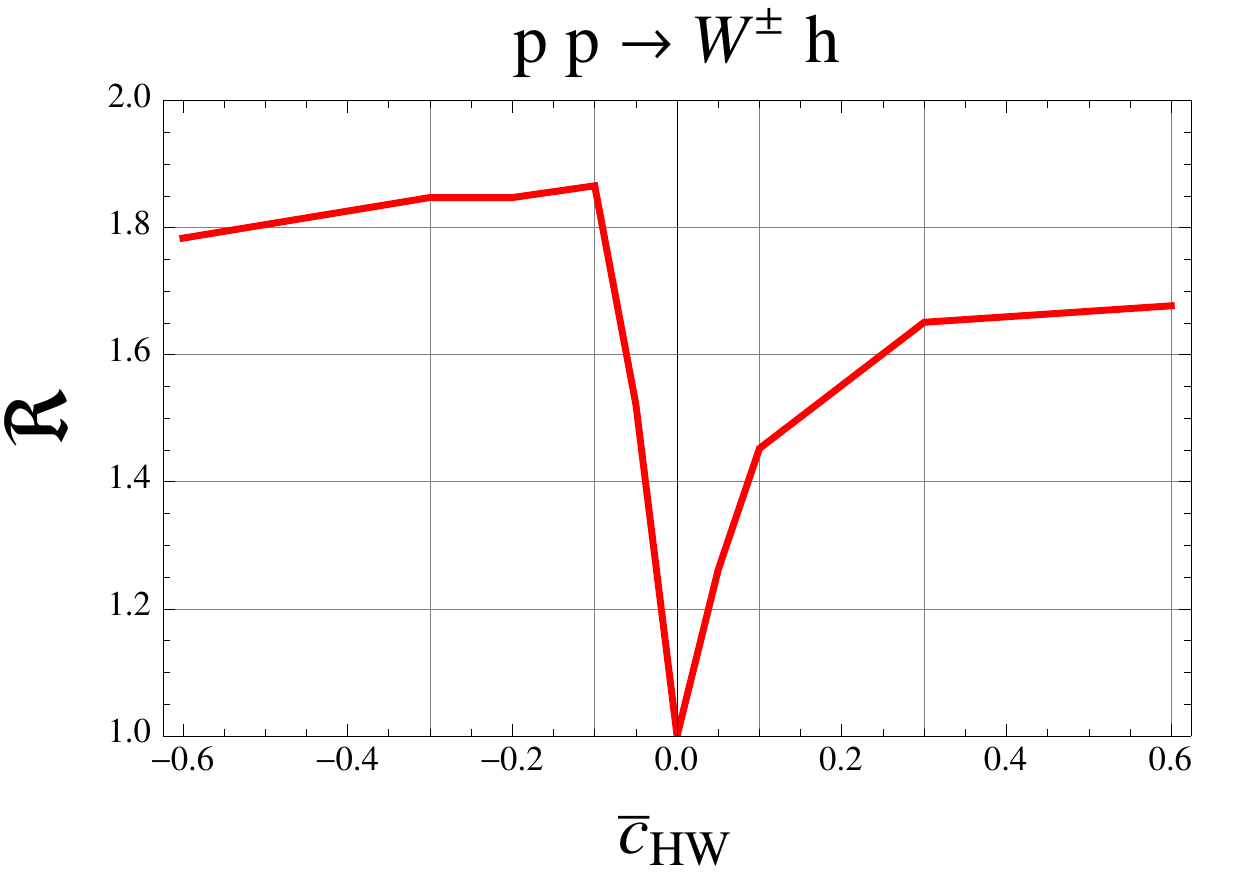}
    \includegraphics[width=.43\columnwidth]{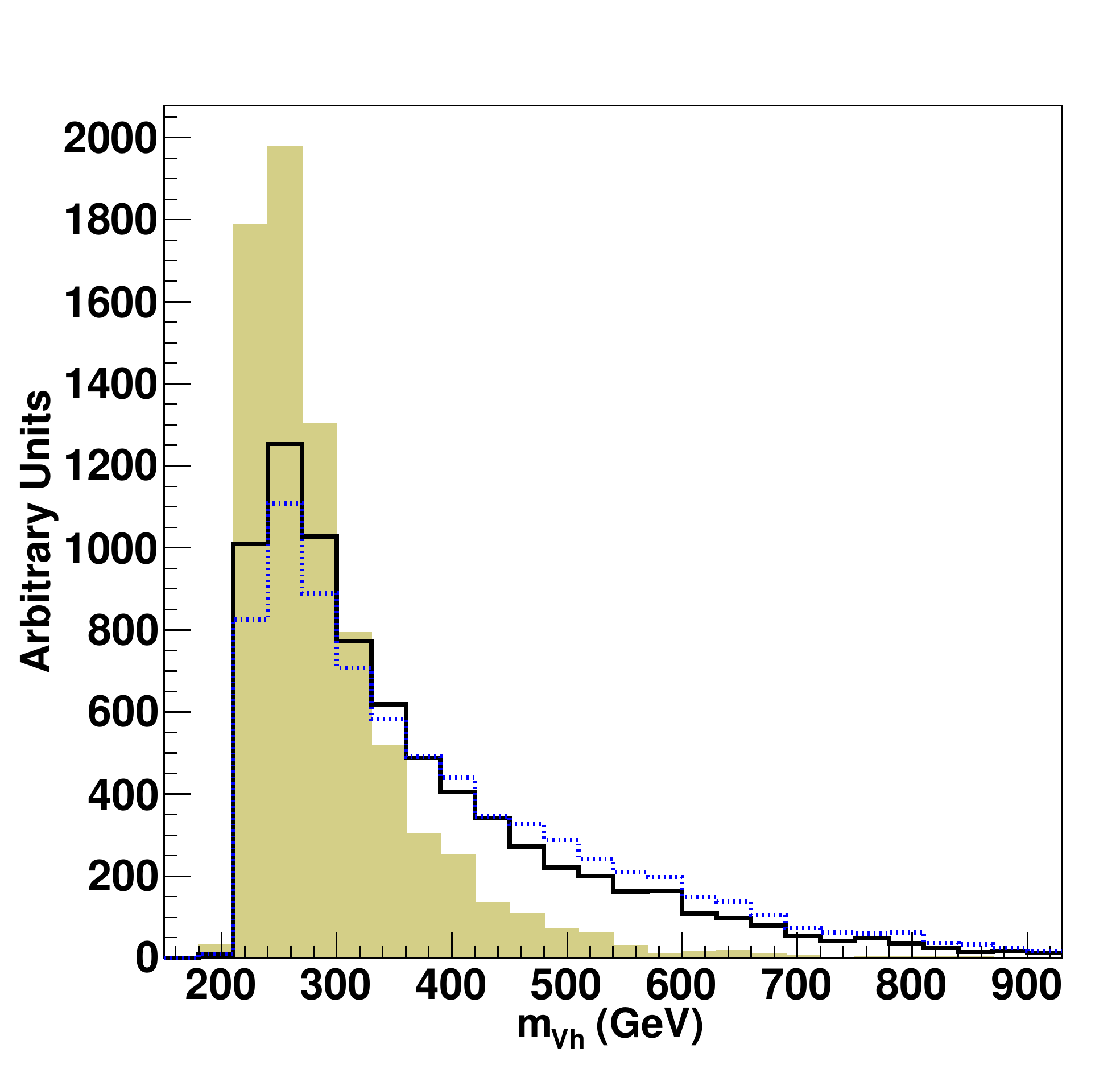}
\caption{
{\it Left panel: the double ratio ${\cal R}$ (\protect\ref{doubleR}) of total cross sections at $\sqrt{S} = 8$~TeV and 14~TeV
for the associated production process $p p \to W^\pm H \to \ell\nu b \bar b$
may provide  a useful constraint on the dimension-6 operator coefficient $\bar{c}_{HW}$.
Right panel: the invariant-mass distribution $m_{VH}$ is displayed for the Standard Model (shaded histogram) 
and with additional couplings $\bar{c}_{HW}=0.1$ (blue-dotted histogram) and $\bar{c}_{W}=0.1$ (black histogram)~\protect\cite{AFS}.}}
\label{fig:AFS2}
\end{figure}

Some of the operators in (\ref{SILH}) and (\ref{SILHCPodd}) may also affect the production
cross sections and kinematic distributions of the $H$ boson. An example is provided by the
double ratio of the cross sections for $H$ production in association with a vector boson at
14 and 8~TeV:
\begin{equation}
  {\cal R} \; \equiv \;
    \left(\frac{\sigma (\sqrt{S}= 14 \textrm{ TeV })}{\sigma (\sqrt{S}= 8 \textrm{ TeV })} \right)_{\bar{c}_{i}} /
    \left(\frac{\sigma (\sqrt{S}= 14 \textrm{ TeV })}{\sigma (\sqrt{S}= 8 \textrm{ TeV })} \right)_{SM} 
\label{doubleR} 
\end{equation}
in the presence of an operator with coefficient $\bar{c}_{i}$, as illustrated in the left panel of
Fig.~\ref{fig:AFS2} for the case of $\bar{c}_{HW}$. The right panel of Fig.~\ref{fig:AFS2}
illustrates the effects on the $WH$ invariant mass distribution for the cases $\bar{c}_{HW}$
(blue dotted histogram) and $\bar{c}_{W} = 0.1$ (black histogram), the shaded histogram
being the prediction of the Standard Model. We see that the double ratio (\ref{doubleR})
and the invariant mass distribution are interesting tools for constraining such operator
coefficients, just as they provide discrimination between the $0^+, 0^-$ and $2^+$
hypotheses for the $H$ spin.

\subsubsection{A or The?}

Now that the $H$ particle has been established 'beyond any reasonable doubt" to be
{\it a} Higgs boson, the questions arise whether it is {\it the} Higgs boson of the Standard Model,
and whether there are any others. Possibilities proposed include models with an extra singlet
field, models with a fermiophobic Higgs boson, and models with two Higgs doublets (2HDM)
such as the minimal supersymmetric extension of the Standard Model (the MSSM). The ATLAS
and CMS experiments have established upper limits on the couplings of possible massive $H^\prime$
boson, as we shall see later in connection with the MSSM.

Also on the agenda of the LHC experiments is to measure $VV$ scattering and make a 
closure test, so as to check the Standard Model $H$ cancellation discussed earlier. 
Does the Higgs boson discovered by ATLAS and CMS cure its
high-energy behaviour so that the theory is indeed renormalizable?

Another strategy is to search for non-Standard Model decays, e.g., into
invisible final states, or into pairs of light (pseudo)scalars $aa$, or into
lepton-flavour-violating final states such as $\mu \tau$ or $e \tau$, as discussed
in the next Section.

\subsubsection{Flavour-Changing $H$ Decays}

\begin{figure}[t]
\begin{center}
\includegraphics[width=0.15\textwidth]{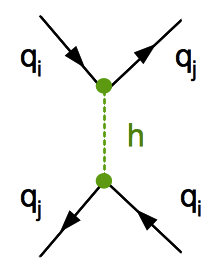}
\hskip 1 cm
\includegraphics[width=0.35\textwidth]{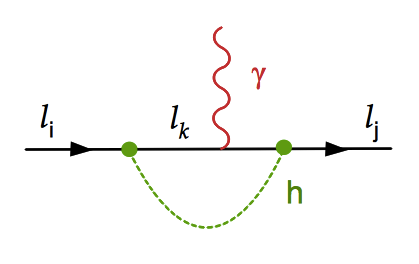}
\vskip -1 cm
\end{center}
\caption{\label{fig:BEI} \it Left: Tree-level diagram contributing to a generic flavour-changing amplitude
via $H$ exchange.
Right: One-loop $H$ loop diagram contributing to anomalous magnetic moments and electric dipole
moments of charged leptons ($i=j$), or radiative LFV decay modes  ($i \ne j$)~\protect\cite{BEI}. }
\end{figure}

In the Standard Model one expects flavour-changing Higgs decays to occur
only far below the sensitivity of present and prospective LHC measurements.
However, this might not be true, e.g., in some composite Higgs models, so
searches for flavour-violating $H$ decays present an interesting opportunity
to look for new physics beyond the Standard Model.
Model-independent constraints on such decays are provided by measurements
of flavour-violating processes at low energies, including the effects of
four-fermion interactions generated by $H$ exchange and $H$ loop contributions
to dipole moments, as illustrated in Fig.~\ref{fig:BEI}~\cite{BEI}.

Constraints from $\Delta F = 2$ processes such as $K - {\bar K}$, $D - {\bar D}$, $B - {\bar B}$
and $B_s - {\bar B_s}$ mixing constrain flavour-changing $H$ couplings so severely that quark-flavour-violating
$H$ decays are too suppressed to be detectable in the foreseeable future. However, the upper
limits on lepton-flavour-changing $H$ couplings are much weaker, and leave open the
possibility that either $H \to \tau \mu$ or $H \to \tau e$ might have a branching ratio as large
as $\sim 10$\%, comparable to the branching ratio for $H \to \tau \tau$. (The upper limit on
$\mu \to e \gamma$ forbids both branching ratios from being large simultaneously.) On the other hand,
the constraints  on the flavour-violating $H \mu e$ coupling from anomalous $\mu \to e$
conversion on nuclei and $\mu \to e \gamma$ are much stronger, and require
BR$(H \to \mu e) \la {\cal O}(10^{-9})$~\cite{BEI}.

\subsubsection{Measuring the Triple-Higgs Coupling}

If the Englert-Brout-Higgs field gives masses to elementary particles, and if this field is itself
elementary, and if the Higgs boson is the particle associated with this field, what gives a mass
to the Higgs boson? The answer within the Standard Model is the Englert-Brout-Higgs field
itself, via the triple-Higgs coupling. Examining the effective Lagrangian (\ref{EqLagHiggsFinal}),
we see the following terms:
\begin{equation}
\mathcal{L}_{\text{Higgs}} \; \ni \; -  \frac{\mu^2}{2} \eta^2 - \lambda v \eta^3 -  \frac{1}{4} \eta^4 \, ,
\label{trilinear}
\end{equation}
where $\eta$ denotes the quantum fluctuation in the Englert-Brout-Higgs field around its classical v.e.v.,
see (\ref{defeta}). 

The triple-Higgs coupling may be measured via $H$ pair production~\cite{DES}, which should be within
reach of the LHC with high luminosity~\footnote{Measuring the quadruple-Higgs coupling would
require measuring triple-$H$ production, which is likely to require a higher-energy collider
such as the VHE-LHC described later.}. The dominant mechanism for $HH$ production is
expected to be $gg$ fusion: $gg \to H^* \to HH$, with an important background from
$t$ and $b$ box diagrams for $gg \to HH$. Another strategy
for measuring the triple-Higgs coupling is indirectly via its effects on the cross section for $e^+ e^- \to Z + H$~\cite{McC}.

\subsection{Supersymmetry}

What else is there beyond the Higgs boson already discovered? Supersymmetry
is my personal favourite candidate for physics beyond the Standard Model~\cite{BCE}. In my view,
the discovery of a/the Higgs boson has strengthened the scientific case for
supersymmetry. In addition to the traditional arguments that low-energy
supersymmetry could resolve the fine-tuning (naturalness) aspect of the electroweak hierarchy
problem, could provide the astrophysical dark matter, could facilitate grand unification
and is essential (?) for string theory, we should remember that simple supersymmetric models
stabilize the electroweak vacuum, predicted
successfully the existence of a Higgs boson weighing $< 130$~GeV~\cite{130}, and also
predict (successfully, so far) that Higgs couplings should be within a few \% of their
Standard Model values. {\it No wonder I wrote the word `supersymmetry' in the largest
possible font on one of my slides!}

\begin{figure}
\vspace{-26cm}
\begin{center}
\includegraphics[height=35cm]{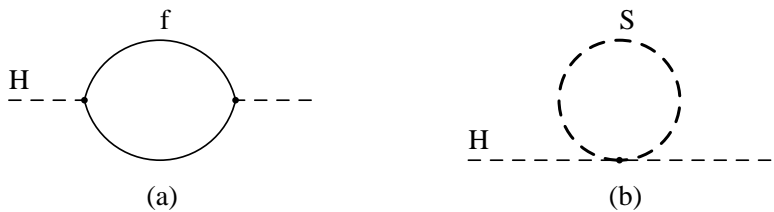}
\vspace{-6cm}
\caption{\it One-loop quantum corrections to the mass-squared of the Higgs boson due to (a) 
the loop of a generic fermion $f$, (b) a generic scalar $S$.
\label{fig:higgscorr1}}
\end{center}
\end{figure}

Historically, the first motivation for supersymmetry at the TeV scale came from considerations of
quantum (loop) corrections to the Higgs mass-squared, $m_H^2$, and thereby to the electroweak
scale~\cite{BCE}. For example, a generic fermionic loop such as that in Fig.~\ref{fig:higgscorr1}(a)
yields a correction:
\begin{equation}
\Delta m_H^2=-\frac{y_f^2}{8\pi^2}[2\Lambda^2 + 6m_f^2 \ln(\Lambda/m_f)+...],
\label{quadf}
\end{equation}
where $y_f$ is the Yukawa coupling: $y_f H\overline{\psi}\psi$, and
$\Lambda$ is an ultraviolet cutoff that represents the scale up to which
the Standard Model remains valid, beyond which new physics appears. This contribution
to the mass of the Higgs diverges quadratically with $\Lambda$. Hence if the Standard Model
were to remain valid up to the Planck scale, $M_P\simeq 10^{19}$ GeV, so that $\Lambda=M_P$,
this correction would be $\simeq 10^{34}$ times larger 
than the physical mass-squared of the Higgs, namely ${\cal}(10^2)$~GeV)$^2$.
Moreover, the loop of a scalar field $S$, shown in Fig.~\ref{fig:higgscorr1}(b), makes a
similarly divergent contribution:
\begin{equation}
\Delta m_H^2=\frac{\lambda_S}{16\pi^2}[\Lambda^2 - 2m_S^2 \ln(\Lambda/m_S)+...],
\label{quadS}
\end{equation}
where $\lambda_S$ is the quartic coupling of $S$ to the Higgs boson.

Comparing (\ref{quadf}) and (\ref{quadS}), we see that the quadratically-divergent terms
$\propto \Lambda^2$ would cancel if, corresponding to every fermion $f$ there is a 
scalar $S$ with quartic coupling
\begin{equation}
\lambda_S \; = \; 2 y_f^2 \, . 
\label{condition}
\end{equation}
This is exactly the relationship imposed by supersymmetry! Therefore, there are no
quadratic divergences in supersymmetric field theories, not just at the one-loop level
discussed above, but also at the multi-loop level~\footnote{Moreover, many
logarithmic corrections to couplings are also cancelled in a supersymmetric theory.}.
This means that if there is some
dynamical mechanism that imposes a large hierarchy
between different physical mass scales at the tree level, supersymmetry anables
it to be maintained in a natural way. 

A different motivation for supersymmetry is provided by the measured mass of the Higgs boson.
As already remarked, the electroweak vacuum would not be stable in the absence of any
new physics, since the (negative) renormalization by top quark loops would drive the
quartic Higgs self-coupling negative at some scale $\ll 10^{19}$~GeV, probably in the range
$10^{10}$ to $10^{13}$~GeV. This could be averted if there were some new physics to counteract
the negative renormalization by the top quark. In order to have the opposite sign to the top loop,
this new physics should be bosonic, much like the stop squark~\cite{EDAR}
But then one must consider all the quartic bosonic couplings
permitted (enforced) by renormalizability, end ensure that none of them blow up or generate an
instability, which requires fine-tuning to one part in $10^3$ in the simplest case studied.
However, this fine-tuning could be made more natural by postulating a new fermion, much like
the Higgsino. Thus, one finishes up with a theory that looks very much like supersymmetry!

Within a supersymmetric theory, the renormalization due to the top quark could prove
to be a blessing in disguise! After cancelling the quadratic divergences in
(\ref{quadf}, \ref{quadS}), one is left with residual logarithmic divergences that
can be resummed using the renormalization-group equations (RGEs). Not knowing
how supersymmetry is broken, one often assumes that this occurs far above the TeV scale,
e.g., around the grand unification or Planck scale, $M_{GUT}$ or $M_P$. In this case, the Higgs and other
supersymmetry-breaking masses for scalars and gauginos are renormalized significantly by time the electroweak scale
is reached. At leading order in the RGEs, which resum the leading one-loop logarithms,
the renormalizations of the soft gaugino masses $M_a$ coincide with the
corresponding gauge couplings:
\begin{equation}
Q \frac{d M_a}{d Q} \; = \; \beta_a M_a ,
\label{rengaugino}
\end{equation}
where $\beta_a$ is the one-loop renormalization coefficient including
supersymmetric particles. Hence, to leading order
\begin{equation}
M_a(Q) \; = \; \frac{\alpha_a(Q)}{\alpha_{GUT}} m_{1/2}
\label{gauginomasses}
\end{equation}
if the gaugino masses are assumed to have a universal value $m_{1/2}$ at the same large mass scale $M_{GUT}$
as the gauge couplings $\alpha_a$. For this reason, one expects the gluino to be
heavier than the wino: $m_{\tilde g}/m_{\tilde W} \simeq \alpha_3/\alpha_2$ and the
bino to be lighter again.

The gaugino masses contribute to the renormalizations
of the soft supersymmetry-breaking scalar masses-squared $m_i^2$  via the gauge couplings, and the scalar masses
and the trilinear soft supersymmetry breaking parameters $A_\lambda$ contribute via the Yukawa couplings:
\begin{equation}
\frac{Q d m^2_{i}}{d Q} \; = \; \frac{1}{16 \pi^2} \left[ - g_a^2 M_a^2 + 
\lambda^2 (m_i^2 + A_\lambda^2) \right] .
\label{renm0}
\end{equation}
For most of the scalar partners of Standard Model fermions, one has at leading order
\begin{equation}
m^2_i(Q)  \; = \; m_i^2 + C_i m_{1/2}^2 ,
\label{scalarmasses}
\end{equation}
where the coefficients $C_i$ depends on the gauge quantum numbers of the
corresponding fermion. Since renormalization by the strong coupling is largest, one expects the squarks to be
heavier than the sleptons. Specifically, if all the $m_i$ and the $M_a$ are each assumed to be universal
at the GUT scale (a scenario known as the CMSSM), at the
electroweak scale one finds:
\begin{eqnarray}
{\rm Squarks}: \; m^2_{\tilde q} \; & \sim & \; m_0^2 + 6 m^2_{1/2} , \\
{\rm Left-handed~sleptons}: \; m^2_{\tilde \ell_L} \; & \sim & \; m_0^2 + 0.5 m^2_{1/2} , \\
{\rm Right-handed~sleptons}: \; m^2_{\tilde \ell_R} \; & \sim & \; m_0^2 + 0.15 m^2_{1/2} .
\label{numm0}
\end{eqnarray}
Typical results of calculations of these renormalization effects
in the CMSSM are shown in Fig.~\ref{fig:masses}. 

\begin{figure}[htbp!]
\begin{center}
{\includegraphics[height=3in]{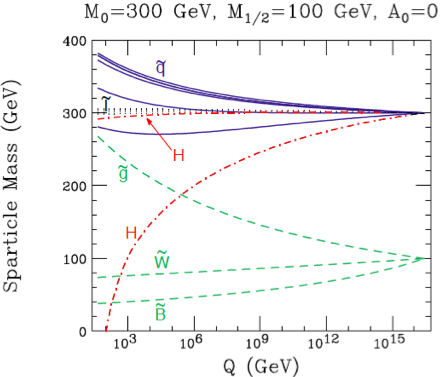}}
\end{center}
\caption[]{\it Results of calculations of the renormalization of soft supersymmetry-breaking
sparticle masses, assuming universal scalar and gaugino masses $m_0, m_{1/2}$
at $M_{GUT}$. In general, strongly-interacting sparticles have larger physical
masses at low scales, and the $m^2_{H_u}$ is driven negative, triggering
electroweak symmetry breaking.} 
\label{fig:RGEmasses}
\end{figure}

Supersymmetry requires at least two Higgs doublets, one to give masses to charge-(+2/3) quarks,
$H_u$, and the other to give masses to charge-(-1/3) quarks and charged leptons, $H_d$, and
we denote the ratio of their v.e.v.s as $\tan \beta$.
As we see in Fig.~\ref{fig:RGEmasses}, renormalization by the top quark Yukawa coupling is important for one of the Higgs
multiplets~\footnote{Renormalization by the other third-generation
sfermions may also be important if $\tan \beta$ is large.},
and may drive $m_{H_u}^2$ negative at the electroweak scale.
This may explain the negative sign of the quadratic term in the
effective Standard Model potential, and would trigger electroweak
symmetry breaking. If the top quark is heavy, it is possible for the electroweak scale to be generated naturally
at a scale $\sim 100$~GeV if $m_t \sim 100$~GeV. For this reason, supersymmetry theorists
actually suggested that the top quark should be heavy, before its discovery.

The two complex Higgs doublets of the MSSM have eight degrees of freedom,
of which three are used by the Higgs mechanism for electroweak breaking to give 
masses to the $W^\pm$ bosons and to the $Z^0$, 
leaving five physical Higgs bosons in the physical spectrum. Of these, two ($h, H$) are
neutral Higgs bosons that are CP-even (scalar), one ($A$) is neutral and CP-odd (pseudoscalar),
and two are charged, the $H^{\pm}$.
At tree level, the masses of the scalar supersymmetric Higgs bosons are:
\begin{equation}
m^2_{h, H} \; = \; \frac{1}{2}\left(m_{A}^2+m_Z^2 \mp 
\sqrt{(m_{A}^2+m_Z^2)^2 -4m_{A}^2m_Z^2\cos^2 2\beta}\,\right) \, ,
\label{hH}
\end{equation}
and the mass of the $h$ is bounded from above by $m_Z$.
This upper limit arises because the quartic Higgs
coupling $\lambda$ is fixed in the MSSM to be equal to the square of the electroweak gauge
coupling (up to numerical factors), so that $\lambda$ and hence $m_{h^0}$
cannot be very large.

\begin{figure}[htbp!]
\begin{center}
\includegraphics[height=7cm]{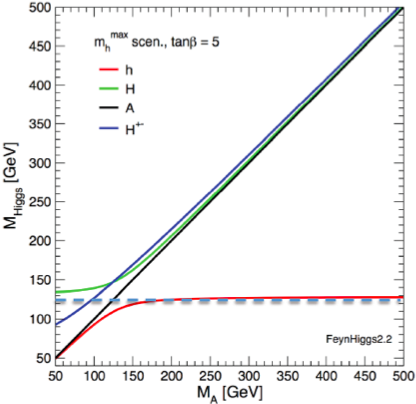}
\caption{\it The masses of the supersymmetric Higgs bosons as functions of $m_{A}$ for 
fixed values of the other MSSM parameters.
\label{fig:mhSUSY}}
\end{center}
\end{figure}

However, there are important radiative corrections to the above relations~\cite{130},
the most important correction for $m_h$ being the one-loop effect of the top quark and stop squark:
\begin{equation}
\Delta m_h^2=\frac{3m_t^4}{4\pi^2 v^2}\ln\left(\frac{m_{\tilde{t}_1}m_{\tilde{t}_2}}{m_t^2}\right) 
+ \dots \, ,  
\label{deltamh}
\end{equation}
where $m_{\tilde{t}_{1,2}}$ are the physical masses of the stops. The correction $\Delta m_h^2$ 
(\ref{deltamh}) depends quartically on the mass of the top, and after including this and higher-order corrections
the mass of the lightest Higgs boson may  be as large as~\cite{130,FH2100}:
\begin{equation}
m_h \lesssim 130\ \mathrm{GeV}.
\label{hmass}
\end{equation}
for stop masses of about a TeV, as seen in Fig.~\ref{fig:mhSUSY}.
The uncertainty in the calculation of $m_h$ for given values of the
supersymmetric model parameters is typically $\sim 1.5$~GeV.
As noted earlier, the range (\ref{hmass}) is perfectly consistent
with the mass measured by ATLAS and CMS,
yet another attractive feature of supersymmetry.

\begin{figure}[htbp!]
\begin{center}
{\includegraphics[height=1.5in]{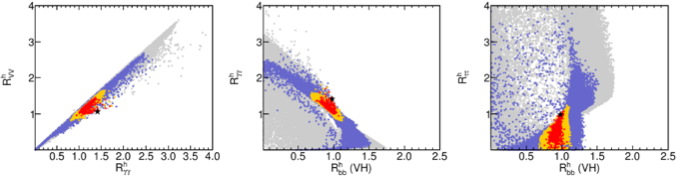}} \\
{\includegraphics[height=1.5in]{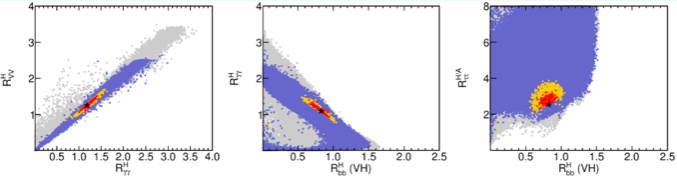}}
\end{center}
\caption[]{\it Results of fits to Higgs data assuming (upper panel) that the Higgs boson discovered
is the lightest supersymmetric Higgs $h$ (upper panel) and in which it is the
heavier scalar $H$ (lower panel)~\protect\cite{lighter}.} 
\label{fig:compare}
\end{figure}

In general, the couplings of the supersymmetric Higgs bosons differ from those in the
Standard Model.
\begin{eqnarray}
g_{hVV} & = & \sin (\beta - \alpha) g^{SM}_{HVV} \, , \\\
g_{HVV} & = & \cos (\beta - \alpha) g^{SM}_{HVV} \, , \\
g_{hAZ} & = & \cos (\beta - \alpha) \frac{g^\prime}{2 \cos \theta_W} \, , \\
g_{h {\bar b} b}, g_{h \tau^+ \tau^-} & = & -\frac{\sin \alpha}{\cos \beta} g^{SM}_{h {\bar b} b}, g^{SM}_{h \tau^+ \tau^-} \, .
\label{hHAcouplings}
\end{eqnarray}
If $m_A \gg m_W$, the masses of the other four Higgs bosons are
very similar: $m_H \sim m_A \sim m_{H^\pm}$. However, there is a different and interesting
possibility of $m_A$ is small, namely that $m_H \sim 125$~GeV, in which case the Higgs
discovered at the LHC might actually be the second-lightest Higgs boson, and there might
be a lighter one waiting to be discovered~\cite{lighter}~\footnote{There are more possibilities for
a lighter Higgs boson in the next-to-minimal supersymmetric extension of the Standard Model (NMSSM).}. 
Fig.~\ref{fig:compare} compares the predictions
for various Higgs decays, relative to their Standard Model values, for fits in which the Higgs
boson discovered is assumed to be the lightest one $h$ (upper panel) and in which it is the
heavier scalar $H$ (lower panel). Overall, the quality of the conventional fit is better, but the
unconventional fit may not yet be excluded. Experiments should continue the search for a
lighter Higgs boson, remembering that it might have different couplings from those in the
Standard Model.

\subsection{Higgs and Supersymmetry}

\begin{figure}[htbp!]
\begin{center}
\includegraphics[height=7cm]{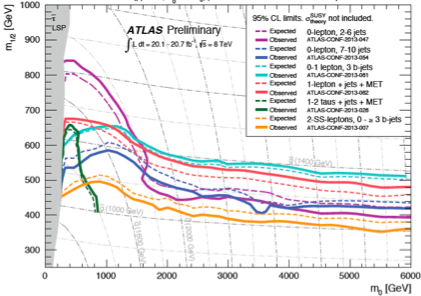}
\caption{\it Constraints on the universal soft supersymmetry-breaking scalar masses $m_0$,
and gaugino masses $m_{1/2}$ of the CMSSM from ATLAS searches for supersymmetry 
with the full Run 1 data set of $\sim 20$/fb of data at 8~TeV, using signatures with missing
transverse energy (MET), jets, leptons and $b$ quarks, assuming trilinear parameters 
$A_0 = - 2 m_0$ at the input GUT scale and $\tan \beta = 30$~\protect\cite{ATLASMET}.
\label{fig:ATLAS}}
\end{center}
\end{figure}

Let us now explore the implications for supersymmetry of the Higgs discovery,
assuming that it is indeed the lighter scalar supersymmetric Higgs boson $h$.
Important constraints on supersymmetric models are imposed by electroweak
precision observables and flavour physics observables, the cosmological
density~\cite{EHNOS} and astrophysical searches for cold dark matter~\cite{LUX}, as well as LHC searches.
In the following, we also take into account the experimental measurement of the
anomalous magnetic moment of the muon, $g_\mu - 2$, which disagrees with
theoretical calculations within the Standard Model by $\sim 3$ standard deviations~\cite{g-2}.
This discrepancy could be explained by supersymmetry at a relatively low mass scale,
although this possibility is disfavoured in simple supersymmetric models by the LHC Higgs
mass measurement and the absence (so far) of direct evidence for supersymmetric
particles at the LHC. Fig.~\ref{fig:ATLAS} displays the relevant constraints provided
by various ATLAS searches for supersymmetry with the full Run 1 data set of
$\sim 20$/fb of data at 8~TeV~\cite{ATLASMET}, using signatures with missing
transverse energy (MET), jets, leptons and $b$ quarks, interpreted within the CMSSM
in which there are universal soft supersymmetry-breaking scalar masses $m_0$,
gaugino masses $m_{1/2}$ and trilinear parameters $A_0 = - 2 m_0$ at the input
GUT scale, assuming that $\tan \beta = 30$. We see that at small $m_0$ the most
important constraint is provided by searches for jets + MET, whereas searches for
leptons, $b$-jets and MET are more important at large $m_0$.

In the following I present some results from a recent analysis~\cite{MC9} of these
constraints made using the {\tt MasterCode} framework~\cite{MasterCode}, which
incorporates a code for the electroweak
observables based on~\cite{Svenetal}, the flavour codes {\tt SuFla}~\cite{SuFla}
and {\tt SuperIso~3.3}~\cite{SuperIso}, {\tt SoftSUSY~3.3.9}~\cite{Allanach:2001kg}
and {\tt FeynHiggs} {\tt 2.10.0}~\cite{FH2100} for spectrum calculations, and the 
{\tt MicrOMEGAs~3.2}~\cite{MicroMegas} code for dark matter, 
which are interfaced using the SUSY Les Houches
Accord~\cite{SLHA}. We use the {\tt MasterCode} framework to construct
a global likelihood function ($\chi^2$) that includes contributions from
all the relevant observables.

\begin{figure}[htbp!]
\begin{center}
\includegraphics[height=7cm]{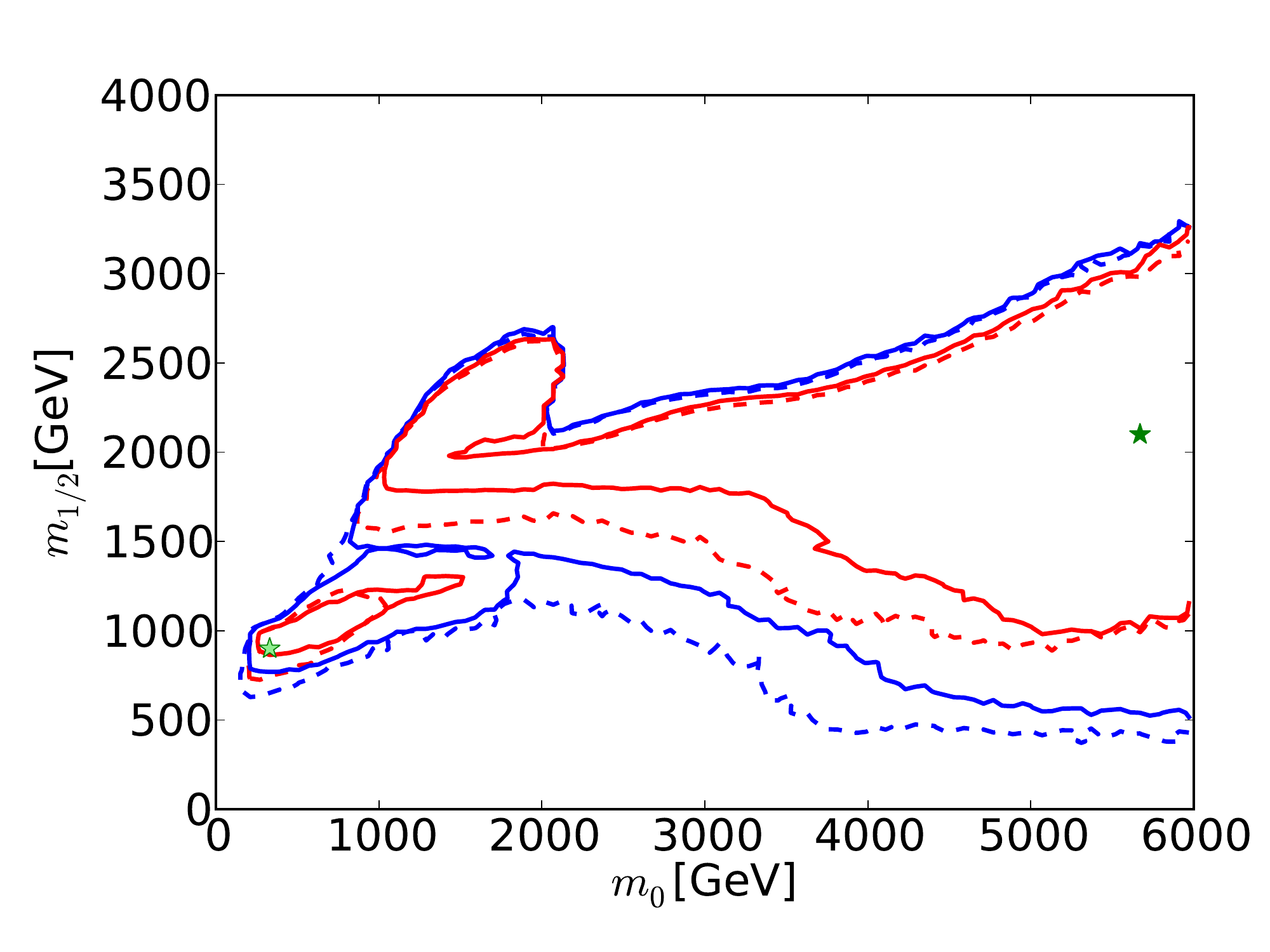}
\caption{\it The $(m_0, m_{1/2})$ plane in the CMSSM
after implementing the constraints from ATLAS MET searches, precision electroweak.data,
flavour physics, $g_\mu - 2$, $m_h$ and dark matter.
The results of the current CMSSM fit are indicated by solid lines
and a filled  star, and a fit to previous data is indicated by dashed lines and open stars.
The red lines denote $\Delta \chi^2 = 2.30$ contours (corresponding approximately to
the  68\%~CL), and the red lines denote $\Delta \chi^2 = 5.99$ (95\%~CL) contours~\protect\cite{MC9}.
\label{fig:m0m12}}
\end{center}
\end{figure}

Fig.~\ref{fig:m0m12} displays the regions of the $(m_0, m_{1/2})$ plane
alowed at the 95\% CL (blue lines) and favoured at the 68\% CL (red lines) after taking
all these constraints into account~\cite{MC9}. The solid lines and filled star are obtained using the
current 20/fb ATLAS constraints, and the dashed lines and open star are based on the
previous constraints from 7/fb of LHC data at 7~TeV. The $m_h$ constraint has the 
effect of favouring relatively large values of $m_{1/2}$ beyond the reach of the direct
LHC searches for supersymmetric particles, which have an impact only at low values
of $m_{1/2}$. We note that the $m_h$
constraint is relatively independent of $m_0$. Large values of $m_{1/2}$ are excluded
by the dark matter density constraint.

\begin{figure*}[htb!]
\resizebox{8cm}{!}{\includegraphics{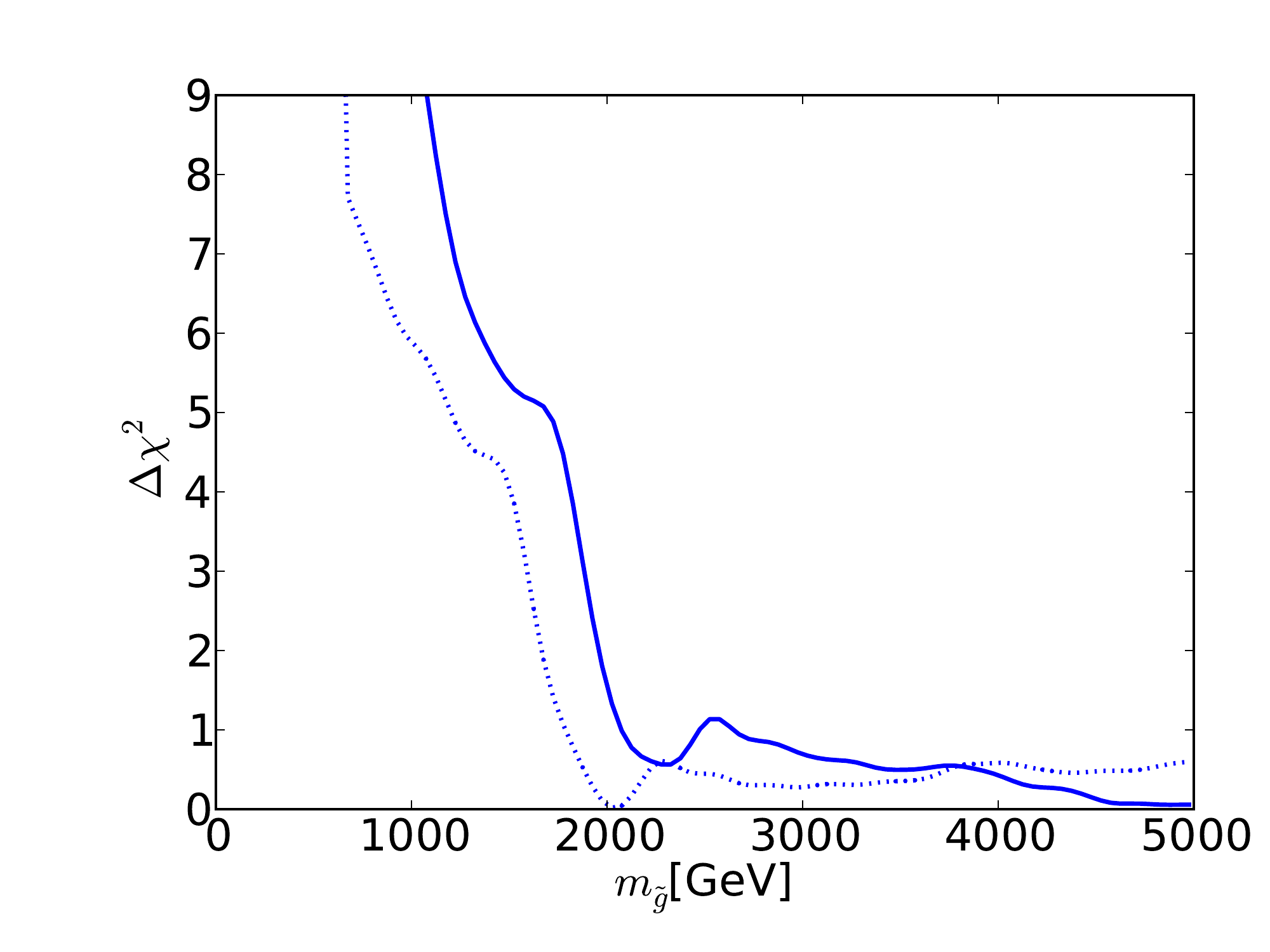}}
\resizebox{8cm}{!}{\includegraphics{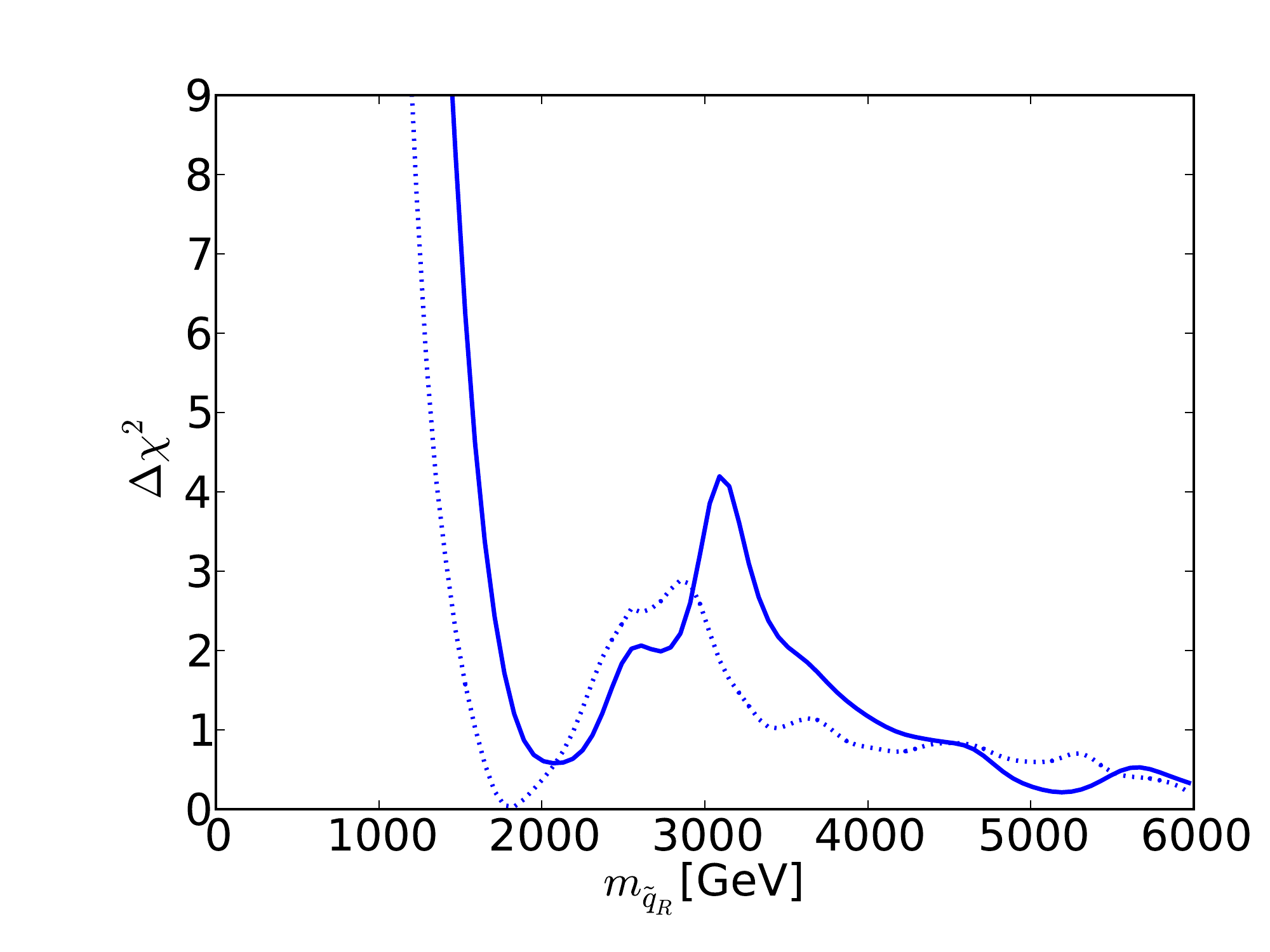}}
\resizebox{8cm}{!}{\includegraphics{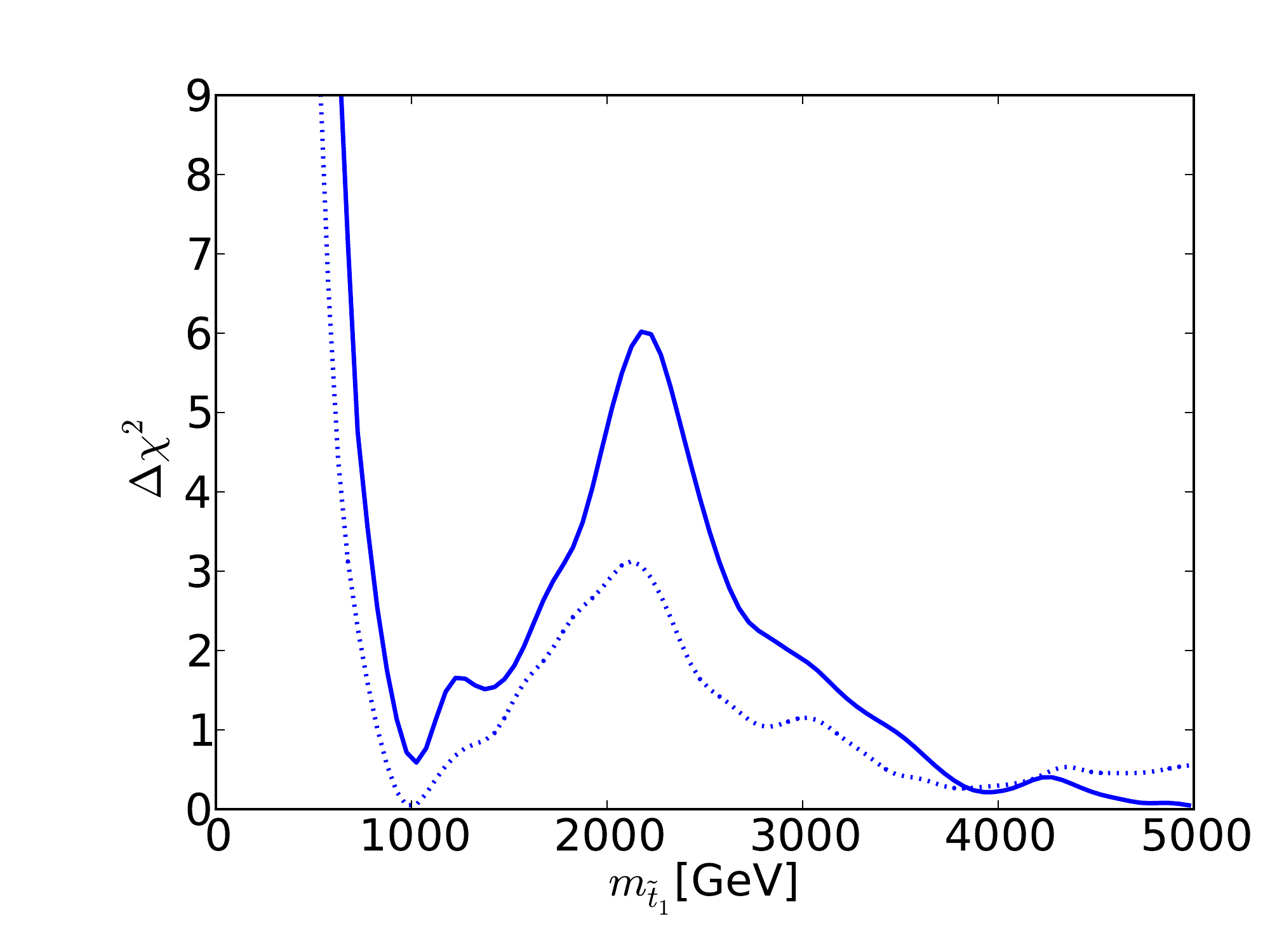}}
\resizebox{8cm}{!}{\includegraphics{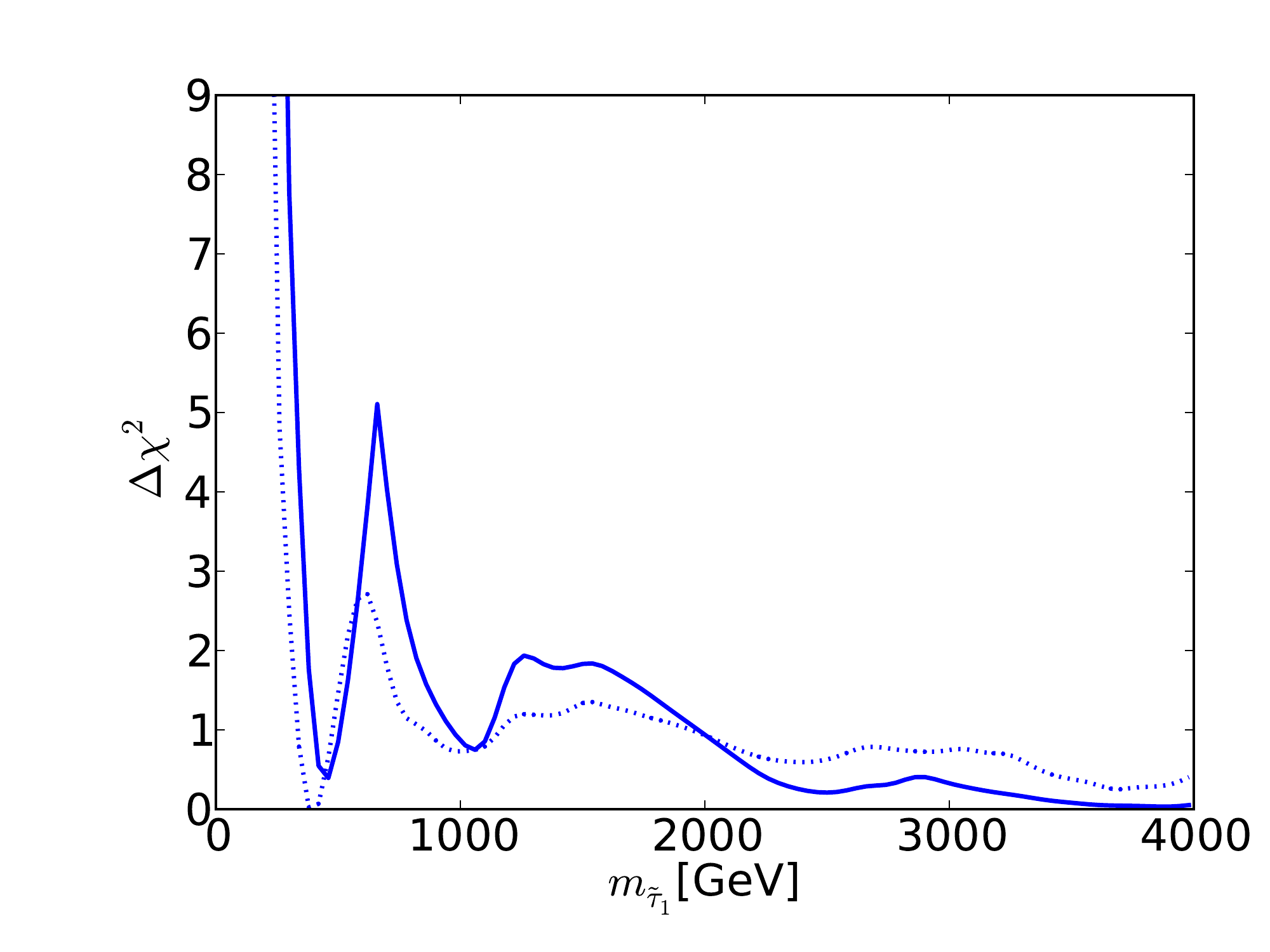}}
\vspace{-1cm}
\caption{\it The one-dimensional $\chi^2$ likelihood functions in the CMSSM for
$m_{\tilde g}$ (upper left), $m_{\tilde q_R}$ (upper right), $m_{\tilde t_1}$ (lower left) and $m_{\tilde \tau_1}$
(lower right). In each panel, the solid line is derived from a global analysis of the
present data, and the dotted line is derived from an analysis of a previous data set, using current implementations of the constraints
applied there~\protect\cite{MC9}.} 
\label{fig:masses}
\end{figure*}

The one-dimensional $\chi^2$ function for the gluino mass $m_{\tilde g}$ resulting
from this analysis of the CMSSM is shown in the upper left panel of Fig.~\ref{fig:masses}~\cite{MC9}. Again, the solid line
is based on the current data set and the dotted line is based on the previous data
set. We see that updating from the 7/fb 7-TeV data to the 20/fb 8-TeV data
does not change the $\chi^2$ function substantially. The current 95\% CL lower
limit on $m_{\tilde g} \sim 1350$~GeV. A similar plot for the mass of a generic supersymmetric
partner of a right-handed quark is shown in the upper right panel of Fig.~\ref{fig:masses}.
In this case, the 95\% CL lower limit is $m_{\tilde q_R} \sim 1650$~GeV. The lighter
supersymmetric partner of the top quark may be significantly lighter, as shown in the
lower left panel of Fig.~\ref{fig:masses}, with a 95\% CL lower limit $m_{\tilde t_1} \sim 750$~GeV.
Finally, the corresponding plot for the lighter supersymmetric partner of the $\tau$ lepton
is shown in the lower right panel of Fig.~\ref{fig:masses}. This is expected to be the
next-to-lightest supersymmetric particle, after the dark matter particle $\chi$, and may
have a mass as low as $330$~GeV at the 95\% CL.

\begin{figure}[htbp!]
\begin{center}
\includegraphics[height=7cm]{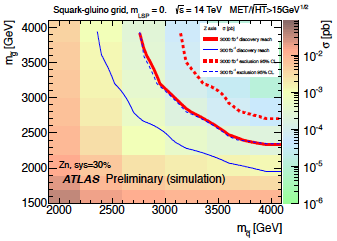}
\caption{\it The physics reach of the LHC in the $(m_0, m_{1/2})$ plane provided by
searches for squarks and gluinos assuming that the LSP mass is negligible. 
the different colours represent the production cross section at 14~TeV. The solid (dashed) lines
display the 5-$\sigma$ discovery reach (95\% CL exclusion limit) with 300/fb and 3000/fb respectively~\protect\cite{ATLAS-HL-LHC}.
\label{fig:ATLASglsq}}
\end{center}
\end{figure}

Ref.~\cite{ATLAS-HL-LHC} provides estimates of the supersymmetry discovery reach of the LHC with 14~TeV,
e.g., the $(m_0, m_{1/2})$ plane displayed in Fig.~\ref{fig:ATLASglsq}. As seen there,
the 5-$\sigma$ discovery reach for squarks and gluinos with 300/fb of luminosity
should be to $m_{\tilde g} \sim 3500$~GeV and $m_{\tilde q_R} \sim 2000$~GeV in the CMSSM,
and the discovery range with 3000/fb of luminosity
would extend a few hundred GeV further. Thus, large parts of the CMSSM parameter space will be
accessible in future runs of the LHC~\footnote{It should be emphasized that the 
likelihood estimates made here are specific to the models studied, as are the estimates of
the physics reaches.}. The priorities and prospects for future colliders will depend
whether the LHC discovers supersymmetry during its runs at 14~TeV, but certainly
more detailed studies of the Higgs boson will be on the agenda of the LHC and future
accelerators, as discussed in the last Section of these lectures.

\subsection{What Accelerator Next: a Higgs Factory?}

One of the possible options for a future accelerator is a `Higgs Factory'
designed to study the Higgs boson in detail. In fact, we already have a
Higgs Factory, namely the LHC, which has already produced millions
of Higgs bosons, though only a small fraction of them have been observed.
High-energy runs of the LHC will start in 2015, and luminosity upgrades are
being planned. Accordingly, the capabilities of the LHC are being rethought
from the Higgs Factory perspective. However, although the upgraded LHC
will produce orders of magnitude more Higgs bosons, its capabilities are
limited by theoretical uncertainties in the production cross section as well as
by the backgrounds that render unobservable some interesting decay modes.

Both these shortcomings would be avoided at a lepton collider, and various
options are being considered. The most mature concept is a
linear $e^+ e^-$ collider, and two projects are being developed: the
ILC that would operate initially at energies up to 500 GeV~\cite{ILC}, and
CLIC that could possibly operate at energies up to 3 TeV~\cite{CLIC}, where the Higgs
production cross section would be larger. 

In the last couple of years, the
alternative possibility of a circular $e^+ e^-$ collider has been revived~\cite{TLEP}.
Proposals include LEP3, a design for a high-luminosity $\sqrt{s} = 240$~GeV
$e^+ e^-$ collider that could be installed in the LHC tunnel, and TLEP, a
$\sqrt{s} = 350$ to 500~GeV collider that could be installed in a larger tunnel
with a circumference of 80 to 100~km. Also under consideration is a $\mu^+ \mu^-$
collider, that would benefit from the (expected) larger coupling of the Higgs to the
muon. Finally, there is the idea of a photon-photon collider, for example SAPPHiRE~\cite{SAPPHiRE}
that would exploit the recirculating linear accelerators proposed for the LHeC
electron-proton collider.

\begin{figure}[htbp!]
\begin{center}
\includegraphics[height=10cm]{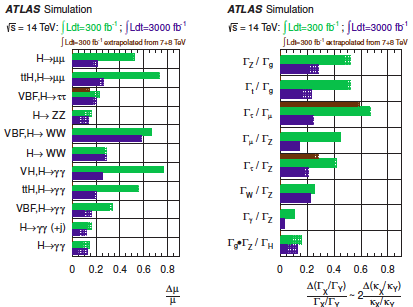}
\caption{\it Summary of ATLAS Higgs analysis sensitivities to signal strengths $\mu$ (left)
and ratios of partial decay widths (right) with integrated luminosities of 300/fb (green)
and 3000/fb at $\sqrt{s} = 14$~TeV for a Standard Model Higgs boson with a mass of 125~GeV~\protect\cite{ATLAS-HL-LHC}.
\label{fig:ATLASH}}
\end{center}
\end{figure}

Fig.~\ref{fig:ATLASH} displays ATLAS estimates of the measurement uncertainties in Higgs
signal strengths $\mu$ (left) and ratios of partial decay widths (right) with integrated
luminosities of 300/fb (green) and 3000/fb (blue)~\cite{ATLAS-HL-LHC}~\footnote{The possible improvements
in $\tau^+ \tau^-$ measurements that could be provided by a more complete analysis are
shown in brown.}. We see good prospects for significant improvements with 300/fb relative to the
current measurements, and for further improvements with 3000/fb that would enable several
Higgs couplings to be measured with accuracies $\lesssim 10$~\%.

\begin{figure}[htbp!]
\begin{center}
\includegraphics[height=7cm]{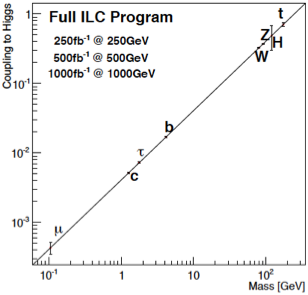}
\caption{\it Summary of the possible ILC accuracies for measurements of the Higgs
couplings to other particles that could be obtained by combining data at 250,
500 and 1000~GeV~\protect\cite{ILC}.
\label{fig:ILCH}}
\end{center}
\end{figure}

Fig.~\ref{fig:ILCH} displays estimates of the accuracies of measurements  of
the Higgs couplings to other particles that
would be possible with the ILC~\cite{ILC}, combining data from $\sqrt{s} = 250, 500$ and 1000~GeV.
This figure exhibits the prospective improvements in testing the linear dependence of the
couplings on the other particle masses expected in the Standard Model.

\begin{figure}[htbp!]
\begin{center}
\includegraphics[height=7cm]{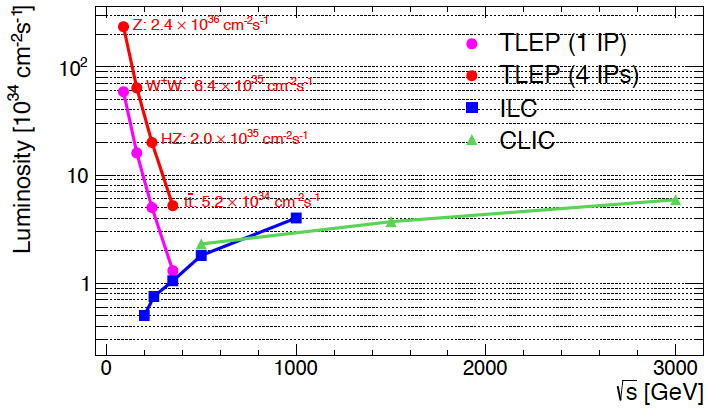}
\caption{\it Comparison of the luminosities estimated for TLEP with either one
or four interaction points (IPs), CLIC and the ILC (which would have only a single
IP) as functions of $\sqrt{s}$~\protect\cite{FirstLook}.
\label{fig:lumis}}
\end{center}
\end{figure}

As compared to linear $e^+ e^-$ colliders, circular colliders possess the
feature that the achievable luminosity increases at lower energies, assuming
that a fixed amount of power can be supplied to the beams. This feature is illustrated in
Fig.~\ref{fig:lumis} for the cases of TLEP, CLIC and the ILC~\cite{FirstLook}. Another feature of a
circular $e^+ e^-$ collider is that it can accommodate multiple interaction points (IPs),
whereas a linear collider has only a single IP, possibly with multiple detectors
operated alternately in push-pull mode.

\begin{figure}[htbp!]
\begin{center}
\includegraphics[height=7cm]{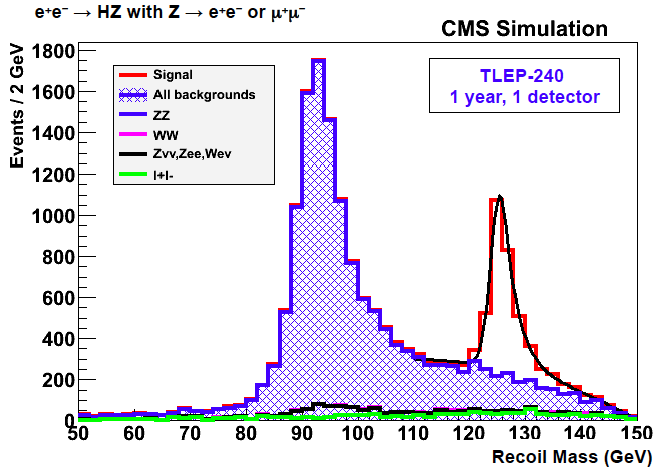}
\caption{\it Example of a simulation of the process $e^+ e^- \to H + Z$
followed by $H \to e^+ e^-, \mu^+ \mu^-$ at TLEP using a simulation of the
CMS detector that was designed for LHC physics~\protect\cite{FirstLook}.
\label{fig:TLEPHZ}}
\end{center}
\end{figure}

The experimental conditions at circular and linear $e^+ e^-$ colliders are
similar, with the difference that the bean energies are spread by beamstrahlung
in the linear case and by synchrotron radiation in the circular case, which
yields a lower probability of large energy loss. Preliminary studies of possible
Higgs measurements at TLEP have been made with simulations of the CMS
detector that was designed for LHC physics~\cite{FirstLook}. Fig.~\ref{fig:TLEPHZ} shows one
example, that of the process $e^+ e^- \to H + Z$ followed by $H \to e^+ e^-, \mu^+ \mu^-$:
a detector specifically designed for $e^+ e^-$ collisions such as those developed for the ILC
or CLIC would undoubtedly improve on these measurements.

\begin{figure}[htbp!]
\begin{center}
\includegraphics[height=7cm]{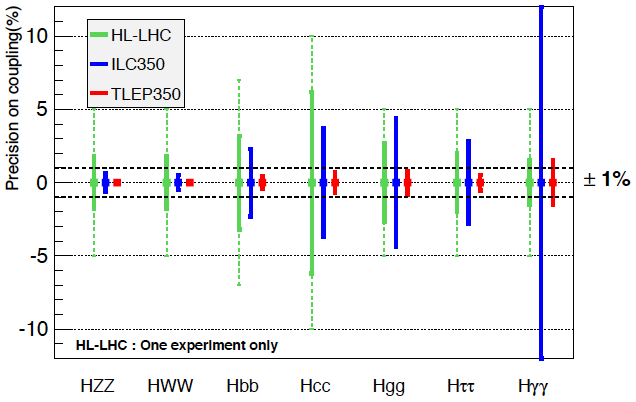}
\caption{\it Comparison of the estimated uncertainties of measurements of
Higgs couplings with the high-luminosity upgrade of the LHC (HL-LHC, green),
the ILC (blue) and TLEP (red) operating at $\sqrt{s} = 350$~GeV~\protect\cite{FirstLook}.
\label{fig:Hcomparison}}
\end{center}
\end{figure}

Fig.~\ref{fig:Hcomparison} shows a comparison of the estimated uncertainties in possible measurements of
Higgs couplings with the high-luminosity upgrade of the LHC at $\sqrt{s} = 14$~TeV (HL-LHC, green),
the ILC (blue) and TLEP (red) operating at $\sqrt{s} = 350$~GeV~\cite{FirstLook}. In the case of the
HL-LHC, just one experiment is included, and the dashed lines neglect the possibility
of improved theoretical calculations leading to reduced theoretical uncertainties. In each case, we see that TLEP
could provide an accuracy considerably superior to that of the ILC, which is traceable
directly to the higher statistics made available by the higher luminosity of TLEP visible in
Fig.~\ref{fig:lumis}.

\begin{figure}[htbp!]
\begin{center}
\includegraphics[height=7cm]{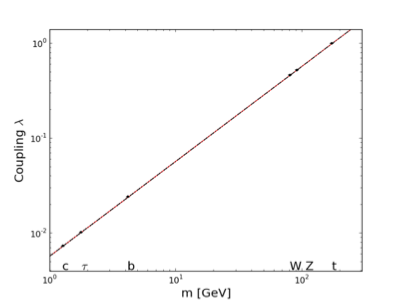}
\caption{\it The result of a two-parameter $(M, \epsilon)$
fit (\protect\ref{Mepsilon}) to the TLEP
coupling measurements listed in Figs.~\protect\ref{fig:Hcomparison},
assuming the same central values as the Standard Model, to be compared with the left panel of Fig.~\ref{fig:Mepsilon}.
\label{fig:TLEPvsmass}}
\end{center}
\end{figure}

Fig.~\ref{fig:TLEPvsmass} shows the result of a two-parameter $(M, \epsilon)$
fit (\ref{Mepsilon}) to the TLEP
coupling measurements listed in Figs.~\ref{fig:Hcomparison},
assuming the same central values as the Standard Model, which yields
\begin{equation}
M \; = \; 246.0 \pm  0.8~{\rm GeV}, \; \epsilon \; = \; 0.0000^{+0.0015}_{-0.0010} \, ,
\label{TLEPMepsilon}
\end{equation}
offering the possibility of probing the Standard Model couplings at the few per-mille
level. From this analysis we see that TLEP offers peerless capabilities for measuring
Higgs properties. It is worth mentioning that TLEP also offers unique possibilities
for other precision measurements of the Standard Model, e.g., at the $Z^0$ peak
and in $W^+ W^-$ production. However, the full exploitation of these accurate
measurements will require a new generation of high-precision calculations within
the Standard Model, posing a challenge to the theoretical community. 

\begin{figure}[htbp!]
\begin{center}
\includegraphics[height=10cm]{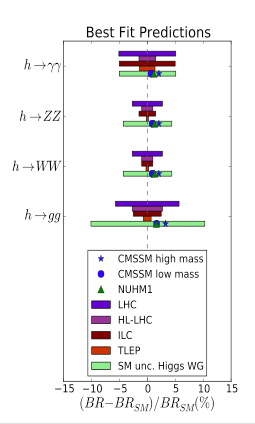}
\caption{\it From top to bottom: uncertainties in the measurements of Higgs branching ratios that may be
made at the LHC with 300/fb, the HL-LHC with 3000/fb, the ILC and TLEP, and finally the
current theoretical uncertainties within the Standard Model. Also shown
are the deviations from Standard Model predictions found in representative fits within
supersymmetric models~\protect\cite{FirstLook}.
\label{fig:uncertainties}}
\end{center}
\end{figure}

This point is exemplified in Fig.~\ref{fig:uncertainties}. The horizontal bars represent
the experimental accuracies with which various Higgs couplings can be measured
at the LHC with 300/fb of luminosity, at the HL-LHC with 3000/fb of luminosity, at the
ILC and at TLEP. Also shown are the deviations from the Standard Model predictions
for various Higgs branching ratios calculated in typical supersymmetric fits within the models described in the previous
Section. The good news is that TLEP would have sufficient precision to distinguish these
models from the Standard Model. Unfortunately, there is some bad news as well.
Fig.~\ref{fig:uncertainties} that shows the current theoretical uncertainties in
these branching ratios quoted by the LHC Higgs cross-section working group~\cite{LHXSWG}
dwarf the TLEP experimental uncertainties. More precise theoretical calculations will
be sorely needed.

\subsection{A Vision for the Future}

It would be premature to decide on the top priority for a possible future large
collider before we see at least some first results from the LHC at 13/14~TeV.
The physics landscape will be completely different if supersymmetric particles
or some other new physics is discovered at the TeV scale. In my opinion, in
that case it would be a mistake to invest the world's particle physics resources in a
collider incapable of studying the new TeV-scale physics.

Beyond any reasonable doubt, the ATLAS and CMS experiments at the
LHC have discovered a Higgs boson, but it remains an open question whether
there may be others. However, the existence of an (apparently) elementary
scalar boson poses a big challenge for theoretical physics. Most of the
responses to this challenge postulate some new physics at the TeV scale.
There is some puzzlement that no such new physics turned up in the
first LHC run at 7 and 8~TeV, but it is too soon for disappointment, still less despair.
The LHC has broad possibilities for discovering new physics beyond the Standard
Model when it restarts at 13/14~TeV. If it does discover new physics at the TeV
scale, the top priority will be to study it, and beyond the LHC,
a very-high-energy $pp$ collider may offer the best prospects for long-term
studies of this new physics. If the LHC does not discover more new physics, 
it would be natural to focus on studies of the Higgs boson that has already been
discovered, in which case TLEP offers the best prospects, also for other 
high-precision physics.

The TLEP project is part of a vision for the future of particle physics that combines
indirect exploration of possible new physics at the 10-TeV scale in $e^+ e^-$
collisions with direct exploration of this energy scale in very-high-energy $pp$
collisions at $\sqrt{s} \lesssim 100$~TeV in the same tunnel with a circumference
of 80 to 100~km~\cite{FCC}. The communities
interested in these complementary exploratory projects should work together
to realize this vision, whose physics case will require a major effort to develop
and convince those who control the global resources for scientific research.

\section*{Acknowledgements}

I thank Tevong You for his collaboration on topics discussed in these lectures. I also
thank fellow members of the MasterCode Collaboration, particularly
Oliver Buchmueller, Sven Heinemeyer, Jad Marrouche, Keith Olive and Kees de Vries
for many discussions. I also thank members of the TLEP Study Group, including
Alain Blondel, Patrick Janot, Mike Koratzinos and Frank Zimmermann for many discussions.
This work was supported in part by the London Centre for Terauniverse Studies (LCTS),
using funding from the European Research Council 
via the Advanced Investigator Grant 267352.

\end{document}